# On Ni-Sb-Sn based skutterudites

W. Paschinger[1], P.F. Rogl[1,2,#], G. Rogl[2], A. Grytsiv[2,3], E. Bauer[2,3], H. Michor[3], Ch. Eisenmenger-Sitter[3], E. Royanian[3], P.R. Heinrich[3], M. Zehetbauer[4], J. Horky[4], S. Puchegger[5], M. Reinecker[6], G. Giester[7], P. Broz[8], A. Bismarck[1,2,9]

[1] Institute of Physical Chemistry, University of Vienna, Währinger Straße 42, A-1090 Vienna, Austria

[2] Christian Doppler Laboratory for Thermoelectricity, TU-Wien, Wiedner Hauptstr. 8, A-1040 Vienna, Austria

[3] Institute for Solid State Physics, TU-Wien, Wiedner Hauptstr. 8, A-1040 Vienna, Austria

[4] Research Group Physics of Nanostructured Materials, University of Vienna, Boltzmanngasse 5, A-1090 Vienna, Austria

[5] Faculty Center for Nanostructure Research, University of Vienna, Boltzmanngasse 5, A-1090 Vienna, Austria

[6] Research Group Physics of Nanostructured Materials, University of Vienna, Boltzmanngasse 5, A-1090 Vienna, Austria

[7] Institute of Mineralogy and Crystallography, University of Vienna, Althanstr. 14 (UZA 2), A-1090 Vienna, Austria

[8] Faculty of Science, Deparment of Chemistry, Masaryk University, Kotlářská 267/2, 611 37 Brno, Czech Republic

[9] Polymer and Composite Engineering (PaCE) group, Department of Material Chemistry, University of Vienna, Währinger Straße 42, A-1090 Wien, Austria

**Abstract**

Novel filled skutterudites $Ep_yNi_4Sb_{12-x}Sn_x$ (Ep = Ba and La) have been prepared by arc melting followed by annealing at 250°C, 350°C and 450°C up to 30 days in sealed quartz vials. A maximum filling level of $y = 0.93$ and $y = 0.65$ was achieved for the Ba and La filled skutterudite, respectively. Extension of the homogeneity region, solidus temperatures and structural investigations were performed for the skutterudite phase in the ternary Ni-Sn-Sb and in the quaternary Ba-Ni-Sb-Sn system. Phase equilibria in the Ni-Sn-Sb system at 450°C were established by means of Electron Probe Microanalysis (EPMA) and X-ray Powder Diffraction (XPD). Single-phase samples with the composition $Ni_4Sb_{8.2}Sn_{3.8}$, $Ba_{0.42}Ni_4Sb_{8.2}Sn_{3.8}$ and $Ba_{0.92}Ni_4Sb_{6.7}Sn_{5.3}$ were employed for measurements of the physical properties i.e. temperature dependent electrical resistivity, Seebeck coefficient and thermal conductivity. Resistivity data showed a crossover from metallic to semiconducting behaviour, which is discussed in terms of a temperature-dependent carrier concentration employing a simple model for a rectangular density of states with the Fermi energy slightly below a narrow gap. The corresponding gap width was extracted from maxima in the Seebeck coefficient data as a function of temperature. Temperature dependent single crystal X-ray structure analyses (at 100 K, 200 K and 300 K) revealed the thermal expansion coefficients, Einstein and Debye temperatures for two selected samples $Ba_{0.73}Ni_4Sb_{8.1}Sn_{3.9}$ and $Ba_{0.95}Ni_4Sb_{6.1}Sn_{5.9}$. These data compare well with Debye temperatures from measurements of specific heat (4.4 K < T < 200 K). Several mechanical properties were measured and evaluated. Elastic moduli, collected from Resonant Ultrasonic Spectroscopy (RUS) measurements, range from 100 GPa for $Ni_4Sb_{8.2}Sn_{3.8}$ to 116 GPa for $Ba_{0.92}Ni_4Sb_{6.7}Sn_{5.3}$. Thermal expansion coefficients (capacitance dilatometry and DMA) are $11.8\times10^{-6}$ $K^{-1}$ for $Ni_4Sb_{8.2}Sn_{3.8}$ to $13.8\times10^{-6}$ $K^{-1}$ for $Ba_{0.92}Ni_4Sb_{6.7}Sn_{5.3}$. Room temperature Vicker's hardness values (up to a load of 24.5 mN) vary within the range of 2.6 GPa to 4.7 GPa. Severe plastic deformation (SPD) via high-pressure torsion (HPT) was used to introduce nanostructuring. Physical properties before and after HPT were compared, showing no significant effect on the material's thermoelectric behaviour.

---

[#] Author to whom any correspondence should be addressed.
 e-mail: peter.franz.rogl@univie.ac.at

**Keywords:** Skutterudites, Physical properties, Mechanical properties, Nanomaterials, Sever plastic deformation (SPD) via High-pressure torrsion

## 1. Introduction

Skutterudite-based materials have been in the field of research for a long time [Uhe1] because they show a large variety of physical properties particularly interesting for commercial thermoelectric applications [Sny1], for which high figures of merit

$$zT = \frac{S^2 \cdot T}{\rho \cdot \kappa} \qquad (1.1)$$

for p-type as well as for n-type configuration are a precondition [Shi1, Rog5, Rog6, Rog7]. Skutterudites crystallize in the cubic $CoAs_3$ structure (space group $Im\bar{3}$) with the general formula $F_xT_4M_{12}$ where T is a transition metal of the VIII$^{th}$ group located in position 8c (¼, ¼, ¼) and M is a pnictogen, chalcogen or an element of the IV$^{th}$ main-group in Wyckoff site 24g (0, y, z). These atoms form a cage like structure with a large icosahedral hole in the 2a site (0, 0, 0), which may accommodate various filler atoms F, including alkaline and alkaline earth metals, lanthanides, actinides, as well as halogens or in particular cases Y, Hf [Hor1, Hor2], Pb and Sn [Nol1, Tak1, Tak2, Tak3]. A ternary skutterudite in the system Ni-Sn-Sb was first reported by Grytsiv *et al.* [Gry1], who defined a wide homogeneity range at 250°C and 350°C by establishing the isothermal sections in the subsystem $Sn-Sb-NiSb-Ni_3Sn_2$ at these temperatures. Investigations in the Sn-rich part of the Ni-Sn-Sb phase diagram were done by Mishra *et al.* [Mis] and suggests that the phase equilibria determined by Grytsiv *et al.* [Gry1] lie within 10°C above or below the declared temperatures.

It has to be noted, that many reports (see for example [Gry1, Rog1, Rog2, Rog3, Rog4, Rog12, Mal1]) describe mixed occupancies for all three crystallographic sites (24g, 8c, 2a). For Ni-Sn-Sb based skutterudites the structure with Ni-atoms fully occupying the 8c site seems to be stabilized by a random distribution of Sb and Sn atoms in the 24g position, because in the binary systems Ni-{Sn, P, As, Sb} only the skutterudite $NiP_3$ exists as a high temperature modification [Llu1, Jei1]. A special situation occurs for Sn-atoms, which may occupy the 24g site, but may simultaneously enter the 2a site of the same compound reaching filling levels of 0.21 in this position [Gry1]. Ternary and isotypic quaternary skutterudites with Eu and Yb as filler elements have been characterised by their physical properties as well as by Raman- and Mössbauer-spectroscopy, unambiguously revealing the fact, that a small amount of Sn is able to enter the 2a (0, 0, 0) site [Gry1]. The filler atom in 2a is loosely bonded in the large icosahedral cage and is generally believed to decrease the thermal conductivity of the material via rattling modes [Tob1, Uhe1]. Not much information is available on the thermal stability of skutterudites. To our knowledge the melting temperatures are determined only for two binary skutterudites $CoAs_3$ [Ish1], $CoSb_3$ [Oka1], however, there are no data either on the influence of a filler on the melting points nor on the thermal stability of ternary or multi-component skutterudites.

The current article will focus on a series of tasks outlined below:

(i) the Ni-Sn-Sb isothermal section at 450°C and the extension of the ternary skutterudite phase $Ni_4Sb_{12-x}Sn_x$ will be discussed in comparison with data reported by Grytsiv *et al.* [Gry1].

(ii) temperature dependent filling levels of the skutterudites $Ep_yNi_4Sb_{12-x}Sn_x$ with $Ep_y$ being Ba and La will be presented.

- (iii) from a combination of literature data and DTA measurements the solidus surface will be established for the ternary and quarternary skutterudites $Ni_4Sb_{12-x}Sn_x$ and $Ba_yNi_4Sb_{12-x}Sn_x$ in dependence of composition.
- (iv) for Ba-filled skutterudite $Ba_yNi_4Sb_{12-x}Sn_x$ the structure will be determined by single crystal X-ray structure analyses.
- (v) physical property measurements such as temperature dependent resistivities, thermal conductivities and Seebeck coefficients will be used to characterise the thermoelectric behaviour of the three single-phase samples $Ni_4Sb_{8.2}Sn_{3.8}$, $Ba_{0.42}Ni_4Sb_{8.24}Sn_{3.8}$ and $Ba_{0.92}Ni_4Sb_{6.7}Sn_{5.3}$.
- (vi) the effect of SPD via HPT on the material will be discussed.
- (vii) mechanical properties such as thermal expansion, elastic moduli and Vicker´s hardness of $Ni_4Sb_{8.2}Sn_{3.8}$, $Ba_{0.42}Ni_4Sb_{8.3}Sn_{3.8}$ and $Ba_{0.92}Ni_4Sb_{6.7}Sn_{5.3}$ will be studied.
- (viii) based on all the data acquired, the thermoelectric behaviour will be discussed in terms of the electron and phonon mean free path.

## 2. Experimental

*Sample preparation*

Starting materials were elemental pieces of Ba, La, Ni, Sb, Sn, all of 99.95 mass% minimum purity. Samples to establish phase equilibria and homogeneity regions as well as for physical property studies were prepared by one of the following optimised melting reactions directly gaining the ternary $Ni_4Sb_xSn_{12-x}$ alloys, whereas for the quaternary skutterudites the filler elements were added in a second reaction step:

i) bulk alloys, each with a total weight of 1-2 g, were synthesized via a $Ni_4Sb_xSn_{12-x}$ master alloy by argon arc melting of metal ingots on a water-cooled copper hearth, adding the filler element to the ternary alloy by the same reaction method.

ii) samples with the nominal composition $Ni_4Sb_xSn_{12-x}$ were prepared from stoichiometric amounts of high purity Sb and Sn pieces and fine Ni wire. After mixing the material was sealed into evacuated quartz tubes, heated to 980°C, kept there in liquid state for half an hour prior to quenching the capsules in air. For quaternary samples the whole procedure was repeated, adding pieces of La or Ba to the ternary master alloys.

Total weight losses of 1-3 mass% that occurred during sample preparation were attributed to the high vapour pressures of Sb and Ba and were compensated by adding additional 3-5 mass% of Ba and Sb.

All samples were then sealed in evacuated quartz tubes, annealed at 250 °C, 350 °C or 450°C for 3 to 30 days to reach equilibrium conditions prior to quenching in cold water.

The material for single-phase samples of 6-8 g was ground below at least 100 μm inside a glove box using a WC mortar followed by ball-milling in a Fritsch planetary mill (Pulverisette 4) with balls of 1.6 mm, rotation speed 250, ratio -2.5 for 2 h to gain a nano crystalline powder. These powders were then loaded into graphite cartridges for hot-pressing under 1 bar of 5N-argon in a FTC uniaxial hotpress system (HP W 200/250 2200-200-KS).

Sb-rich specimens, $Ni_4Sb_{8.2}Sn_{3.8}$ and $Ba_{0.42}Ni_4Sb_{8.2}Sn_{3.8}$, were directly prepared by hot pressing at 450°C resulting in a densely compacted single-phase material.

For the Sn-rich sample $Ba_{0.92}Ni_4Sb_{6.7}Sn_{5.3}$ a different preparation method had to be chosen. 5-10 mass% of extra Sb and Sn were added from the very beginning of the syntheses. The ball-milled powder was pre-compacted and annealed under argon at 450°C over night inside the hotpress, followed by hotpressing at 430°C in order to squeeze out the surplus of $Sb_2Sn_3$.

Sn-rich single crystals $Ba_{0.95}Ni_4Sb_{6.1}Sn_{5.9}$ were prepared from Sn flux, Sb-rich $Ba_{0.73}Ni_4Sb_{8.1}Sn_{3.9}$ single crystals from $Sb_2Sn_3$ flux by heating the starting materials with compositions 2Ba-12Ni-Sb22.5-Sn63.5 and 2Ba-8Ni-45Sb-45Sn, respectively, to 950°C to reach the liquid state followed by cooling at a rate of 12°C h$^{-1}$ to 450°C. Alloys were kept at this temperature for 3 days to reach thermodynamic equilibrium. Subsequently the samples were removed from the furnace and the Sn-rich sample was treated in concentrated hot HCl acid to dissolve the Sn-rich matrix. In case of the Sb-rich single crystals the $Sb_2Sn_3$-rich matrix was removed by treating the sample at first with hot concentred $HNO_3$ and afterwards washing the crystals with cold concentrated HCl.

*Sample characterisation*

X-ray powder diffraction (XPD) data were collected using a Huber Guinier camera with monochromatic Cu $K_{\alpha_1}$ radiation ($\lambda$ = 0.154056 nm) and an image plate recording system (model G670). Pure Si ($a_{Si}$ = 0.5431065 nm) was used as internal standard to determine precise lattice parameters via least-squares fitting to the indexed 2θ values employing the program STRUKTUR [Wac1]. For quantitative Rietveld refinements and to calculate the filling levels y we applied the FULLPROF program [Rod1]. Chemical compositions of the different specimen were extracted from electron probe microanalyses (EPMA) using energy-dispersive X-ray (EDX) spectroscopy with an INCA Penta FETx3-Zeiss SUPRA55VP equipment (Oxford Instruments).

X-ray intensity data for the two single crystals, after inspection on an AXS-GADDS texture goniometer for quality and crystal symmetry, were collected on a four-circle Nonius Kappa diffractometer with a charge-coupled device (CCD) area detector and graphite-monochromated Mo $K_\alpha$ radiation ($\lambda$ = 0.07107 nm), at three different temperatures (100 K, 200 K and 300 K), under a flow of equilibrated nitrogen gas from a cryostat. Orientation matrix and unit cell parameters were derived using the program DENZO (Nonius Kappa CCD program package; Nonius, Delft, The Netherland). Employing the SHELXS-97 and SHELXL-97 software [She1] for single-crystal x-ray diffraction data the structures were solved by direct methods and refined successfully.

Using Archimedes' principle and distilled water the density $\rho_S$ of the hot pressed samples were determined and compared with the calculated x-ray densities

$$\rho_{X-ray} = \frac{Z \cdot M}{N \cdot V} \quad (2.1)$$

where M is the molar mass, Z is the number of formula units per cell, N is Avogadros's number and V is the volume of the unit cell.

*Physical property measurements*

Electrical resistivity, Seebeck coefficient, thermal conductivity were measured (4 K < T < 300 K) using homemade equipments cooled by liquid He as described in detail in ref. [Bau1]. The Seebeck coefficient and electrical resistivity above room temperature were measured simultaneously with an ULVAC-ZEM 3 (Riko, Japan). The thermal conductivity above room temperature was calculated from the thermal diffusivity $D_t$ (measured by a laser flash method FLashline-3000, ANTER, USA), the specific heat $C_p$ and the density $\rho_s$ employing the relationship $\lambda = D_t \cdot C_p \cdot \rho_s$.

*Specific heat measurement*
Employing an adiabatic step heating technique, specific heat measurements from 2 to 140 K were performed on single phase samples with masses between 2.5 g and 5 g cooled with liquid He.

*Thermal expansion measurements*
The thermal expansion from 4.2 K to 300 K was measured with a miniature capacitance dilatometer [Rot1], using the tilted plate principle [Brä1, Gen1]. For this measurement, the sample is placed in a hole of the lower ring like capacitance plate made of silver, which is separated from the silver upper capacitor plate by two needle bearings. For the measurement of thermal expansion at a temperature range from 80 K to 420 K, a dynamic mechanical analyzer DMA7 (Perkin Elmer Inc.) was employed. The sample is positioned in a parallel plate mode with a quartz rod on top of the sample and data are gained using the thermodilatometric analysis (TDA). For further details see Refs. [Kop1, Sch1, Sch2, Wys1].

*Elastic Property measurements*
Resonant ultrasound spectroscopy (RUS) developed by Migliori *et al.* [Mig1] was used to determine elastic properties via the eigenfrequencies of the sample and the knowledge of the sample's mass and dimension. For this measurement the cylindrically shaped samples of 10 mm diameter and masses between 2.5 g and 5 g were mounted "edge-to-edge" between two piezo-transducers and were excited via a network analyser in the frequency range of 100 kHz to 500 kHz at room temperature. As the average symmetry of all skutterudite samples is isotropic, the so gained frequency spectrum is then fitted with calculated spectra and provides the values of Young's modulus E and Poisson's ratio ν.

*Hardness measurements*
The load-independent Vickers hardness HV was determined from measurements on an Anton Paar MHT 4 microhardness tester mounted on a Reichert POLYVAR microscope evaluating all the indention data using Eqn. 2.2.

$$\mathrm{HV} = 2 \cdot \mathrm{F} \cdot \frac{\sin\left(\frac{136°}{2}\right)}{d^2} \qquad (2.2)$$

F gives the indention load and d is defined as

$$d = \frac{d_1 + d_2}{2} \qquad (2.3)$$

with $d_1$ and $d_2$ being the resulting diagonal lengths of the indent.

*SPD via HPT*
The HPT technique is based on the use of a Bridgman anvil-type device. A thin disk-shaped sample is subjected to torsional straining under high pressure between two anvils at room temperature or at elevated temperatures via induction heating. The shear strain γ is dependent on the number of revolutions n, the radius r, and the thickness t of the specimen in the following way:

$$\gamma = \frac{2 \cdot \pi \cdot n \cdot r}{t} \tag{2.4}$$

Therefore the cylindrical single-phase samples were cut into slices of a thickness of about ~ 1 mm with diameter 10 mm. The Sb-rich alloys were processed at 400°C under a pressure of 4 GPa with 1 revolution, whereas the temperature was decreased to 300°C for the Sn-rich samples.

*DTA measurements*

Melting points were determined from DTA measurements in a NETSCH STA 409 C/CD equipment. Pieces of single-phase samples weighing 500 mg to 600 mg were sealed in evacuated quartz glass crucibles. Generally three heating and three cooling curves were recorded for each sample using a scanning rate of 5 K min$^{-1}$.

### 3. Filling levels, phase equilibria and homogeneity region

*Phase equilibria in ternary system Ni-Sn-Sb at 450°C*

The ternary skutterudite phase $Sn_yNi_4Sb_{12-x}Sn_x$ ($\tau$) exhibits a wide range of homogeneity at 250°C (2.4 ≤ x ≤ 5.6, 0 ≤ y ≤ 0.31) and at 350°C (2.7 ≤ x ≤ 5.0, 0 ≤ y ≤ 0.27) [Gry1]. It was shown that phase equilibria change drastically by increasing the temperature from 250°C to 350°C and the authors [Gry1] suggested that at least two invariant reactions exist in this temperature range. Mishra [Mis1] presented a detailed reaction scheme for the Sn rich part of the system and this reaction scheme involved three invariant equilibria in the temperature range from 250°C to 350°C. In order to derive the phase equilibria at 450°C several samples were annealed at this temperature for at least 30 days. Three-phase equilibria (Sb)+NiSb$_2$+$\tau$ and NiSb$_2$+$\alpha$+$\tau$ at 450°C (Fig. 3.1a,b) were found to be similar to those observed by Grytsiv *et al.* at 250°C and 350°C [Gry1]. We furthermore observed three-phase regions L+(Sb)+$\tau$, L+$\alpha$+$\tau$, L+Ni$_3$Sn$_2$+$\alpha$ (Fig. 3.1c,d,e), which agree well with the Schultz-Scheil diagram derived by Mishra [Mis1].

The solid solution $\alpha$ (NiSb-Ni$_3$Sn$_2$) was observed in all as-cast samples and has the widest liquidus field likely being the most stable phase in the system. Thus the as-cast alloy 2.7Ni-24.7Sn-72.6Sb shows primary crystallisation of the alpha-phase, which is surrounded by NiSb$_2$, and the last portion of liquid crystallizes with the composition 40Sn-60Sb. At 450°C the sample is in solid-liquid state (Fig. 3.1e): huge grains of the skutterudite (Ni$_4$Sb$_{9.1}$Sn$_{2.9}$) and antimony (Sb) were grown in equilibrium with the liquid with the composition 0.3Ni-47.75Sn-51.95Sb. Three-phase equilibria with the liquid phase L+Ni$_3$Sn$_2$+$\alpha$ and L+$\alpha$+$\tau$ are also well confirmed via investigation of the samples 21.5Ni-25.5Sn-53Sb and 35Ni-52Sn-13Sb (Fig. 3.1d,e). Between these two phase triangles a huge two-phase field arises containing Sn-rich liquid and the $\alpha$-phase (Fig. 3.1f). The latter solid solution, formed by congruent melting compounds, separates the phase diagram in two subsystems for which phase equilibria may then be investigated independently. The equilibria in the subsystem Ni-NiSb-Ni$_3$Sn$_2$ have not been reinvestigated but were introduced after Burkhardt and Schubert [Bur1] (Fig 3.2).

Composition and lattice parameters of the skutterudite phase $\tau$ coexisting in the various three-phase equilibria at 450°C are listed in Tab. 3.1. The isothermal section at 450°C is shown in Fig. 3.2.

At 450°C the homogeneity region of the ternary skutterudite phase $\tau$-$Sn_yNi_4Sb_{12-x}Sn_x$ extends for 2.4 ≤ x ≤ 3.2. Comparing the compositions of the ternary skutterudites $Ni_4Sb_{12-x}Sn_x$ as a function of the Sn-content from the homogeneity regions at 250°C and 350°C [Gry1] as well as at 450°C with those after DTA, a good agreement was found (see Fig 3.3). One may see a weak temperature dependence for the Sb-rich end of the skutterudite solid solution that coexists in three-phase solid-state equilibria: $(Sb)+NiSb_2+\tau$ and $NiSb_2+\alpha+\tau$. However, the Sn-rich side of the homogeneity region in equilibrium with the liquid phase shows a significant temperature dependence. With rising temperatures, the composition of the liquid gets strongly depleted on Sn, resulting in a pronounced phase segregation during the crystallization of the samples. Thus DTA of the single phase sample $Ni_4Sb_{8.2}Sn_{3.8}$ (Fig. 3.4a) reveals a solidus temperature of 410°C during the first heating (Fig. 3.5), however, the temperature was increased to 453°C after the second and third heating cycles (Fig 3.5). Furthermore additional thermal effects appear in the temperature range from 415°C to 423°C. EPMA of the sample after DTA (Fig. 3.6a) shows three phases α, β (SnSb) and skutterudite τ at an Sb enriched composition ($Ni_4Sb_{8.9}Sn_{3.1}$).

*Homogeneity region for the filled skutterudites $Ep_yNi_4Sb_{12-x}Sn_x$ (Ep = Ba and La)*

The filling of the 2a site in the skutterudite structure has a maximal effect on the reduction of the lattice thermal conductivity [Tob1, Uhe1]. In order to define the solubility of the electropositive (Ep) fillers in $Ni_4Sb_{12-x}Sn_x$ two filler atom species with different valences (Ba and La) were selected and samples with compositions 1Ba-4Ni-5Sb-7Sn, 1Ba-4Ni-8Sb-4Sn 1La-4Ni-4Sb-8Sn and 1La-4Ni-8Sb-4Sn were investigated in as cast state and after annealing at 250°C, 350°C and 450°C. Results of the quantitative Rietveld refinements for the occupancy in the 2a site in combination with EPMA measurements are shown in Fig. 3.6a,b. No skutterudites were observed in the as-cast alloys. The skutterudite phase is formed after annealing and appears in the microstructure as the main phase for one of the three temperatures 250°C, 350°C or 450°C (Fig 3.8). In all cases we observed an increase of the filling level with increasing Sn content reaching y = 0.93 for the Ba filled skutterudite, which additionally shows a clear dependence of the filling level on the temperature. Furthermore, for Sn-rich compositions EPMA reveals two sets of skutterudite compositions in the samples annealed at 350°C and 450°C. Such phase segregation likely is not connected with diffusion suppressed at these temperatures because the heat treatment at 250°C yields the formation of a skutterudite with uniform composition. In contrast to the Ba filled skutterudite the solubility of La in $Ep_yNi_4Sb_{12-x}Sn_x$ is lower and no temperature dependence was detected. Considering the interesting behaviour of the Ba filled skutterudite, this compound was selected for further detailed investigations at 450°C.

For several quaternary samples, annealed at 450°C, the Ba filling level in dependence of the Sn content was determined by Rietveld refinements combined with microprobe measurements and yielded a large homogeneity region of $Ba_yNi_4Sb_{12-x}Sn_x$ as shown in Fig. 3.9 in a two-dimensional (2D) projection. Obviously the maximal solubility of Ba in the skutterudite (y > 0.93) responds to two cases: (i) the equilibrium with the Sn-rich liquid and (ii) to the afore-mentioned crystallization of skutterudites with two compositions ($Ba_{0.94}Ni_4Sb_{6.1}Sn_{5.9}$ and $Ba_{0.93}Ni_4Sb_{5.1}Sn_{6.9}$) that coexist with $Ni_3Sn_4$. With increasing Sb-content the solubility decreases to y = 0 in the three-phase equilibra with (Sb) and $NiSb_2$ as well as with α and $NiSb_2$.

DTA measurements were performed to gather further information on the stability of $Ba_yNi_4Sb_{12-x}Sn_x$. For this investigation three single-phase samples were used: $Ba_{0.42}Ni_4Sb_{8.2}Sn_{3.8}$ and $Ba_{0.92}Ni_4Sb_{6.7}Sn_{5.3}$ (Fig. 3.4c,d) after ball milling and hot pressing, and the compound $Ba_{0.29}Ni_4Sb_{9.1}Sn_{2.9}$, which appeared to be single-phase after annealing at 450°C (Fig. 3.4b). For these alloys three heating and three cooling curves were recorded using a scanning rate of 5 K min$^{-1}$ (Fig. 3.10 to Fig. 3.12). In all first heating curves only one signal could be detected. However, the subsequent heating and cooling curves (s. Fig. 3.10 to Fig. 3.12) revealed additional thermal effects at lower temperatures that are associated with the incongruent melting of the alloys. This was confirmed by the microstructure of the samples after DTA (Fig 3.6b) to d)). In all cases we observe the formation of a high melting α-phase with subsequent crystallization of the β-phase and $BaSb_3$. A Tin rich liquid was found to form during crystallisation of $Ba_{0.92}Ni_4Sb_{6.7}Sn_{5.3}$. Due to incongruent crystallization of $Ba_yNi_4Sb_{12-x}Sn_x$ the composition of the skutterudite phase changes similarly to that observed for the ternary Ni-Sn-Sb skutterudite (see chapter above). The melting points, $T_m$, for the samples before and after DTA are listed in Tab. 3.2 and the compositional dependences of the solidus temperature are compared with data for the Ni-Sb-Sn skutterudite in Fig. 3.13. It is clearly seen that the filling of the skutterudite lattice with Ba-atoms results in a significant increase of the melting temperature and this influence is particularly visible for Sn-rich samples.

The knowledge on the extension of the homogeneity region for $Ba_yNi_4Sb_{12-x}Sn_x$ as a function of temperature was used to prepare single-phase samples for measurements of physical properties (Fig 3.4a,c,d and Tab. 3.3) as well as to grow single crystals for structural investigations.

## 4. Crystal structure and vibration modes of $Ba_yNi_4Sb_{12-x}Sn_x$

Rietveld refinements of X-ray powder data for $Ba_yNi_4Sb_{12-x}Sn_x$ were fully consistent with the skutterudite structure $CoAs_3$ in the unfilled case and $LaFe_4Sb_{12}$ in the filled case. The refinements combined with EPMA defined the degree of filling in the 2a site and the Sb/Sn ratio in the 24g site. Compositional data and corresponding lattice parameters of all new phases $Ba_yNi_4Sb_{12-x}Sn_x$ are summarized in Tab. 4.1. Lattice parameters revealed neither a dependence on the filling level of Ba nor on the Sb/Sn ratio.

For two flux-grown single crystals, $Ba_yNi_4Sb_{12-x}Sn_x$ (for details see micrographs in Fig. 4.1) X-ray diffraction intensities were recorded at three different temperatures (100 K, 200 K and 300 K). The refinement in all cases proved isotypism with the $LaFe_4Sb_{12}$ type (filled skutterudite; space group $Im\overline{3}$) with Ni atoms occupying the 8c site while Sb and Sn atoms randomly share the 24g site. The residual density at the 2a site was assigned to Ba atoms. The compositions derived from structure refinement, namely $Ba_{0.73}Ni_4Sb_{8.1}Sn_{3.9}$ and $Ba_{0.95}Ni_4Sb_{6.1}Sn_{5.9}$, agree well with the microprobe measurements. Final refinements with fixed occupancies (Table 4.2a) and anisotropic atom displacement parameters (ADPs) led to a reliability factor $R_F$ below 2%. The maximum residual electron density of ~3 e/Å$^3$ appears at a distance of 1.5 Å from the centre of the 24g site. This density can be interpreted as a "diffraction ripple" of the Fourier series around the heavy Sb and Sn atoms located at this site. Interatomic distances for Ni and Sb lie within range of values known for $CoSb_3$ [Tak1] and $LaFe_4Sb_{12}$ [Bra1].

Figure 4.2 shows the temperature dependence of ADPs for $Ba_{0.73}Ni_4Sb_{8.1}Sn_{3.9}$ and $Ba_{0.95}Ni_4Sb_{6.1}Sn_{5.9}$. Filler atoms in cage compounds like skutterudites generally exhibit ADPs at RT, which are about three to four times higher than those of the framework atoms [Uhe1]. In case of $Ba_yNi_4Sb_{12-x}Sn_x$ this factor is only slightly above 1.2. Despite their low ADPs, the Ba atoms may be described as harmonic Einstein oscillators uncoupled from a framework that behaves as a Debye solid. Applying Eqn. 4.1 for Einstein oscillators on the temperature dependent atom ADP of the Ba-atom, $U_{eq}$ (isotropic by symmetry), yields an Einstein temperature of 104 K for $Ba_{0.73}Ni_4Sb_{8.1}Sn_{3.9}$ and 111 K for $Ba_{0.95}Ni_4Sb_{6.1}Sn_{5.9}$. Similarly the Debye model (Eqn. 4.2) applied to the framework atoms (Ni, Sb and Sn) revealed only slight differencess between $\theta_D$ for Ni and the random Sn/Sn mixture.

$$U_{eq} = \frac{\hbar}{2 \cdot m \cdot k_B \cdot \theta_{E,eq}} \cdot \coth\left(\frac{\theta_{E,eq}}{2 \cdot T}\right) \quad (4.1)$$

$$U_{eq} = \frac{3 \cdot \hbar^2 \cdot T}{m \cdot k_B \cdot \theta_D^2} \cdot \left[\frac{T}{\theta_D} \cdot \int_0^{\theta_D/T} \frac{x}{e^x - 1} dx + \frac{\theta_D}{4 \cdot T}\right] \quad (4.2)$$

The results for $\theta_E$ and $\theta_D$ are listed at Tab 4.3. As the skutterudite framework is built by the Sn and Sb atoms forming octahedra with Ni atoms in the octahedral centres, one may explain these results as following: because of its position in an octahedral cage formed by Sb and Sn the vibration of the Ni atom is not independent from that of the framework and therefore lead to values close to those obtained for Sb/Sn. While Debye temperatures fit well (especially for Sb/Sn), Einstein temperatures are in most cases apparently higher than those of other skutterudites reported in the literature (see for example [Rog4]). By fitting the temperature dependent lattice parameters via a linear function of the form $a \cdot x + b$ (with $a = 1.02 \times 10^{-5}$ and $b = 0.92$ in case of $Ba_{0.73}Ni_4Sb_{8.1}Sn_{3.9}$, but $a = 1.15 \times 10^{-5}$ and $b = 0.92$ for $Ba_{0.95}Ni_4Sb_{6.1}Sn_{5.9}$) also the thermal expansion coefficients for the two compositions could be determined (Tab. 4.3). All values for α all lie within the range discovered for other skutterudites and increase with higher filling levels. ([Rog4], [Rog3]).

## 5. Specific heat

To gather further information of the vibrational behavior of the filled and unfilled $Ba_yNi_4Sb_{12-x}Sn_x$ skutterudites, specific heat measurements in the temperature range from 3 K to 140 K were applied to the single-phase specimens $Ni_4Sb_{8.2}Sn_{3.8}$, $Ba_{0.42}Ni_4Sb_{8.2}Sn_{3.8}$ and $Ba_{0.92}Ni_4Sb_{6.7}Sn_{5.3}$. The data are displayed in Fig. 5.1 in the form of $C_p/T$ versus $T$. According to Eq. 5.1 $C_p$ consists of two independent parts of which the Sommerfeld electronic specific heat coefficient γ (listed in Tab. 7.1) was obtained from a linear regression to the low temperature range (< 5 K). Results are shown as insert to Fig. 5.1.

$$C_p = C_{p,el} + C_{p,ph} = \gamma \cdot T + \beta \cdot T^3 \quad (5.1)$$

$$\beta = \frac{12 \cdot \pi^4 \cdot N_A \cdot k_B}{5 \quad \theta_D^3} = \frac{1994 \cdot n}{\theta_D^3} \quad (5.2)$$

$N_A$ is Avogadros's number, $k_B$ is the Boltzmann factor and n stands for the number of atoms per formula unit. Fitting $C_p$ according to Eq. 5.1 yields β, which in turn via Eq. 5.2 delivers the Debye temperature $\theta_D$ (listed in Tab. 7.1). Fig. 5.1 suggests a moderate overall lattice softening with increasing Ba and Sn content. To extract the corresponding Einstein temperatures and Einstein frequencies, the specific heat data were analysed by applying two different methods. The first approach is based on an additive combination of Debye and Einstein models. It is assumed that the phonon spectrum of a polyatomic compound contains three acoustic branches and 3n-3 optical branches, where the acoustic part of the phonon specific heat can be described via the Debye model (Eq. 5.3) with R being the gas constant and $\omega_D = \theta_D/T$.

$$C_{ph,D} = \frac{9 \cdot R}{\omega_D^3} \cdot \int_0^{\omega_D} \frac{\omega^2 \cdot \left(\frac{\omega}{2 \cdot T}\right)}{\sinh^2\left(\frac{\omega}{2 \cdot T}\right)} d\omega \qquad (5.3)$$

Here the three acoustic branches are taken as one triply degenerated branch. In a similar way the Einstein model describes the optical branches

$$C_{ph,Ei} = c_i \cdot R \cdot \frac{\left(\frac{\theta_{Ei}}{2 \cdot T}\right)^2}{\sinh^2\left(\frac{\theta_{Ei}}{2 \cdot T}\right)} \qquad (5.4)$$

with $c_i$ referring to a degeneracy of the corresponding Einstein mode. The three acoustic and nine optical branches of the phonon dispersion of the $Ni_4Sb_{12-x}Sn_x$ framework is then represented by (12 times) one Debye and two Einstein functions ($27 \cdot f_E + 9 \cdot f_E$) with $c_1 = 27$ and $c_2 = 9$. To account for the filler atom in the icosahedral voids, a further Einstein function is added to describe the increase of phonon modes, with $c_3 = 3 \cdot y$, where y represents the filling level. Including the electronic part from Eqn. 7.1, this leads to

$$C_p = C_{p,el} + C_{ph,D} + \sum_{i=1}^{n=2,3} C_{ph,Ei} \qquad (5.5)$$

with the sum running over two or three Einstein modes. Least squares fits according to this model are presented as dashed lines in Fig. 5.1 and the extracted data for Einstein and Debye temperature are also listed in Tab. 4.3.

The second approach uses the model introduced by Junod *et al.* [Jun1, Jun2], to get some insight in the complex phonon spectrum. Special functionals of the phonon specific heat can take the form of convolutions of the phonon spectrum $F(\omega)$. Therefore the electronic part of the specific heat is subtracted from $C_p(T)$ ($C_{ph}(T) = C_p(T) - \gamma \cdot T$) and least squares fits with two estimated Einstein-modes are applied to the phonon specific heat as shown in Fig. 5.2 in the form $C_{ph}/T^3$ vs. T. Further details on this method are described for example in [Mel1]. In Fig. 5.2 the corresponding phonon spectra are plotted as solid lines scaled to the right axis. The simple Debye function based on Eqn. 5.4 is presented as dotted line using the $\theta_D$-data extracted together with the Sommerfeld coefficients. All data referring to specific heat are listed in Tab. 4.3. Debye and Einstein temperatures extracted by the two different methods described above match well with each other for all three samples investigated: $Ni_4Sb_{8.2}Sn_{3.8}$, $Ba_{0.42}Ni_4Sb_{8.2}Sn_{3.8}$ and $Ba_{0.92}Ni_4Sb_{6.7}Sn_{5.3}$.

## 6. Physical properties

As the $Ba_yNi_4Sb_{12-x}Sn_x$ compound exists within a wide homogeneity region, detailed investigations of physical properties as a function of the composition were performed.

*Electrical resistivity*

Figure 6.1a shows the temperature dependent electrical resistivities $\rho(T)$ of the unfilled single-phase skutterudite $Ni_4Sb_{8.2}Sn_{3.8}$ and the two Ba-filled skutterudites $Ba_{0.42}Ni_4Sb_{8.2}Sn_{3.8}$ and $Ba_{0.92}Ni_4Sb_{6.7}Sn_{5.3}$. These alloys show relatively low resistivities comparable to those of $Sn_yNi_4Sb_{12-x}Sn_x$ reported by Grytsiv *et al.* [Gry1]. Whilst the decrease of the overall resistivity by filling $Ni_4Sb_{8.2}Sn_{3.8}$ with Ba atoms

seems to be correlated with an increase of the free carrier concentration, increasing Sn-content seems to exert the inverse effect. Although the concept introduced by E. Zintl [Kau1, Sev1] is based on simple crystal chemistry, it seems to be able to properly describe the changes in $\rho(T)$ caused by changes in the composition of $Ba_yNi_4Sb_{12-x}Sn_x$ alloys on the basis of a simple counting of charge carriers $n_z$ by Equ. 6.1:

$$n_z = y \cdot 2(Ba) + 4 \cdot 10(Ni) - (12-x) \cdot 3(Sb) - x \cdot 4(Sn) \qquad (6.1)$$

This composition dependent change of carriers can also be seen from the Seebeck coefficients S(T) of the three samples, as discussed below.

For all the three samples the temperature dependent electrical resistivity exhibits two different regimes. At low temperatures a metallic-like behaviour is obtained, which changes to a semiconducting behaviour at higher temperatures, at least up to 700 K. This change cannot be explained by a simple activation-type conductivity mechanism. In order to take care of this rather complicated temperature dependence, a model of density of states was considered, with a narrow gap lying slightly below the Fermi energy $\varepsilon_F$ [Ber1], which successfully described resistivities $\rho(T)$ of various clathrate and skutterudite systems (see for example [Mel1] and [Gry1]). In this model unoccupied states above $E = \varepsilon_F$ are available at $T = 0$ K. This is possible only at low temperatures because of metallic conductivity as long as the limited number of states slightly above $\varepsilon_F$ become involved in the transport process. Further transport processes are only possible, if electrons are exited across the energy gap of width $E_{gap}$. The total number of carriers (electrons and holes) following from general statistical laws is strongly dependent on the absolute temperature as well as on $E_{gap}$. On the basis of this assumptions, $\rho(T)$ can be calculated by

$$\rho(T) = \frac{\rho_0 \cdot \rho_{ph}}{n(T)} \qquad (6.2)$$

where $\rho_0$ gives the residual resistivity and $\rho_{ph}$ describes the scattering of electrons on phonons, taken into account the Bloch Grüneisen law (Eqn. 6.3).

$$\rho(T) = \rho_0 + 4 \cdot \Re \cdot \theta_D \cdot \left(\frac{T}{\theta_D}\right)^5 \cdot \int_0^{\theta_D/T} \frac{x^5}{(e^x-1) \cdot (1-e^{-x})} dx \qquad (6.3)$$

where $\Re$ stands for a temperature independent electron-phonon interaction constant. Least square fits according to Eqn. 6.2 were performed for $Ba_{0.92}Ni_4Sb_{6.7}Sn_{5.3}$ extracting $\rho_0$ and $E_{gap}$, which are listed in Tab. 6.1. For $Ni_4Sb_{8.2}Sn_{3.8}$ and $Ba_{0.42}Ni_4Sb_{8.2}Sn_{3.8}$ least square fits on the basis of Eqn. 6.2 led to no success, but describing $\rho(T)$ for this two compounds by Mott's variable-range hopping conductivity mechanism [Mot1] resulted in fine least squares fits presented as dashed lines in Fig. 6.1a. Corresponding fit parameters $\rho_0$ and $E_{gap}$ can be found in Tab. 6.1. Debye temperatures $\theta_D$ are listed in Tab. 4.3 and can be compared to data extracted by various other methods (see below).

*Thermal conductivity*

Figure 6.2 presents the temperature-dependent thermal conductivities $\kappa(T)$ of the three compounds $Ni_4Sb_{8.2}Sn_{3.8}$, $Ba_{0.42}Ni_4Sb_{8.2}Sn_{3.8}$ and $Ba_{0.92}Ni_4Sb_{6.7}Sn_{5.3}$. Radiation losses in the low-temperature steady-state heat flow measurement were corrected subtracting a Stefan-Boltzmann $T^3$ term. Generally, the thermal conductivities are quite high compared to those of other skutterudites (see for example [Shi1, Rog5,

Rog6, Rog7]), but they lie in the same range as those obtained by Grytsiv *et al.* [Gry1] for their Sn filled and unfilled $Sn_yNi_4Sb_{12-x}Sn_x$ compounds. For more detailed analysis only data between 4 and 300 K were taken into account shown in Fig. 6.3. The total thermal conductivity can be written as

$$\kappa = \kappa_{el} + \kappa_{ph} \tag{6.4}$$

where $\kappa_{el}$ presents the electronic part and $\kappa_{ph}$ the phonon part. A number of scattering processes limit both contributions, so that a finite thermal conductivity will result in any case. For simple metals $\kappa_{el}$ can be calculated from the temperature dependent electrical resistivity via the Wiedemann Franz (Eq. 6.5), where $L_0$ presents the Lorentz number ($L_0 = 2.45 \cdot 10^{-8}\ W \cdot \Omega \cdot K^{-2}$).

$$k_{el} = \frac{L_0 \cdot T}{\rho(T)} \tag{6.5}$$

Subtracting this term $\kappa_{el}$ form the measurement data results in $\kappa_{ph}$ shown in Fig. 6.3a,c,e. For all the three samples the electronic part is quite low but gets enhanced at higher temperatures. According to the Matthiessen rule, the electronic thermal resistivity $W_{el}$ of a simple metal can be written in terms of a thermal resistivity $W_{el,0}$ caused by electron scattering due to impurities and defects as well as caused by electrons scattered due thermally excited phonons $W_{el,ph}$ (Eqn. 6.6).

$$W_{el} \equiv \frac{1}{\kappa_{el}} = W_{el,0} + W_{el,ph} \tag{6.6}$$

$W_{el,0}$ can then be described by

$$W_{el,0} = \frac{\rho_0}{L_0 \cdot T}. \tag{6.7}$$

Using the Wilson equation [Wil1] $W_{el,ph}$ can be expressed as:

$$W_{el,ph} = \frac{A}{L_0 \cdot T} \cdot \left(\frac{T}{\theta_D}\right)^5 \cdot J_5\left(\frac{\theta_D}{T}\right) \cdot \left[1 + \frac{3}{\pi^2} \cdot \left(\frac{k_F}{q_D}\right)^2 \cdot \left(\frac{\theta_D}{T}\right)^2 - \frac{1}{2 \cdot \pi^2} \cdot \frac{J_7(\theta_D/T)}{J_5(\theta_D/T)}\right] \tag{6.8}$$

with $k_F$ the wave vector at the Fermi surface, $q_D$ the phonon Debye wave vector and A a material constant, which depends on the strength of the electron-phonon interaction, Debye temperature, the effective mass of an electron, the number of unit cells per unit volume, Fermi velocity and on the electron wave number at the Fermi surface. The Debye integrals $J_n$ have the form

$$J_n(x) = \int_0^x \frac{z^n}{(e^z - 1) \cdot (1 - e^{-z})} dz \tag{6.9}$$

with $z = \frac{\theta_D}{T}$.

In non-metallic systems lattice thermal conductivity is the dominant part of the thermal conduction mechanism, which can be described by a model introduced by Callaway [Cal1, Cal2, Cal3]. According to his model heat carrying lattice vibrations can be described by

$$k_{ph} = \frac{k_B}{2 \cdot \pi^2 \cdot v_s} \cdot \left(\frac{k_B}{\hbar}\right)^3 \cdot T^3 \cdot \int \frac{\tau_c \cdot x^4 \cdot e^x}{(e^x - 1)^2} dx + I_2 \tag{6.10}$$

with the velocity of sound (in Debye model)

$$v_s = \frac{k_B \cdot \theta_D}{\hbar \cdot (6 \cdot \pi^2 \cdot n)^{1/3}} \tag{6.11}$$

and
$$x = \frac{\hbar \cdot \omega}{k_B \cdot T} \qquad (6.12)$$

where n is the number of atoms per unit volume and $\omega$ the phonon frequency. The second integral $I_2$ in Eqn. 6.10 can be expressed as

$$I_2 = \left[ \int_0^{\theta_D/T} \frac{\tau_c}{\tau_N} \cdot \frac{x^4 \cdot e^x}{(e^x-1)^2} dx \right] \Big/ \int \frac{1}{\tau_N} \cdot \left(1 - \frac{\tau_c}{\tau_N}\right) \cdot \frac{x^4 \cdot e^x}{(e^x-1)^2} dx \qquad (6.13)$$

where $\tau_c^{-1}$ is the sum of the reciprocal relaxation times for point defect scattering (Eqn. 6.14) and $\tau_N^{-1}$ is relaxation time for the normal 3-phonon scattering process.

$$\tau_c^{-1} = \tau_N^{-1} + \tau_D^{-1} + \tau_U^{-1} + \tau_B^{-1} + \tau_E^{-1} \qquad (6.14)$$

Here $\tau_D^{-1}, \tau_U^{-1}, \tau_B^{-1}$ and $\tau_E^{-1}$ denote point defect scattering, Umklapp processes, boundary scattering and scattering of phonons by electrons.

The dashed lines in 6.3a,c,e refer to least squares fits according to Eqn. 6.4 assuming a combination of the Wilson and Callaway model for $\kappa_{ph}$, taking into account the crossover from metallic to semi-metallic behaviour for all three specimens. The extracted values for the Debye temperatures $\theta_D$ are listed to Tab. 4.3 and show a good agreement with values gained from various other methods described in this article and compare well with literature data.

The minimum thermal conductivity $\kappa_{min}$, presented as shaded area in Fig. 6.3a,c,e, was estimated by the model of Cahill and Pohl [Cah1, Gia1]:

$$\kappa_{min} = \left(\frac{3 \cdot n}{4 \cdot \pi}\right)^{\frac{1}{3}} \cdot \frac{k_B^2 \cdot T^2}{\hbar \cdot \theta_D} \cdot \int_0^{\theta_D/T} \frac{x^3 \cdot e^x}{(e^x-1)^2} dx \qquad . \qquad (6.15)$$

Here n is the number of atoms per unit volume and x is a dimensionless parameter connected to the phonon frequency $\omega$ via $x = \frac{\hbar \cdot \omega}{k_B \cdot T}$. Figures 6.3b,d,f compare the situation for temperatures above 300 K. For the two compounds $Ni_4Sb_{8.2}Sn_{3.8}$ and $Ba_{0.42}Ni_4Sb_{8.2}Sn_{3.8}$ $\kappa_{el}$ is the dominant part of thermal conductivity while $\kappa_{ph}$ is of the order of $\kappa_{min}$. For $Ba_{0.92}Ni_4Sb_{6.7}Sn_{5.3}$ both parts, $\kappa_{ph}$ as well as $\kappa_{el}$, are in the same order in magnitude but 2 to 3 times higher than for the previous samples. This may be caused by two effects: i) substitution of Sb with Sn-atoms in the 2a site is combined with an increase in thermal conductivity, ii) or more likely, as reported in various articles ([Chr1, Koz1]), the vibrations of the filler atom are not independent from those of the network atoms. In this case a higher Ba content enhances the thermal conductivity of the material.

*Thermopower*

The temperature-dependent Seebeck coefficients $S(T)$ for samples with the composition $Ni_4Sb_{8.2}Sn_{3.8}$, $Ba_{0.42}Ni_4Sb_{8.2}Sn_{3.8}$ and $Ba_{0.92}Ni_4Sb_{6.7}Sn_{5.3}$ are displayed in Fig. 6.4a. S(T) is negative at low temperatures up to at least 400 K, suggesting that the transport phenomena are dominated by electrons as charge carriers indicating n-type behaviour. This can be explained by the Zintl concept in parallel to the electrical resistivity, as discussed above. Below about 400 K, $S(T)$ behaves almost linearly before reaching a minimum between 400 K and 500 K. The Seebeck coefficient obeys Mott´s formula [Bla1]

$$S(T > \theta_D) = \frac{2 \cdot \pi^2 \cdot k_B^2 \cdot m^*}{e \cdot \hbar^2 \cdot (3 \cdot n \cdot \pi^2)^{\frac{2}{3}}} \quad (6.16)$$

with $m^*$ being the effective mass and e the respective charge and of the involved carrier which is accepted to be valid for systems without significant electronic correlations. As shown previously [Rog1, Rog2] one can extract the charge carrier density n in the temperature range from 0 K to 300 K from the slope of a linear fit extrapolated to 0 K of the measured $S(T)$ data, assuming $m^*$ to be $m_e = 9.1094 \cdot 10^{-31}$ kg (Tab 6.1). From the calculated number of charge carriers, n, two effects can be distinguished: (i) increasing Ba content seems to increase the number of charge carriers, and (ii) n is reduced with increasing Sn content, as the sample $Ba_{0.92}Ni_4Sb_{6.7}Sn_{5.3}$ contains double the amount of Ba in comparison to $Ba_{0.42}Ni_4Sb_{8.2}Sn_{3.8}$, but the number of charge carriers is only higher by 12%.

From the knowledge of the charge carrier concentration n, the mobility in the temperature range 0-300 K can be calculated by Eqn. 6.17, and is shown in Fig. 6.5. Mobilities reveal an inverse but similar trend as the charge carrier densities discussed above.

$$\mu = \frac{1}{\rho \cdot e \cdot n} \quad (6.17)$$

With the density n, the Hall coefficient, $R_H$, can be calculated using Eqn. 6.18 with an accuracy of less than 10% in comparison to measured data (for details see [Rog1, Rog11, Rog12]).

$$n = -\frac{f}{R_H \cdot e} \quad (6.18)$$

The parameter f depends generally on the temperature, the scattering mechanism of the charge carriers and the Fermi level, but as for degenerated systems f does not deviate from unity by more than 10%, it was assumed to be 1. The listed $R_H$ values in Tab. 6.1 are all negative, confirming n-type conductivity, and they behave in the same way as their corresponding charge carrier mobilities. The Fermi energy $\varepsilon_F$ of the three systems $Ni_4Sb_{8.2}Sn_{3.8}$, $Ba_{0.42}Ni_4Sb_{8.2}Sn_{3.8}$ and $Ba_{0.92}Ni_4Sb_{6.7}Sn_{5.3}$ can be estimated by

$$\varepsilon_F = \frac{\hbar^2}{2 \cdot m^*} \cdot (3 \cdot \pi^2 \cdot n)^{\frac{2}{3}}. \quad (6.19)$$

Corresponding vales for $\varepsilon_F$ can be extracted form Tab. 6.1. As they are proportional to the charge carrier density n, they mirror their current behaviour. Furthermore Goldsmid and Sharp [Gol1] showed the possibility to estimate the gap energy $E_{gap}$ from the maximum/minimum of the Seebeck-temperature curve by applying Eqn. 6.20.

$$S_{max} = \frac{E_g}{2 \cdot e \cdot T_{max/min}} \quad (6.20)$$

As in the present case $S(T)$ is negative over the whole temperature range investigated, minimum data were inserted in Eqn. 6.20 leading to the results summarized in Tab. 6.1. From Eqn. 6.16 the reduced effective mass $m^*/m_e$ can be calculated. Fig. 6.6 shows the values of the effective mass $m^*(T)$. The results for room temperature are listed in Tab. 6.1. Values vary within 0.940 and 0.953. In the corresponding Pisarenko plot relating the Seebeck coefficients to the charge carrier densities for the respective effective masses (Fig. 6.7) therefore only one solid line was chosen at 300 K. Also the data at 423.15 K are shown, which provide a correlation of the maximum of zT with the corresponding charge carrier densities (dotted band in Fig. 6.7). Comparison of all the data leads to the fact that with higher Ba and Sn content the Seebeck coefficient is lowered by ⅓ in case of $Ba_{0.42}Ni_4Sb_{8.2}Sn_{3.8}$ up to ½ for $Ba_{0.92}Ni_4Sb_{6.7}Sn_{5.3}$ compared to $Ni_4Sb_{8.2}Sn_{3.8}$. Therefore out of the Pisarenko plot (Fig. 6.7) the highest possible figure of merit can be achieved for the unfilled skutterudite at 423.15 K. Estimated gap energies $E_g$ compare well with those obtained form temperature dependent electrical resistivity and show a significant decrease from $Ba_{0.42}Ni_4Sb_{8.2}Sn_{3.8}$ to $Ba_{0.92}Ni_4Sb_{6.7}Sn_{5.3}$. From this behaviour one may assume that a higher Sn content in the network (24g site) influences the band structure more than an increased Ba filling level in the icosahedral voids (2a site).

*Figure of merit*

The dimensionless figure of merit zT characterizes the thermoelectric ability of a single material concerning power generation or cooling. For commercial applications a zT above 1 is required. The temperature dependent figure of merit can be calculated by Eqn. 1.1 and is shown in Fig. 6.8a.
Although the electrical resistivity for all the three investigated compounds $Ni_4Sb_{8.2}Sn_{3.8}$, $Ba_{0.42}Ni_4Sb_{8.2}Sn_{3.8}$ and $Ba_{0.92}Ni_4Sb_{6.7}Sn_{5.3}$ is quite low, low absolute values of the Seebeck coefficients $S(T)$ ($<70\ \mu V \cdot K^{-1}$) and high thermal conductivities $\kappa(T)$ ($>20\ mW \cdot cm^{-1} \cdot K^{-1}$) prevent zT from reaching values above 0.11 at RT.

### 7. Physical properties after SPD via HPT

Severe plastic deformation (SPD) via HPT is one technique to significantly increase the performance of thermoelectric materials [Rog8, Rog9, Rog10], with various effects being responsible for this behaviour. Because of the interdependence of $\rho$, S and $\kappa_{el}$ one strategy to rise zT is to reduce the $\kappa_{ph}$ by increased scattering of the heat carrying phonons via various mechanisms as already discussed in chapter 6 in terms of their corresponding relaxation times $\tau_n$ (Eqn. 6.14). In general $I_2$ in Eqn. 6.10 is zero for $\tau_N \gg \tau_U$. A possible route to decrease $\kappa_{ph}$ therefore is either to reduce the grain size $d_g$, because $\tau_B^{-1} = v_s/d_g$, or to increase the density of dislocations, because $\tau_D^{-1}$ includes the contribution of the dislocation core

$$\tau_{core}^{-1} \propto \frac{N_D \cdot r^4 \cdot \omega^3}{v_c} \quad (7.1)$$

with $N_D$ being the dislocation density and the effects of the surrounding strain field

$$\tau_{str}^{-1} \propto \frac{N_D \cdot \gamma^2 \cdot b \cdot \omega}{2 \cdot \pi} \quad (7.2)$$

whereas b represents the Burgers vector of the dislocation. Therefore a smaller grain size results in raised phonon scattering on electrons as well as on lattice defects and this way decreases $\kappa_{ph}$. Furthermore Hicks and Dresselhaus [Hic1, Dre1] demonstrated that grain sizes, approaching nanometer length scales (favourable < 10 nm) are able to influence the Seebeck coefficient. Looking at Mott´s formula (Eqn. 6.16) it can be seen that the Seebeck coefficient is primarily dependent on the energy derivation of the electronic density of states (DOS) at the Fermi-level. So any method, able to increase the slope at $E_F$ for a given number of states N, will rise S. HPT uses this concept by the transition from a parabolic electronic DOS curve to a spike like while going from macroscopic bulk structures to nanosized ones. With these observations a way is opened to influence S independently from ρ and $\kappa_{el}$. With this knowledge HPT was performed at 400°C, applying a pressure of 4 GPa and one revolution on the two skutterudite samples, $Ni_4Sb_{8.2}Sn_{3.8}$ and $Ba_{0.42}Ni_4Sb_{8.2}Sn_{3.8}$, whereas for the compound $Ba_{0.92}Ni_4Sb_{6.7}Sn_{5.3}$ room temperature was used. During the deformation the sample's geometry stays constant and unlimited plastic strain occurs without early failure and crack formation [Zeh1, Zha1]. Therefore with the hydrostatic pressure not simply defects are created but also grain boundaries are built up from these defects [Saf1, Ung1, Zeh2, Zeh3, Zeh4, Zha1]. The resulting shear strain corresponds to Eqn. 2.4. The deformation affects the samples' rim more than the centre. Physical properties of processed specimens therefore always cover more and less deformed parts. Pieces of all parts of the HPT processed samples were collected and used for X-ray powder analysis. In general, an increase in the half width of the X-ray patterns as well as of the lattice parameters was observed, indicating a reduction of the crystallite size and an increase of the dislocation density. For the samples $Ni_4Sb_{8.2}Sn_{3.8}$ and $Ba_{0.42}Ni_4Sb_{8.2}Sn_{3.8}$ this behaviour is distinctive while for the compound $Ba_{0.92}Ni_4Sb_{6.7}Sn_{5.3}$ almost no change was visible (Fig. 7.1). By EPMA micrographs before and after HPT (Fig. 7.2) reveal no changes in the microstructure. Due to microcracks the density of the materials after HPT processing is lower than before (Tab. 7.1). Figures 6.1b, Fig. 6.3b,d,f and Fig. 6.4b summarize the effect of HPT treatment on the physical properties ρ(T), κ(T) and S(T). As the samples are in most cases very brittle only data for measurements above room temperature are available. The electrical resistivity for all samples (Fig. 6.1b) after HPT is higher. Whereas a crossover from metallic to semiconducting behaviour was observed for $Ni_4Sb_{8.2}Sn_{3.8}$ and $Ba_{0.42}Ni_4Sb_{8.2}Sn_{3.8}$, $Ba_{0.92}Ni_4Sb_{6.7}Sn_{5.3}$ was semi conducting in the entire temperature range investigated. Thermal conductivity in case of the Sb-rich samples (Fig. 6.3b,d) seems to stay nearly the same. Determining $\kappa_{el}$ and $\kappa_{ph}$ as shown in chapter 5 leads to the fact that $\kappa_{el}$ is increased by the same amount as $\kappa_{ph}$ is decreased resulting in no evident change in the thermal conductivity κ. For the Sn-rich sample (Fig. 6.3f) both components $\kappa_{el}$ as well as $\kappa_{ph}$ are lowered leading to a decreased overall thermal conductivity κ. The data for the Seebeck coefficients after HPT (Fig. 6.4b) stay in the same range as those before HPT or are even slightly lower. In contrary to filled skutterudites without Sb-Sn substitution, all these observations show no significant change in the figure of merit above room temperature (Fig. 6.8b).

### 8. Thermal expansion

Figure 8.1 shows the thermal expansion $\left(\Delta l/l_0\right)$ for all the three single-phase samples $Ni_4Sb_{8.2}Sn_{3.8}$, $Ba_{0.42}Ni_4Sb_{8.2}Sn_{3.8}$ and $Ba_{0.92}Ni_4Sb_{6.7}Sn_{5.3}$ measured with the capacitance dilatometer in the temperature range 4.2-300 K. $\Delta l/l_0$ decreases linearly within the temperature from room temperature to about 150 K, whereas for temperatures below 150 K, a non-linear behaviour is observed. Data gained from measurements with a dynamic mechanical analyser with the sample first cooled with liquid nitrogen and afterwards heated up or simply heated from room temperature to 400 K show a linear increase with increasing temperature (Fig. 8.2). Temperature derivation of the length change defines the thermal expansion coefficient α via Eqn. 8.1.

$$\alpha = \left(\frac{\partial \Delta l}{\partial l}\right) \cdot \frac{1}{l_0} \tag{8.1}$$

Applying this model, α was calculated in the temperature range above 150 K up to 300 K (see dashed lines in the insertion to Fig. 8.1), and also from combination of low and high temperature measurement data. Corresponding thermal expansion coefficients are listed in Tab. 4.3. The values for the current samples lead to a good agreement between high and low temperature measurements proven by combination of the data according to Fig. 8.2. From Tab. 4.3 follows that the thermal expansion coefficients α for $Ba_yNi_4Sb_{12-x}Sn_x$ based skutterudites lie within the range obtained for other skutterudites [Rog3]. A higher α occurs with a higher Ba filling level, visible at higher temperatures, which can be associated to the rattling behaviour of Ba on the 2a site inside the structural cage. For a cubic material the lattice parameter varies with the temperature in parallel to the thermal expansion coefficient; therefore the lattice parameters $a_n$ at various temperatures can be derived if $a_n$ at a certain temperature is known, applying Eqn. 8.2.

$$\alpha = \frac{a_2 - a_1}{a_1} \Big/ \Delta T \tag{8.2}$$

Values according to all three specimens are added to Fig. 8.2. The higher Ba filling level results in bigger lattice parameters for $Ba_{0.42}Ni_4Sb_{8.2}Sn_{3.8}$ and $Ba_{0.92}Ni_4Sb_{6.7}Sn_{5.3}$ as compared to $Ni_4Sb_{8.2}Sn_{3.8}$. To analyse the thermal expansion as a function of temperature in the whole temperature range from 4.2 to 300 K, the semi-classical model, introduced by Mukherjee [Muk1], was used. This model takes into account three- and four-phonon interactions, considering anharmonic potentials, and uses both the Debye model for the acoustic phonons and the Einstein model for the optical modes. The length change $\Delta l/l(T_0)$ is then given by

$$\frac{\Delta l}{l(T_0)} = \frac{\langle x \rangle_T - \langle x \rangle_{T_0}}{x_0} \tag{8.3}$$

with

$$\langle x \rangle_T = \frac{\gamma}{2} \cdot T^2 + \frac{3 \cdot g}{4 \cdot c^2} \cdot \left( \varepsilon - G \cdot \varepsilon^2 - F \cdot \varepsilon^3 \right) \tag{8.4}$$

and ε of the form

$$\varepsilon = \left\{ \left(\frac{3}{p}\right) \cdot 3 \cdot k_B \cdot T \cdot \left(\frac{T}{\theta_D}\right)^3 \cdot \int \frac{z^3}{e^z - 1} dz + \left(\frac{p-3}{p}\right) \cdot \frac{k_B \cdot \theta_E}{e^{\theta_D/T} - 1} \right\} \tag{8.5}$$

Here γ stands for the electronic contribution to the average lattice displacement, $\theta_D$ is the Debye temperature, $\theta_E$ is the Einstein temperature and p is the average number of phonon branches actually exited over the temperature range with G, F, c, and g being material constants. Least squares fits to the experimental data according to Eqn. 8.3 are added to Fig. 8.1 and Fig. 8.2 as dashed lines. The obtained values for $\theta_D$ and $\theta_E$ are listed to Table 4.3 and can be compared with the data gained from fits to the electrical resistivity, thermal conductivity and specific heat. This comparison shows good agreement among data from the various methods as well as with those available in theliterature.

## 9. Elastic properties

As the average symmetry of skutterudites is isotropic from RUS measurements of all three single-phase samples $Ni_4Sb_{8.2}Sn_{3.8}$, $Ba_{0.42}Ni_4Sb_{8.2}Sn_{3.8}$ and $Ba_{0.92}Ni_4Sb_{6.7}Sn_{5.3}$ Young´s modulus (E) and Poisson´s ratio (ν) could be gained as fitting variables of the materials eigenfrequencies. Based on these data all other elastic properties could be calculated. The shear modulus G and the bulk modulus B can be obtained from Eqn. 9.1 and Eqn. 9.2:

$$G = \frac{E}{2 \cdot (\upsilon + 1)} \qquad (9.1)$$

$$B = \frac{E}{3 \cdot (1 - 2 \cdot \upsilon)} \qquad (9.2)$$

For an isotropic material the elastic constant $C_{11}$ is given by

$$C_{11} = 3 \cdot B - \frac{6 \cdot B \cdot \upsilon}{1 + \upsilon}. \qquad (9.3)$$

The longitudinal ($v_L$) and the transversal ($v_T$) sound velocities can be derived from one of the following relations:

$$C_{44} = G = \rho_s \cdot v_T^2 \qquad (9.4)$$

$$C_{11} = \rho_s \cdot v_L^2 \qquad (9.5)$$

Here $\rho_s$ gives the samples' density measured via Archimedes' principle. From the knowledge of $v_L$ and $v_T$ the materials mean sound velocity $v_m$ can be calculated by Eqn. 9.6.

$$v_m = \left[ \frac{1}{3} \cdot \left( \frac{2}{v_T^3} + \frac{1}{v_L^3} \right) \right]^{-1/3} \qquad (9.6)$$

Anderson's relation [And1] (Eqn. 9.7) yields the Debye temperature $\theta_D$ from elastic properties measurements.

$$\theta_D = \frac{h}{k_B} \cdot \left( \frac{3 \cdot n \cdot N_A \cdot \rho_s}{4 \cdot M \cdot \pi} \right)^{1/3} \cdot v_m \qquad (9.7)$$

In Eqn. 9.7 h denotes Plank´s constant, $k_B$ is the Boltzmann constant, $N_A$ is Avogadro´s number, M is the molecular weight and n stand for the number of atoms in the asymmetric unit. The extracted data for the compounds $Ni_4Sb_{8.2}Sn_{3.8}$, $Ba_{0.42}Ni_4Sb_{8.2}Sn_{3.8}$ and $Ba_{0.92}Ni_4Sb_{6.7}Sn_{5.3}$ are listed in Tab. 9.1 and are shown in Fig. 9.1. Comparison with elastic property data available in the literature [see for example Rog4] leads to the fact that $Ba_yNi_4Sb_{12-x}Sn_x$ based skutterudites are characterized by a rather low Young´s modulus and high Poisson´s ratios continuing this trend for the bulk modulus, longitudinal ($v_L$), the transversal ($v_T$) and mean sound velocity. Extracted Debye temperatures lie in the same range as values gathered by other methods, listed in tables of chapters before.

### 10. Vickers hardness (HV)

Hardness is a measure for a solid material's resistance against permanent shape change under applied force and is dependent on ductility, elastic stiffness, plasticity, strain, strength, toughness, visco-elasticity and viscosity. This was the reason to perform Vickers hardness tests on a series of alloys including the three single-phase bulk samples $Ni_4Sb_{8.2}Sn_{3.8}$, $Ba_{0.42}Ni_4Sb_{8.2}Sn_{3.8}$ and $Ba_{0.92}Ni_4Sb_{6.7}Sn_{5.3}$ particularly on skutterudite crystallites above 200 nm in size. As the Vickers method is an indention measurement, the specimen's resistance against deformation due to a constant compression load from a sharp object is determined. Figure 10.1 shows the results for skutterudites in multiphase samples, whilst Fig. 10.2 presents data for the single-phase ones. Comparison with data available in the literature (see for example [Rog4]) shows that HV of $Ba_yNi_4Sb_{12-x}Sn_x$ skutterudites lay perfectly within the range of other skutterudites. From the measurements performed on single phase sample two trends can be derived [Fig. 10.2]: hardness seems to decrease with i) increasing Ba and Sn content, ii) increasing Young´s modulus and shear modulus. For the compounds $Ni_4Sb_{8.49}Sn_{3.87}$, $Ba_{0.42}Ni_4Sb_{8.44}Sn_{3.88}$ and $Ba_{0.92}Ni_4Sb_{6.78}Sn_{5.41}$ Vickers hardness after HPT processing (Fig. 10.3) shows no measurable changes.

## 11. Electron and phonon mean free path

To learn more about the transport properties caused by electrons and phonons, their mean free paths will be calculated by simple models and then discussed in comparison with resistivity and thermal conductivity data. Assuming a three dimensional system with a spherical Fermi surface [Gun1], the mean free path $\ell_{el}$ of the electrons can be calculated from the measured resistivity data using Eqn. 11.1

$$\ell_{el} = \frac{3 \cdot \pi^2 \cdot \hbar}{e^2 \cdot k_F^2 \cdot \rho(T)}. \tag{11.1}$$

Here e is the electronic charge and ℏ is the reduced Planck´s constant. The Fermi wave vector $k_F$ can be approximated by

$$k_F = \frac{\sqrt{2}}{a} \tag{11.2}$$

with a as the lattice parameter. Figure 11.1 displays the temperature dependence of the electronic mean free path for the three single-phase samples $Ni_4Sb_{8.2}Sn_{3.8}$, $Ba_{0.42}Ni_4Sb_{8.2}Sn_{3.8}$ and $Ba_{0.92}Ni_4Sb_{6.7}Sn_{5.3}$. For all three specimen $\ell_{el}$ is five to ten times higher than the lattice parameter a. This confirms the quite low resistivities $\rho(T)$ found in these alloys. It seems that Ba filling in the 2a site has a positive effect on $\ell_{el}$ while increasing Sn content in site 24g decreases $\ell_{el}$.

From a simple kinetic theory one can also estimate the phonon mean free path (Eqn. 12.3), which is an important parameter for thermal conductivity [Tob1]:

$$\kappa_{ph} = \frac{1}{3} \cdot v_g \cdot \ell_{ph} \cdot \frac{C_v}{V_m} \tag{11.3}$$

Besides the electronic part $\kappa_{el}$, heat transport is performed by phonons of group velocity $v_g$ and mean free path lengths $\ell_{ph}$ between the scattering processes. The amount of transported heat is also proportional to the molar heat capacity $C_v$. For the group velocity $v_g$ a common approximation is, to take the mean sound velocity $v_m$. $C_v$ can then be calculated by

$$C_v = C_p - V_m \cdot T \cdot \frac{\alpha^2}{\beta} \tag{11.4}$$

with the measured heat capacity $C_p$, the thermal expansion $\alpha$ and the molar volume $V_m$. $\beta$ is the isothermal compressibility, which can be taken as 1/B with B being the bulk modulus. As for a solid $C_p \approx C_v$, the comparison between the calculated values for $C_v$ and the measured values for $C_p$ (Fig. 5.1 and Fig. 11.2) confirms that Eqn. 11.4 gives a good approximation in the affected temperature range. The results for the phonon mean free path lengths in the temperature range from 4 to 110 K are shown in Fig. 11.3. As one of the many strategies in designing high performance thermoelectric materials is to decrease $\kappa_{ph}$ by the increase of the phonon scattering, one can see from Fig. 13.3, that for $Ba_yNi_4Sb_{12-x}Sn_x$ skutterudites this cannot be done via filling the 2a site with Ba as a rattling atom. As demonstrated in various papers [Chr1, Koz1], the reason may stem from the fact that the vibrational modes of the rattling atom are not entirely independent of the framework vibrations.

## 12. Summary

Ba filled Ni-Sn-Sb based skutterudites have been successfully prepared and characterised by their physical and mechanical properties with respect to formation and crystal structure. Although the resistivity $\rho(T)$ for the investigated compounds $Ni_4Sb_{8.2}Sn_{3.8}$, $Ba_{0.42}Ni_4Sb_{8.2}Sn_{3.8}$ and $Ba_{0.92}Ni_4Sb_{6.7}Sn_{5.3}$ is quite low, rather high thermal conductivity $\kappa(T)$ prevents the material from reaching attractive figures of merit zT (zT < 1.1). As one of the reasons, Ba-atoms in the 2a position are unable to significantly decrease the phonon mean free paths.

Debye and Einstein temperatures gathered by various methods lie within the same range and are therefore rather comparable to those of other skutterudites. Phase equilibria in the ternary system Ni-Sn-Sb have been determined for the isothermal section at 450°C revealing a rather large extension of the $Ni_4Sn_{12-x}Sn_x$ homogeneity region from $2.4 \leq x \leq 3.2$. A combination of literature data and DTA measurements enabled us to determine the solidus surface for the unfilled $Ni_4Sn_{12-x}Sn_x$ phase as well as for the Ba-filled $Ba_yNi_4Sn_{12-x}Sn_x$ skutterudite phase. The Ba- and Sn-concentration dependent homogeneity region of the phase $Ba_yNi_4Sn_{12-x}Sn_x$ was established. Although the homogeneity region seems to extend into a p-type region, so far only n-type material was obtained.

**Acknowledgments**
The current work was funded by the Austrian FWF as part of the project No. P22295-N20 and FWF P24380.

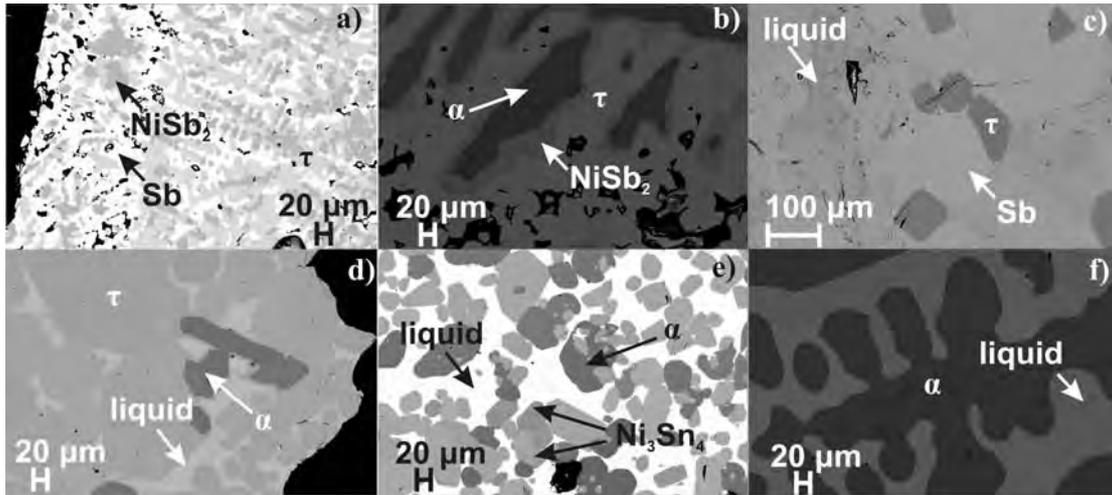

**Figure 3.1:** Microstructure of Ni-Sn-Sb alloys annealed at 450°C belonging to phase equilibria of the isothermal section at 450°C. Nominal composition (at% from EPMA) and X-ray phase analysis are given in Tab. 3.1.

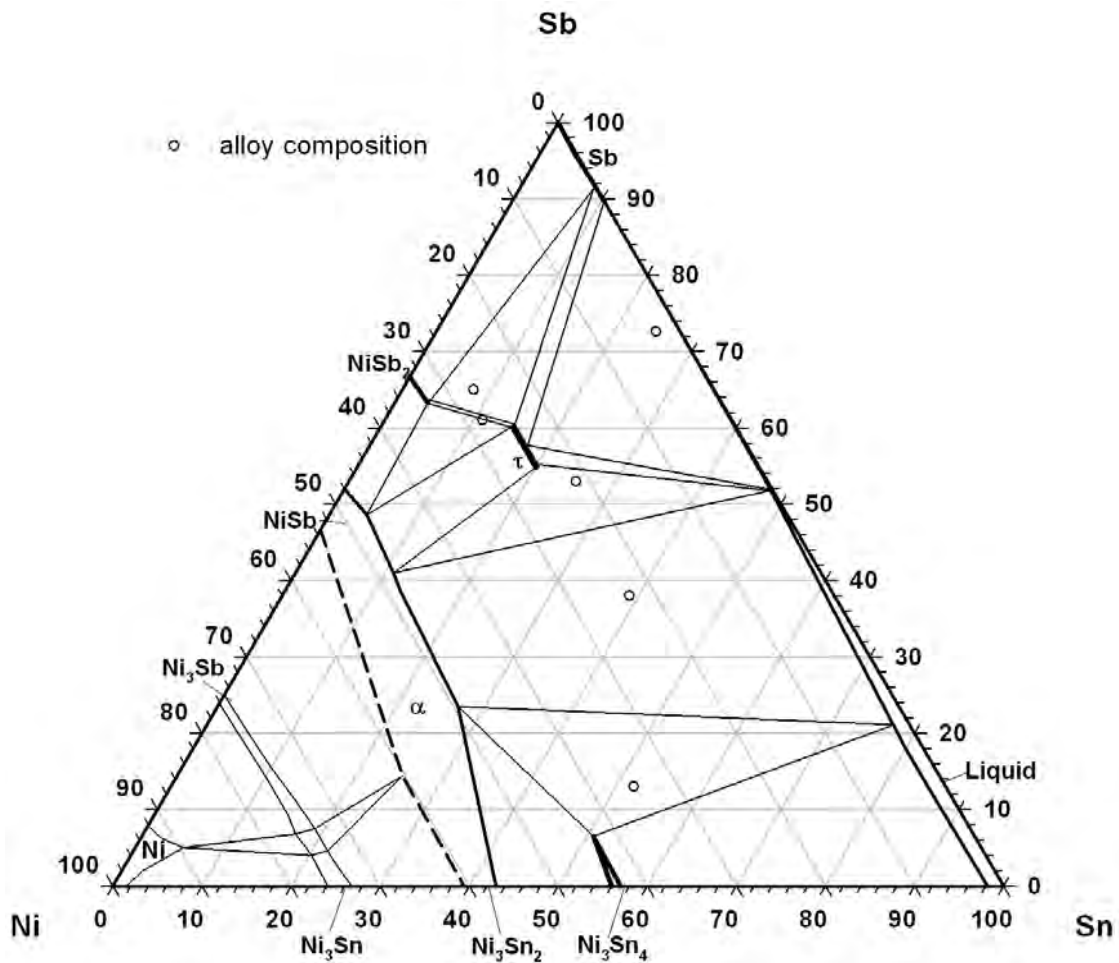

**Figure 3.2:** Isothermal section of the system Ni-Sn-Sb at 450°C. Microstructure of investigated alloys are give in Fig. 3.1.

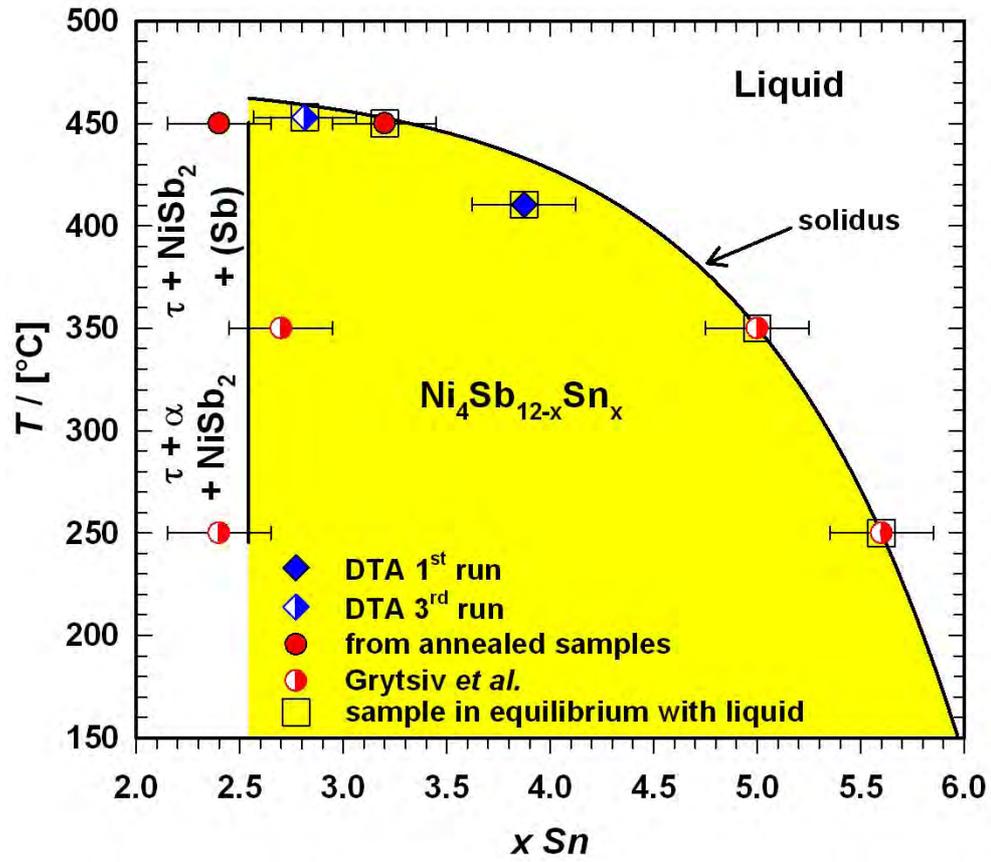

**Figure 3.3:** Temperature dependent extend and solidus curve of the ternary skutterudite phase τ (Ni$_4$Sb$_{12-x}$Sn$_x$).

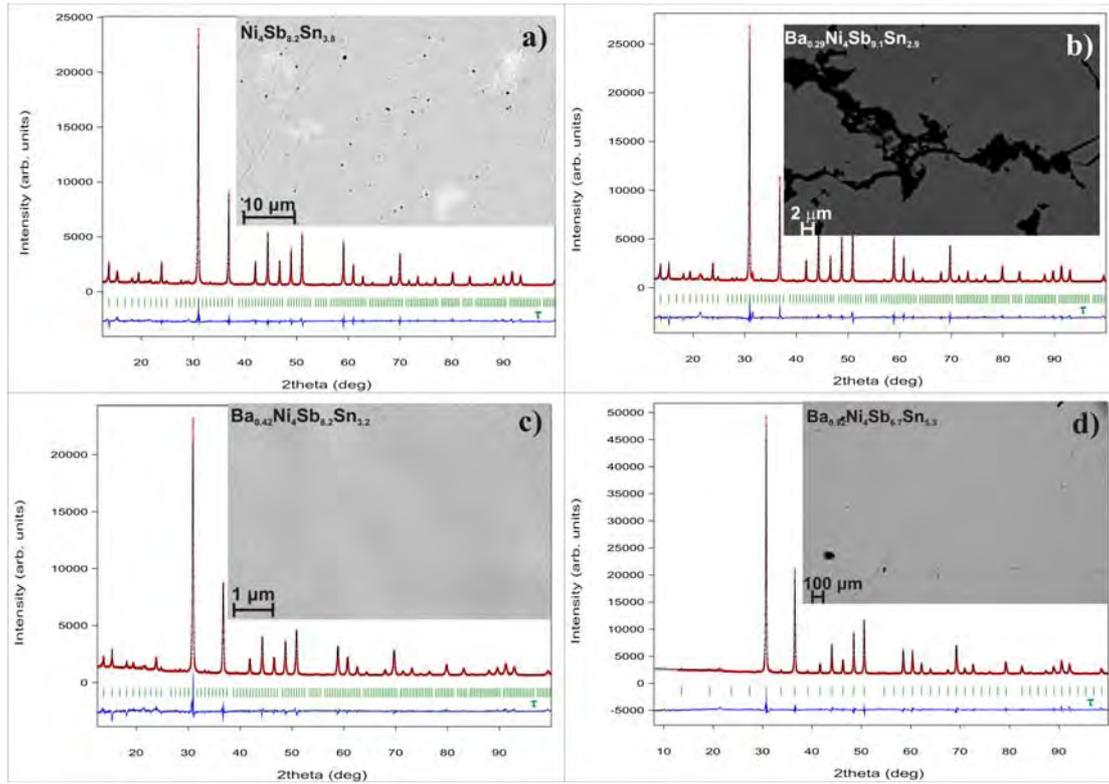

**Figure 3.4:** Microstructure and Rietveld refinements for the single-phase samples a) $Ni_4Sb_{8.2}Sn_{3.8}$, b) $Ba_{0.29}Ni_4Sb_{9.1}Sn_{2.9}$, c) $Ba_{0.42}Ni_4Sb_{8.2}Sn_{3.8}$ and d) $Ba_{0.92}Ni_4Sb_{6.7}Sn_{5.3}$.

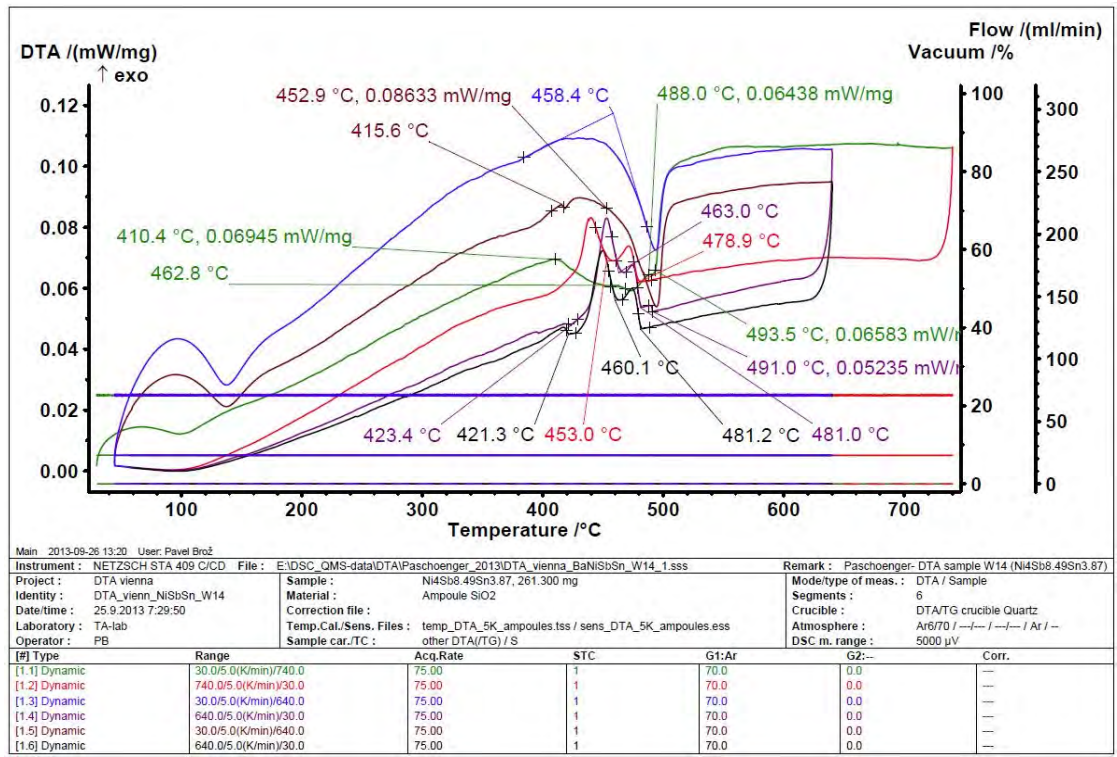

**Figure 3.5:** DTA-measurement curves of the single phase-sample $Ni_4Sb_{8.2}Sn_{3.8}$.

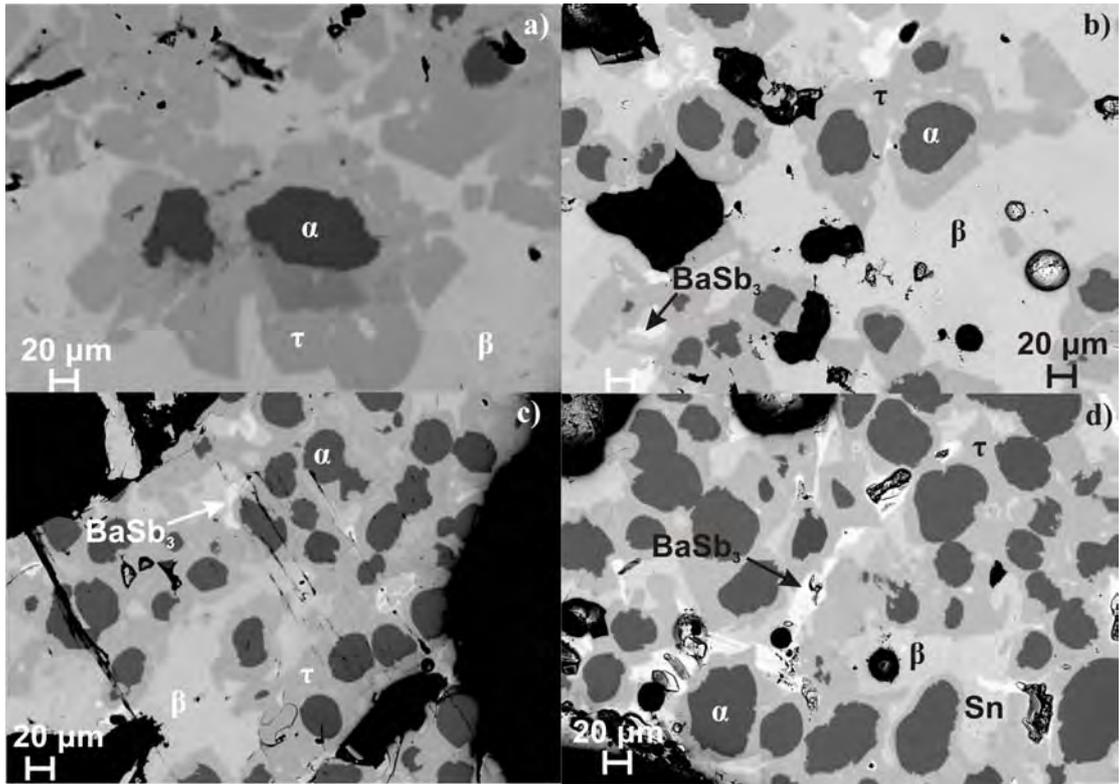

**Figure 3.6:** Microstructure of the former single-phase samples a) Ni$_4$Sb$_{8.2}$Sn$_{3.8}$, b) Ba$_{0.29}$Ni$_4$Sb$_{9.1}$Sn$_{2.9}$, c) Ba$_{0.42}$Ni$_4$Sb$_{8.2}$Sn$_{3.8}$ and d) Ba$_{0.92}$Ni$_4$Sb$_{6.7}$Sn$_{5.3}$ after DTA.

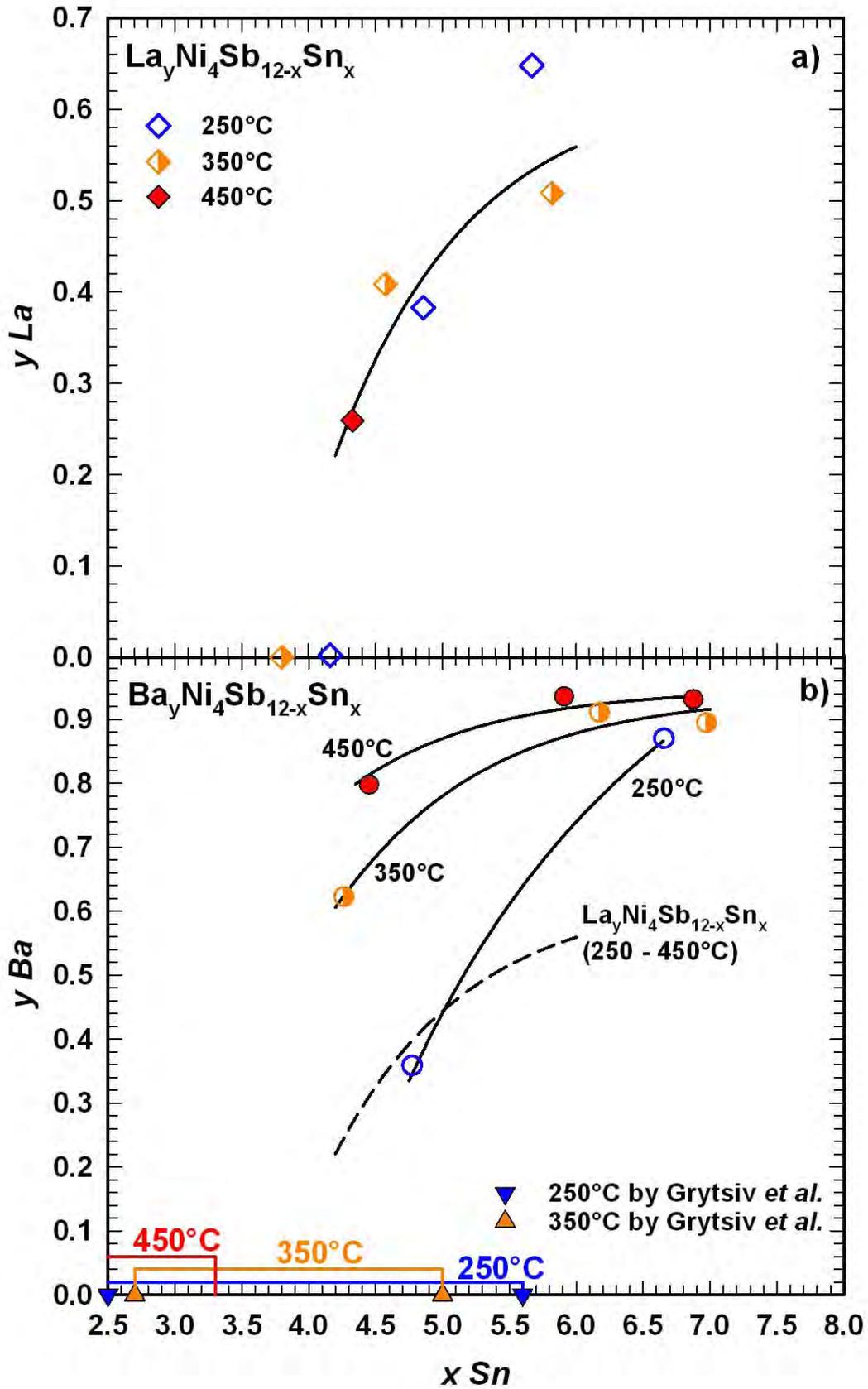

**Figure 3.7**: Compositional dependence of filling level for Ba and La in $Ep_yNi_4Sb_{12-x}Sn_x$ at 250, 350 and 450°C on Sn-rich boundary of the skutterudite solid solution, as determined in the samples 1Ba-4Ni-5Sb-7Sn, 1Ba-4Ni-8Sb- 1La-4Ni-4Sb-8Sn and 1La-4Ni-8Sb-4Sn (for phase composition see Fig. 3.8).

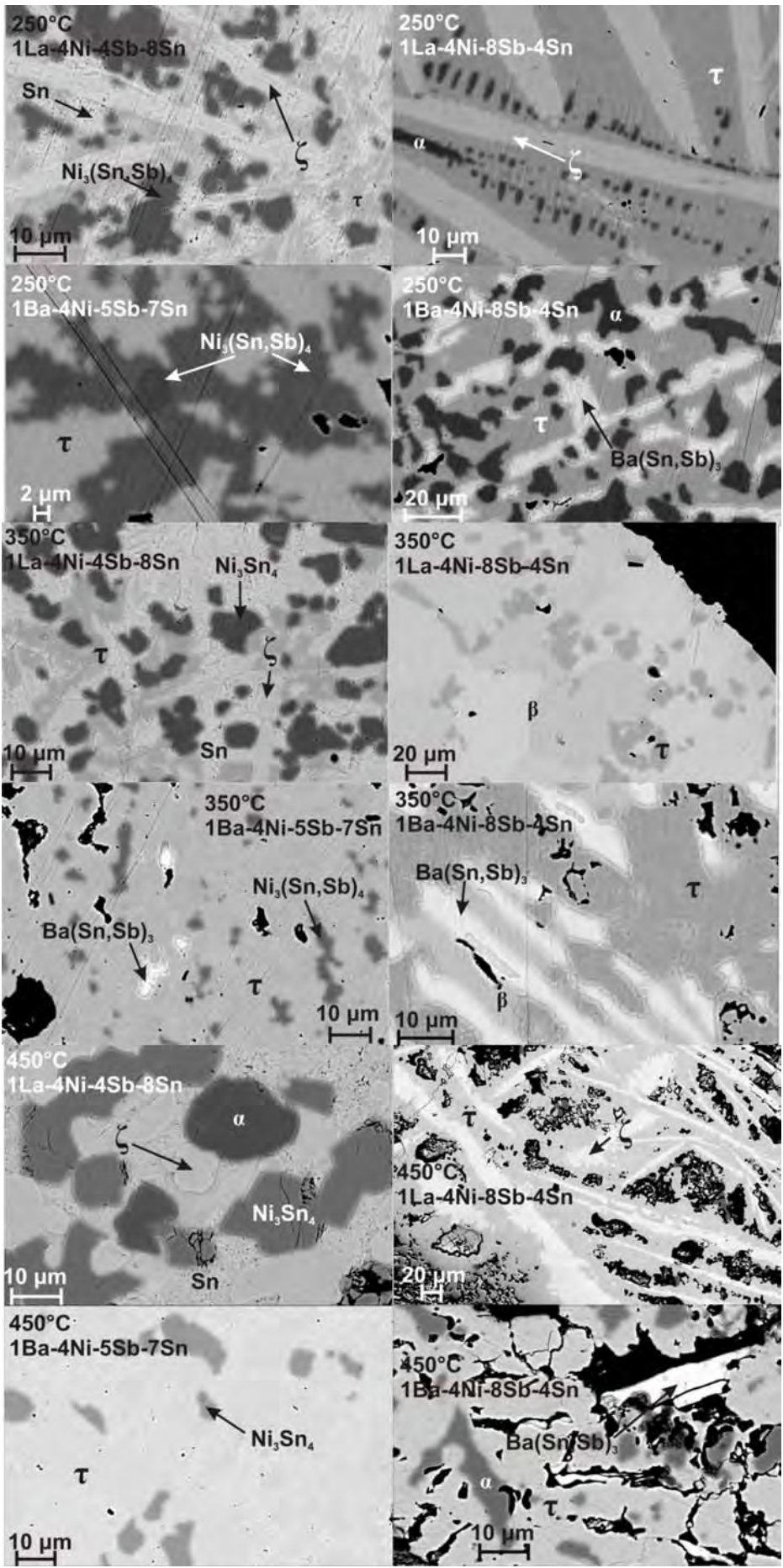

**Figure 3.8:** Microstructure of the alloys 1Ba-4Ni-5Sb-7Sn, 1Ba-4Ni-8Sb-4Sn, 1La-4Ni-4Sb-8Sn and 1La-4Ni-8Sb-4Sn annealed at 250°C, 350°C and 450°C. (ζ denotes for a phase with composition La19-Ni20-Sn20-Sb41 in at% derived by EPMA

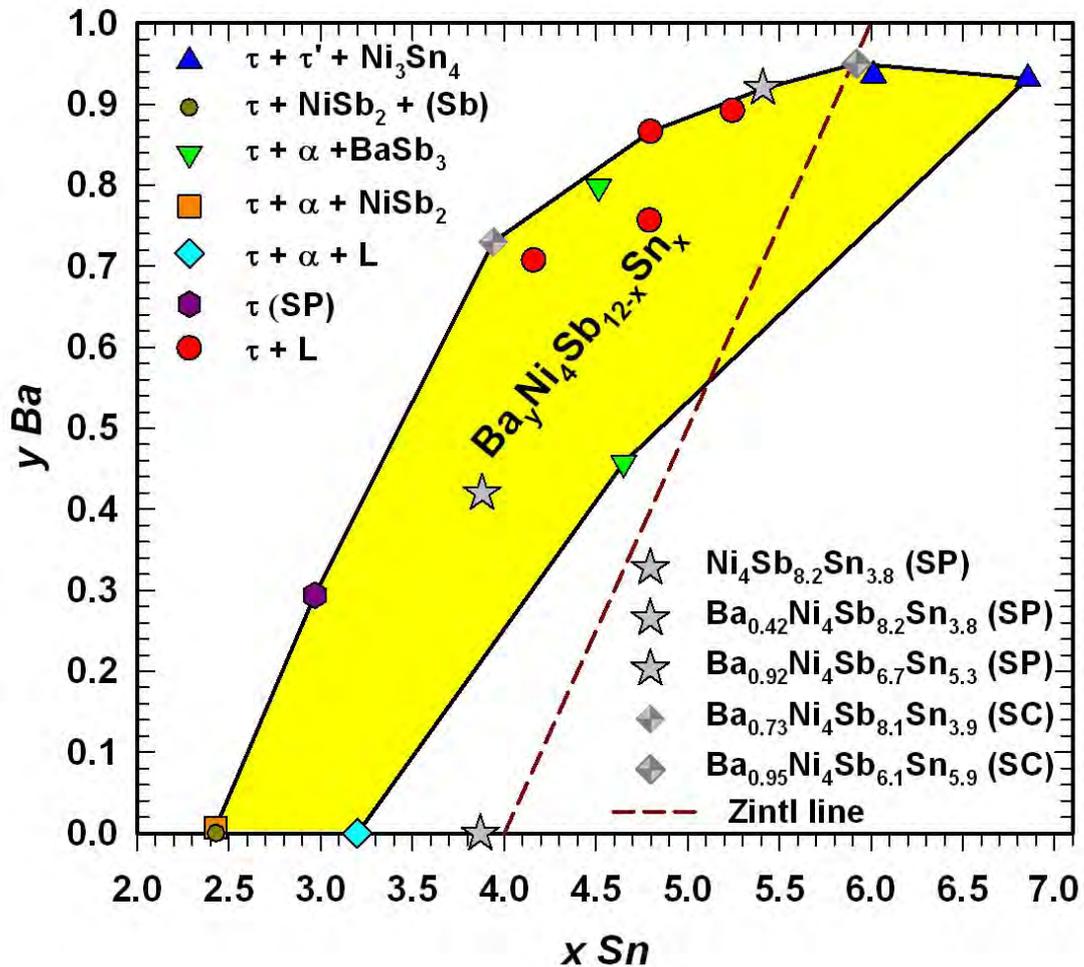

**Figure 3.9:** Homogeneity region for $Ba_yNi_4Sb_{12-x}Sn_x$ at 450°C as a function of the Sn-content and Ba filling level. Composition of single crystals (SC) and samples prepared for investigation of physical properties (SP) are added. (Ternary single-phase sample $Ni_4Sb_{8.2}Sn_{3.8}$ is lying beside the estimated homogeneity region because temperature during preparation was around 10°C lower than 450°C).

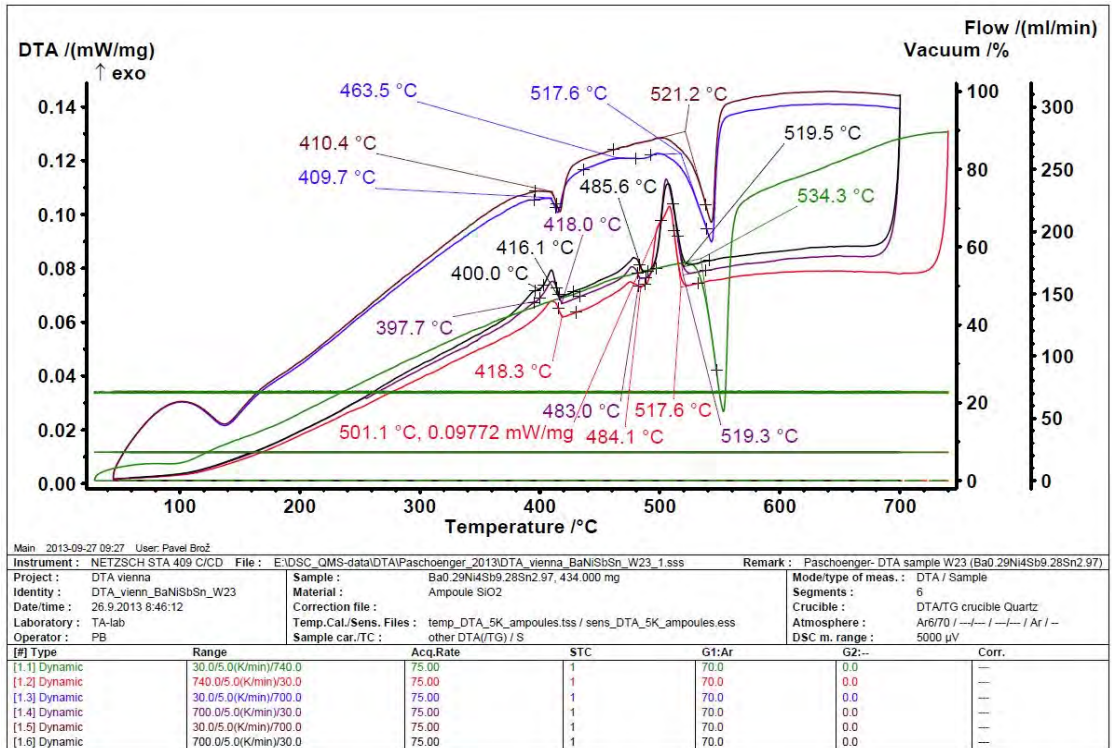

**Figure 3.10:** DTA-measurement curves of the single-phase sample $Ba_{0.29}Ni_4Sb_{9.1}Sn_{2.9}$.

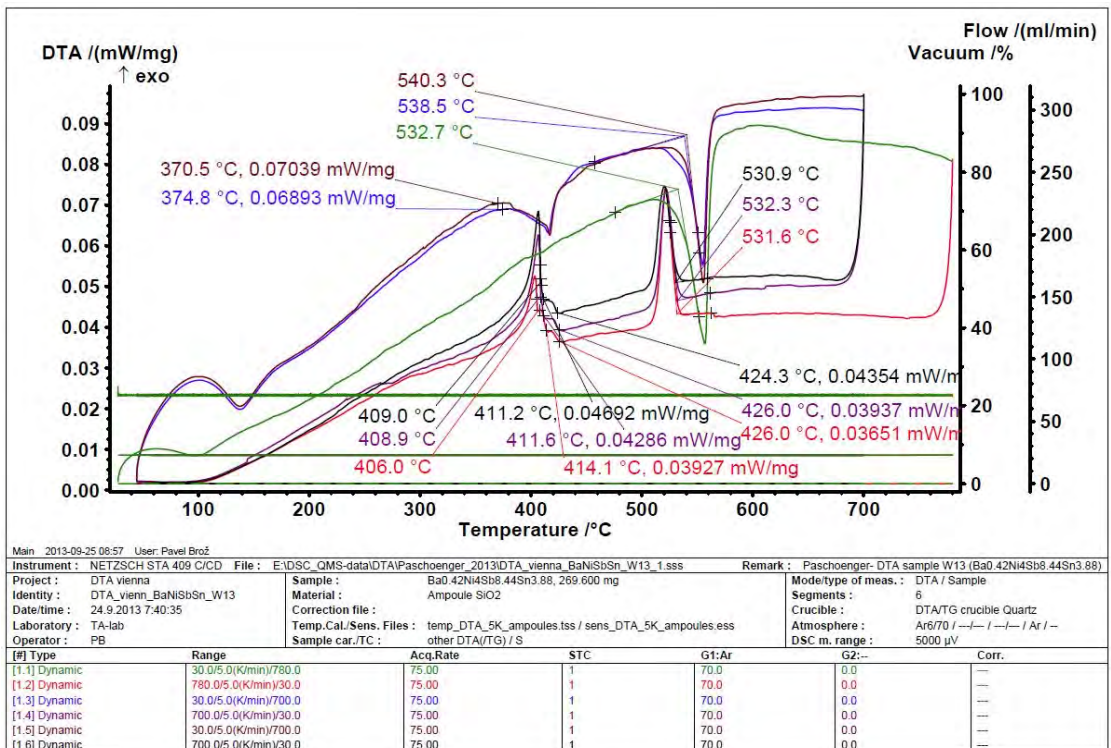

**Figure 3.11:** DTA-measurement curves of the single-phase sample $Ba_{0.42}Ni_4Sb_{8.2}Sn_{3.8}$.

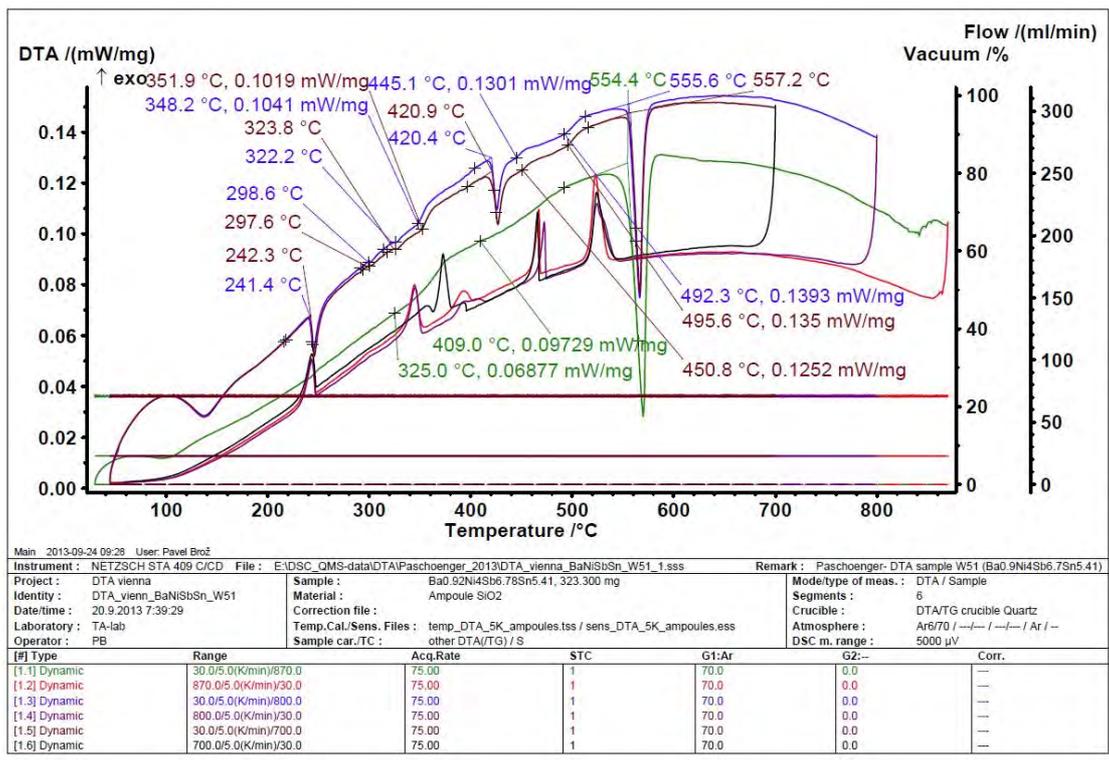

**Figure 3.12:** DTA-measurement curves of the single-phase sample $Ba_{0.92}Ni_4Sb_{6.7}Sn_{5.3}$.

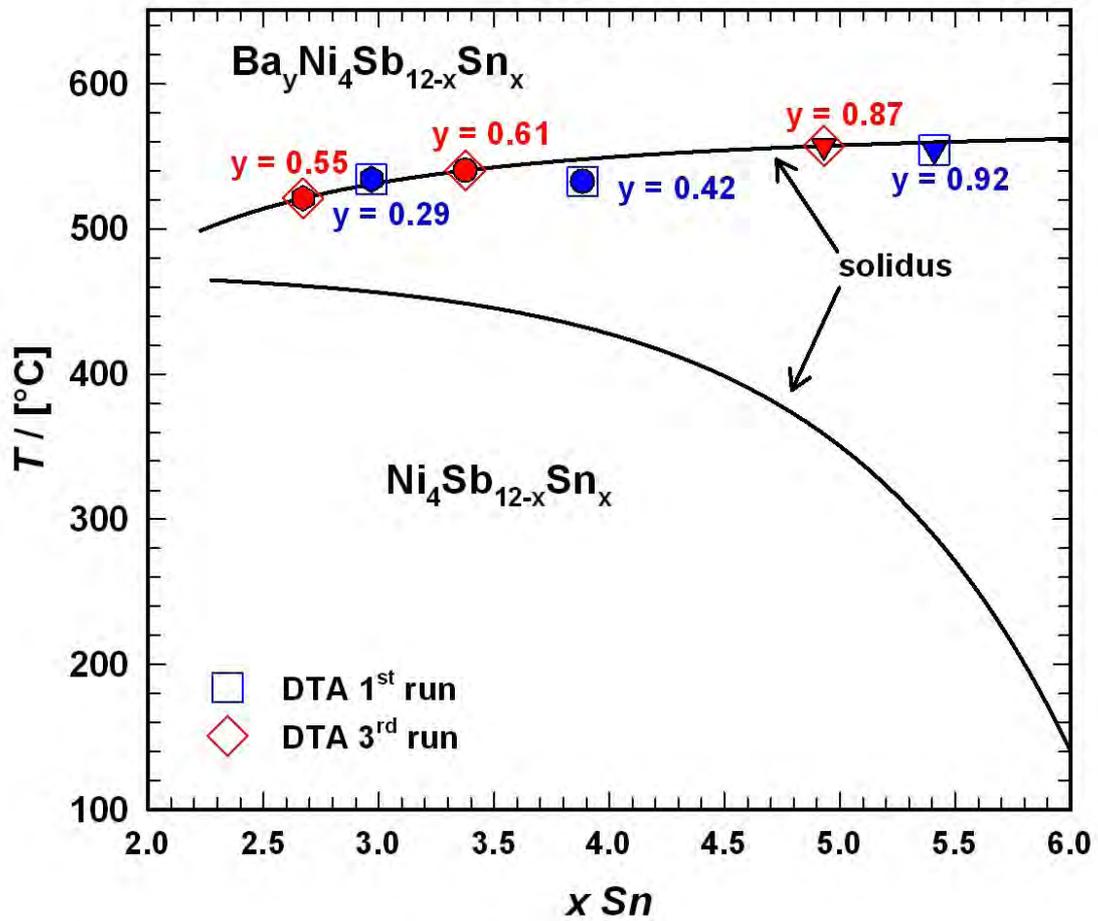

**Figure 3.12:** Compositional dependence of the solidus curve for $Ba_yNi_4Sb_{12-x}Sn_x$ in comparison to that of $Ni_4Sb_{12-x}Sn_x$ (for details see Fig. 3.3).

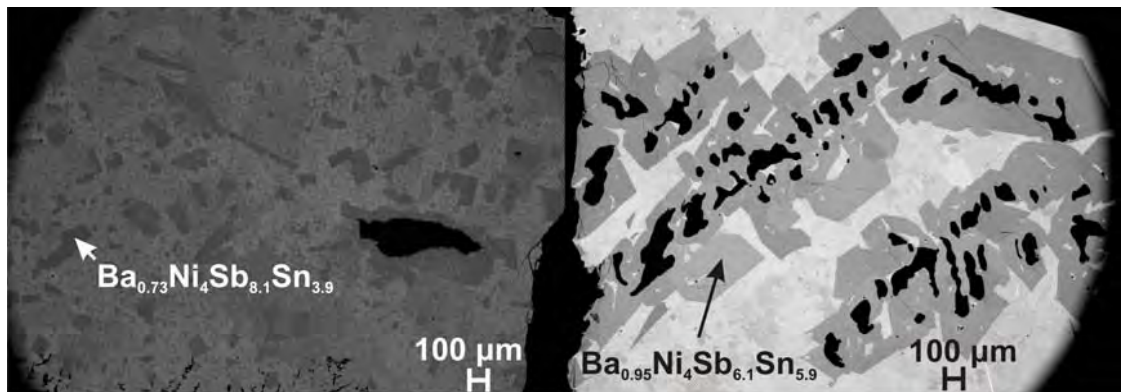

**Figure 4.1:** Microstructure of the alloys used to gain single crystals.

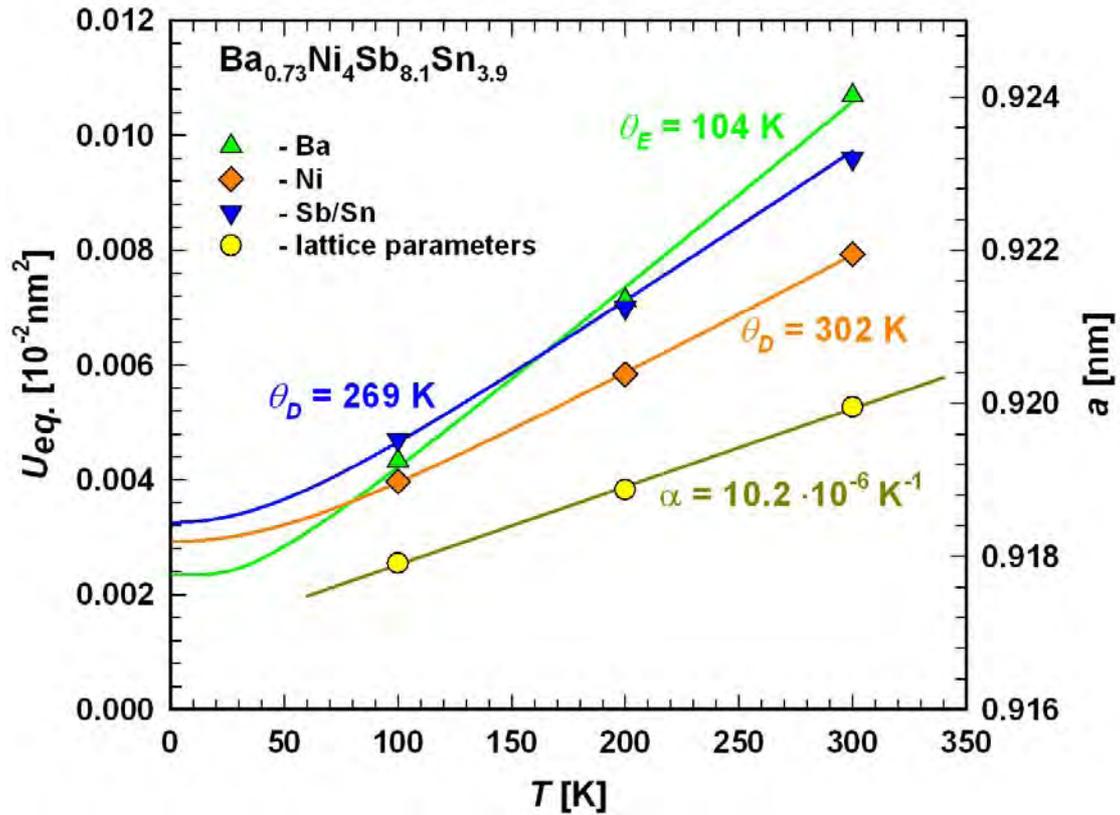

**Figure 4.2a):**

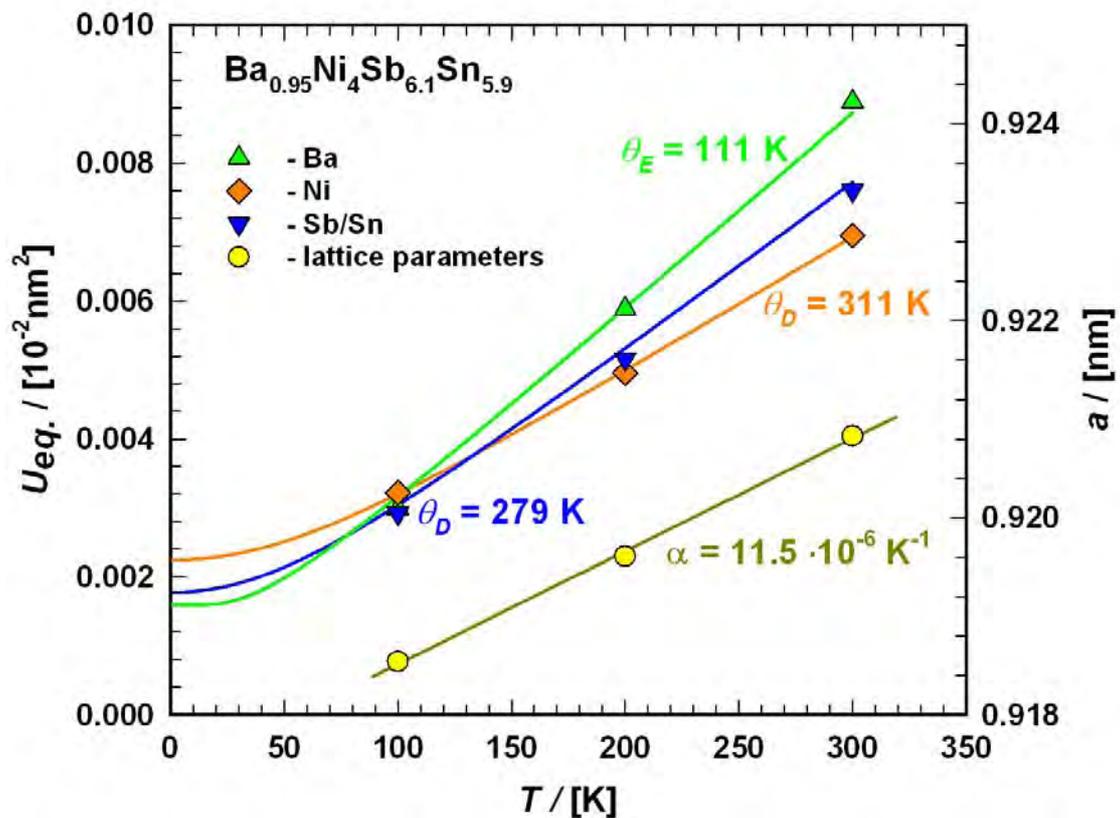

**Figure 4.2:** Temperature dependence of the lattice parameter a and the atomic displacement parameters (ADP) of a) $Ba_{0.73}Ni_4Sb_{8.1}Sn_{3.9}$ and b) $Ba_{0.95}Ni_4Sb_{6.1}Sn_{5.9}$

obtained from X-ray single-crystal measurements at 100 K, 200 K and 300 K. Solid lines correspond to least square fits according to Eqn. 4.1 and 4.2 for the ADP and a simple linear fit for the thermal expansion coefficient.

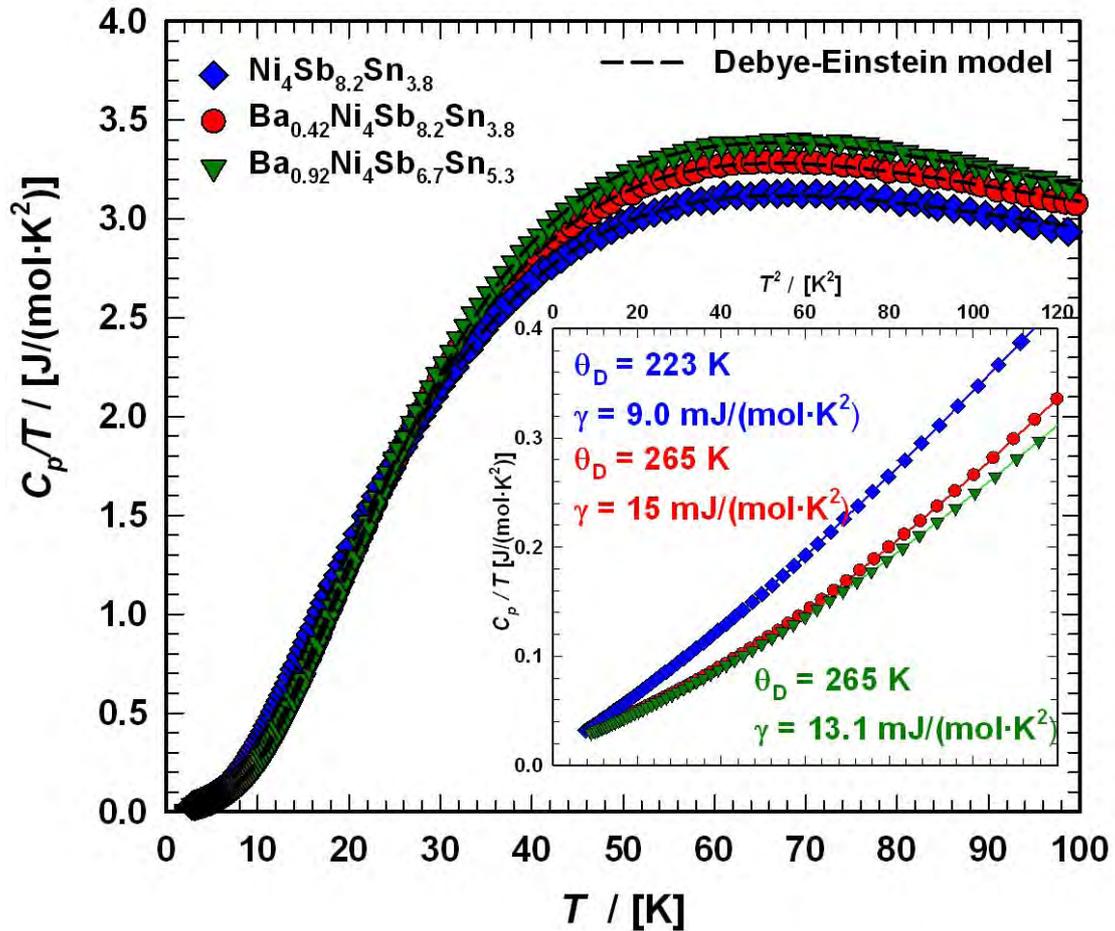

**Figure 5.1:** Temperature-dependent specific heat divided by temperature $C_p/T$ of $Ni_4Sb_{8.2}Sn_{3.8}$, $Ba_{0.42}Ni_4Sb_{8.2}Sn_{3.8}$ and $Ba_{0.92}Ni_4Sb_{6.78}Sn_{5.41}$. Dashed lines correspond to least squares fits according to Eqn. 5.5.

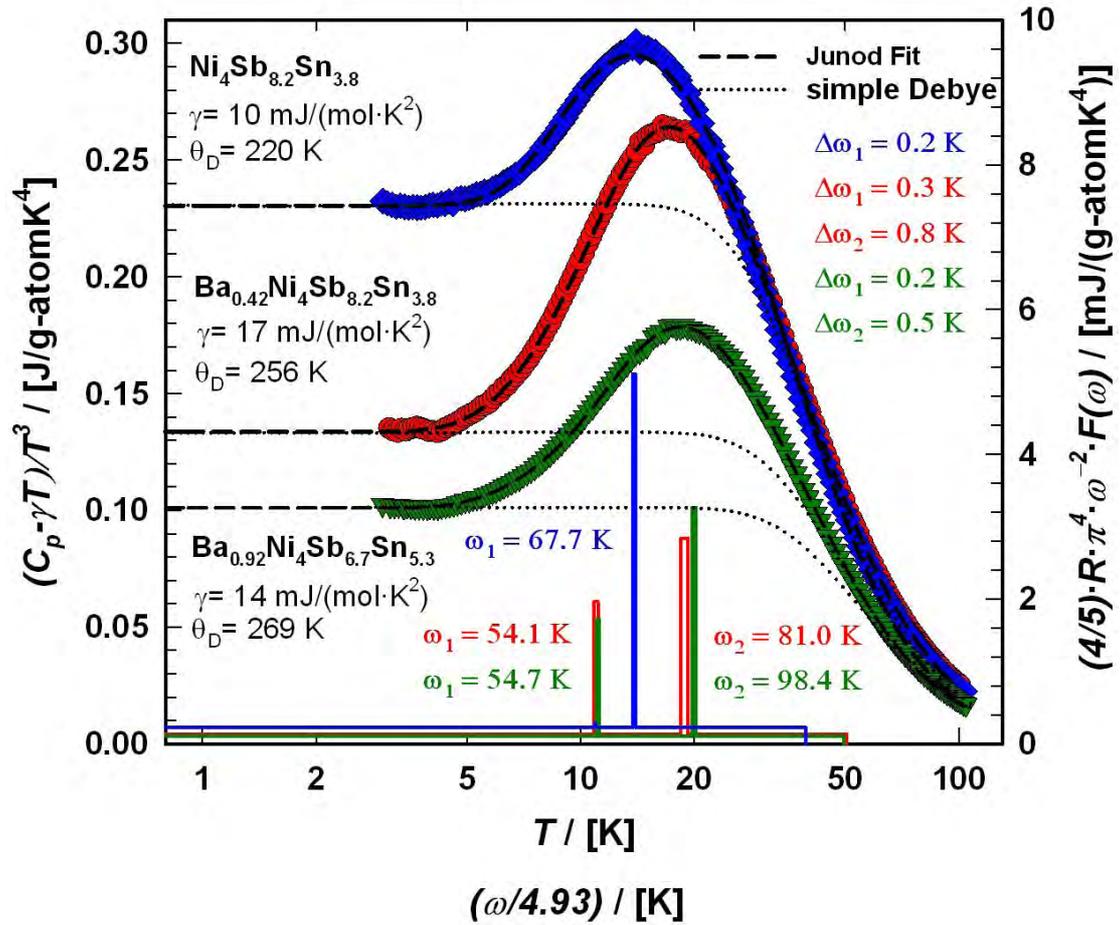

**Figure 5.2:** Temperature-dependent specific heat of $Ni_4Sb_{8.2}Sn_{3.8}$, $Ba_{0.42}Ni_4Sb_{8.2}Sn_{3.8}$ and $Ba_{0.92}Ni_4Sb_{6.7}Sn_{5.3}$ plotted as $(C_p-\gamma \cdot T)/T^3$ versus $\ln(T)$. Least square fits were made to the experimental data using the model indriduced by Junod et al. [Jun1, Jun2]. The blue, red and green lines (referring to the right axis) sketch te corresponding phonon spectral functions $F(\omega)$ plotted as $(4/5) \cdot R \cdot \pi^4 \cdot \omega^{-2} \cdot F(\omega)$ versus $\omega/4.93$ with $\omega$ in Kelvin. In comparison a simple Debye model is added as dotted lines.

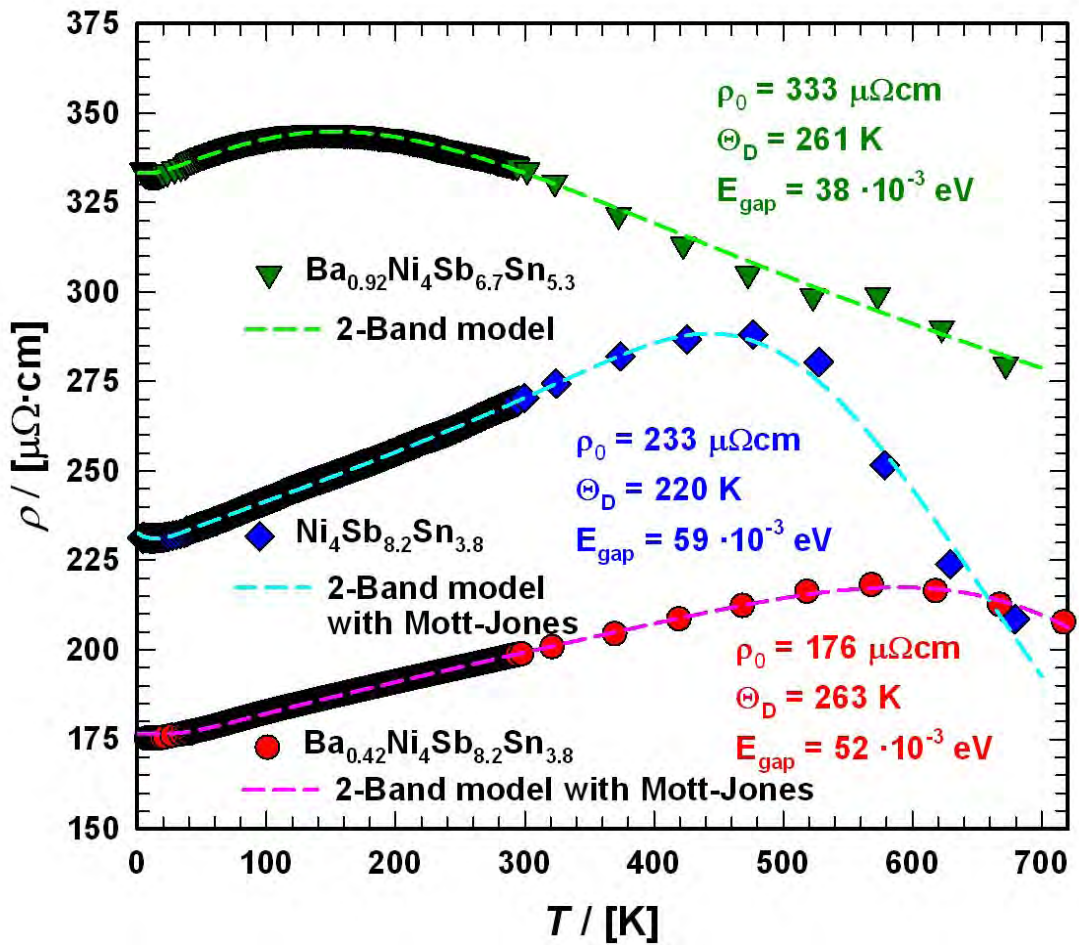

**Figure 6.1a:** Temperature-dependent electrical resistivities ρ of the skutterudites $Ni_4Sb_{8.2}Sn_{3.8}$, $Ba_{0.42}Ni_4Sb_{8.2}Sn_{3.8}$ and $Ba_{0.92}Ni_4Sb_{6.7}Sn_{5.3}$ a) in the temperature range from 4 K to 700 K and b) in comparison before and after SPD via HPT above room temperature. Dashed lines correspond to least squares fits according to a two band model.

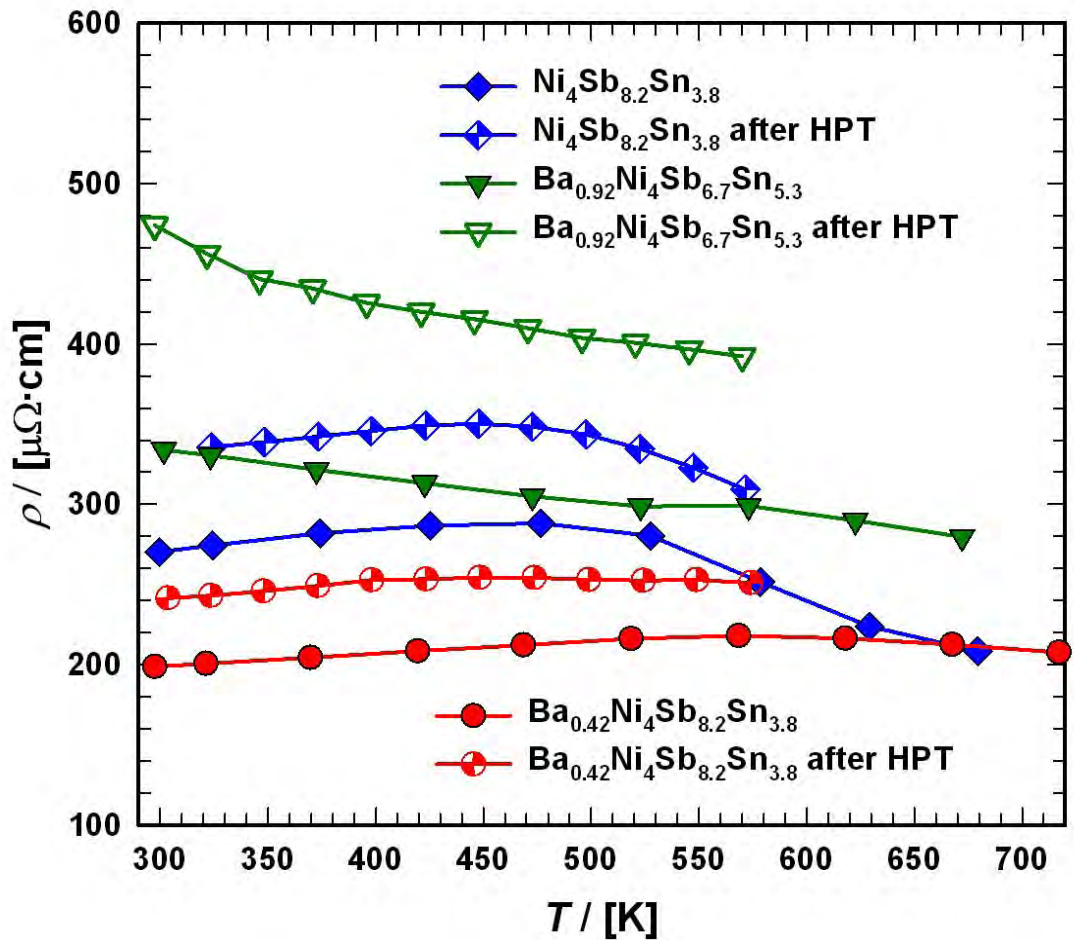

**Figure 6.1b:** Temperature-dependent electrical resistivities ρ of the skutterudites $Ni_4Sb_{8.2}Sn_{3.8}$, $Ba_{0.42}Ni_4Sb_{8.2}Sn_{3.8}$ and $Ba_{0.92}Ni_4Sb_{6.7}Sn_{5.3}$ a) in the temperature range from 4 K to 700 K and b) in comparison before and after SPD via HPT above room temperature. Dashed lines correspond to least squares fits according to a two band model.

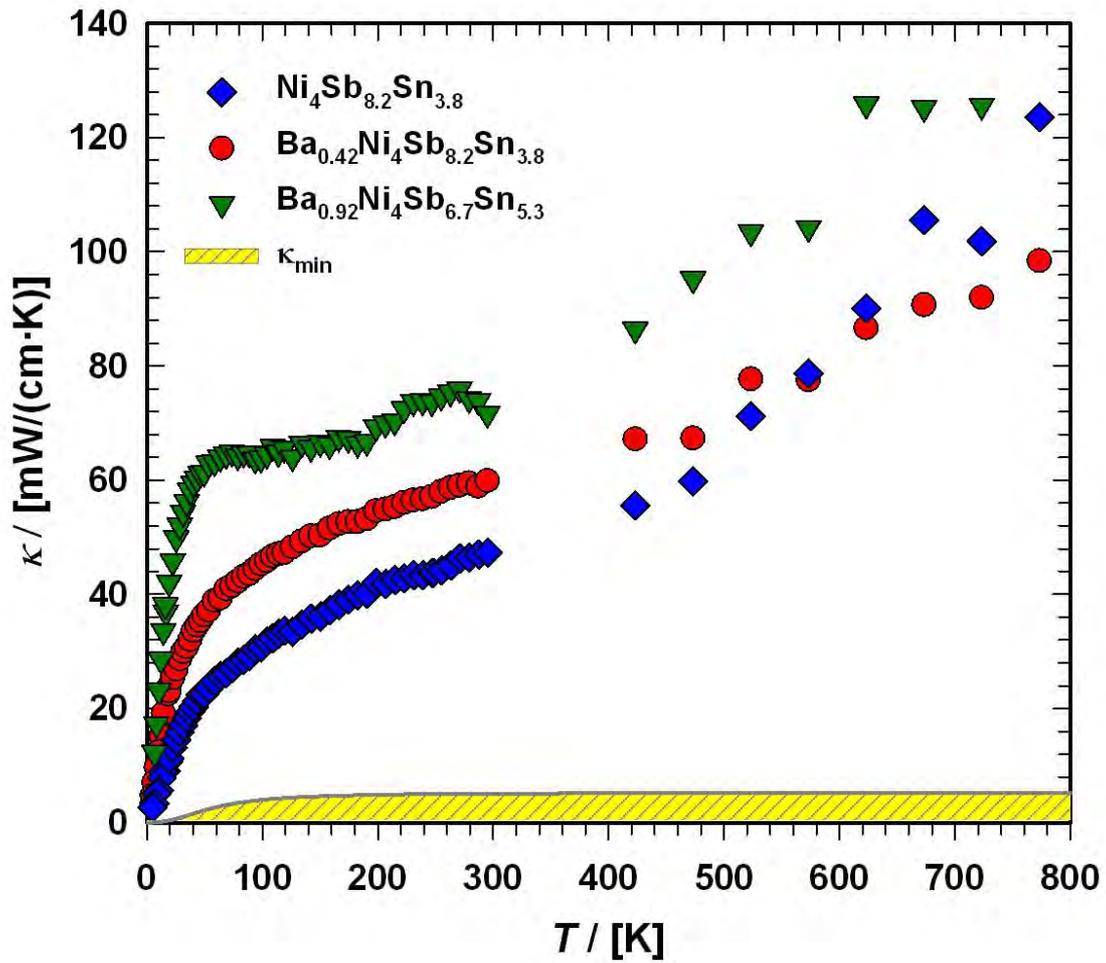

**Figure 6.2:** Temperature dependent thermal conductivity κ of the skutterudites Ni$_4$Sb$_{8.2}$Sn$_{3.8}$, Ba$_{0.42}$Ni$_4$Sb$_{8.2}$Sn$_{3.8}$ and Ba$_{0.92}$Ni$_4$Sb$_{6.7}$Sn$_{5.3}$ The shaded area represents the minimum thermal conductivity κ$_{min}$ for Ni-Sb-Sn based skutterudites.

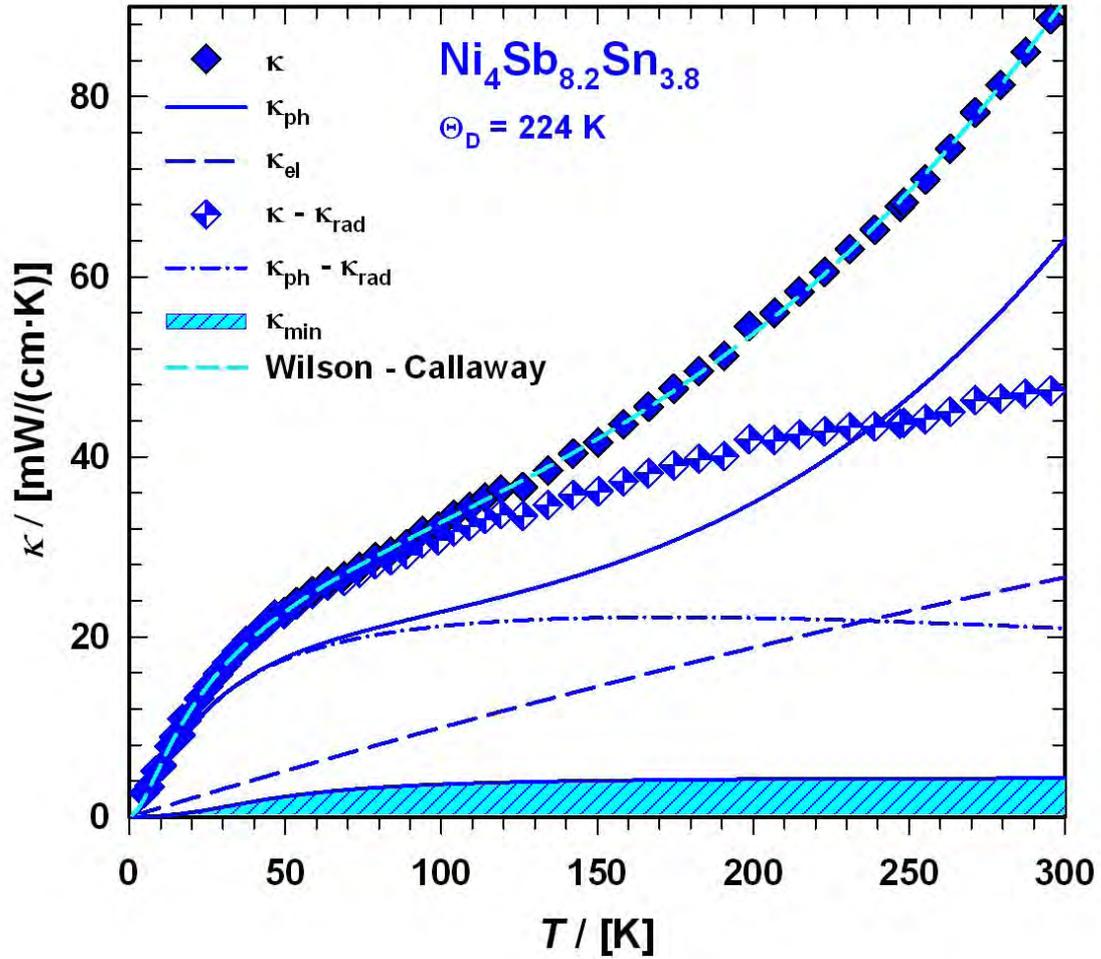

**Figure 6.3a:** Temperature dependent thermal conductivity κ of the skutterudites $Ni_4Sb_{8.2}Sn_{3.8}$, $Ba_{0.42}Ni_4Sb_{8.2}Sn_{3.8}$ and $Ba_{0.92}Ni_4Sb_{6.7}Sn_{5.3}$ at a), c), e) low temperatures and b), d), f) above room temperatures in comparison to the values measured after SPD via HPT. Temperature dependent thermal conductivity κ of the skutterudites $Ni_4Sb_{8.2}Sn_{3.8}$, $Ba_{0.42}Ni_4Sb_{8.2}Sn_{3.8}$ and $Ba_{0.92}Ni_4Sb_{6.7}Sn_{5.3}$ and the dashed lines are least squares fit according to a combination of the Wilson- and the Callaway-model.

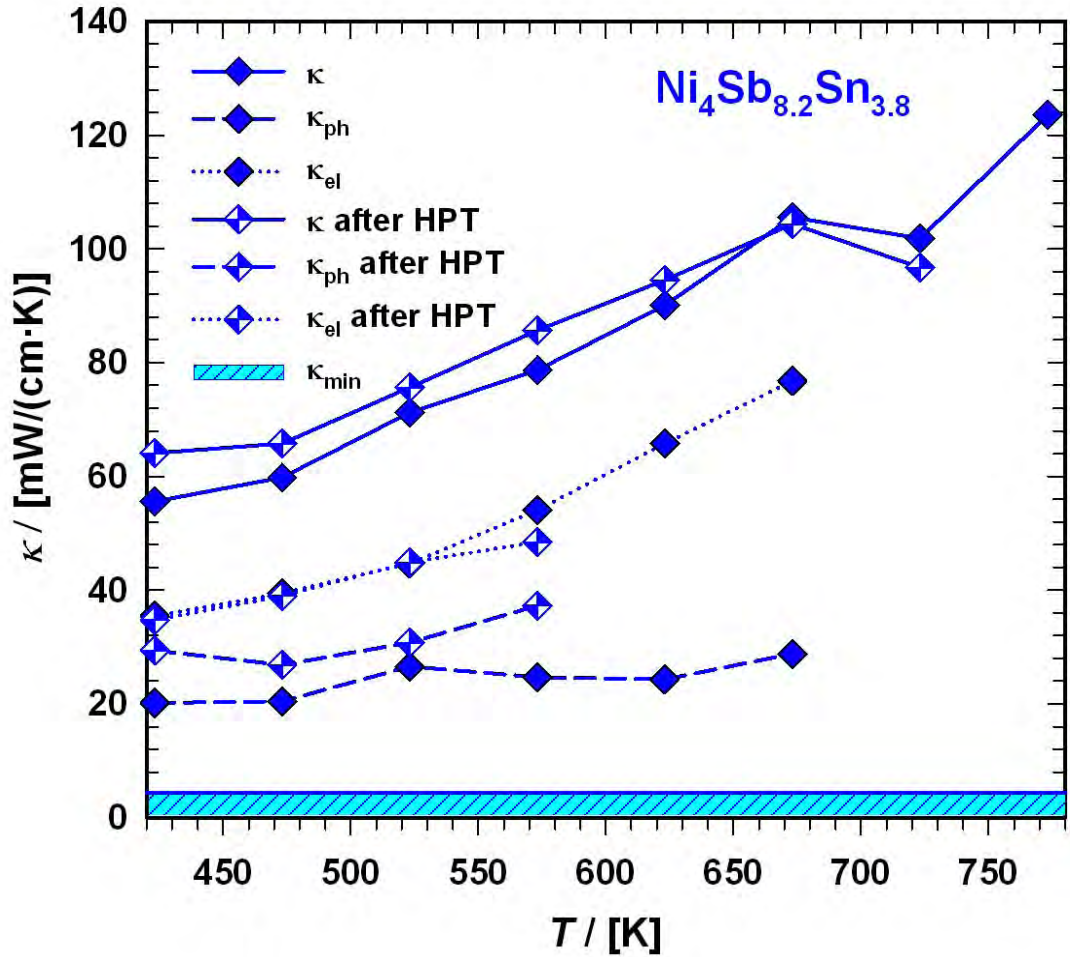

**Figure 6.3b:** Temperature dependent thermal conductivity κ of the skutterudites $Ni_4Sb_{8.2}Sn_{3.8}$, $Ba_{0.42}Ni_4Sb_{8.2}Sn_{3.8}$ and $Ba_{0.92}Ni_4Sb_{6.7}Sn_{5.3}$ at a), c), e) low temperatures and b), d), f) above room temperatures in comparison to the values measured after SPD via HPT. Temperature dependent thermal conductivity κ of the skutterudites $Ni_4Sb_{8.2}Sn_{3.8}$, $Ba_{0.42}Ni_4Sb_{8.2}Sn_{3.8}$ and $Ba_{0.92}Ni_4Sb_{6.7}Sn_{5.3}$ and the dashed lines are least squares fit according to a combination of the Wilson- and the Callaway-model.

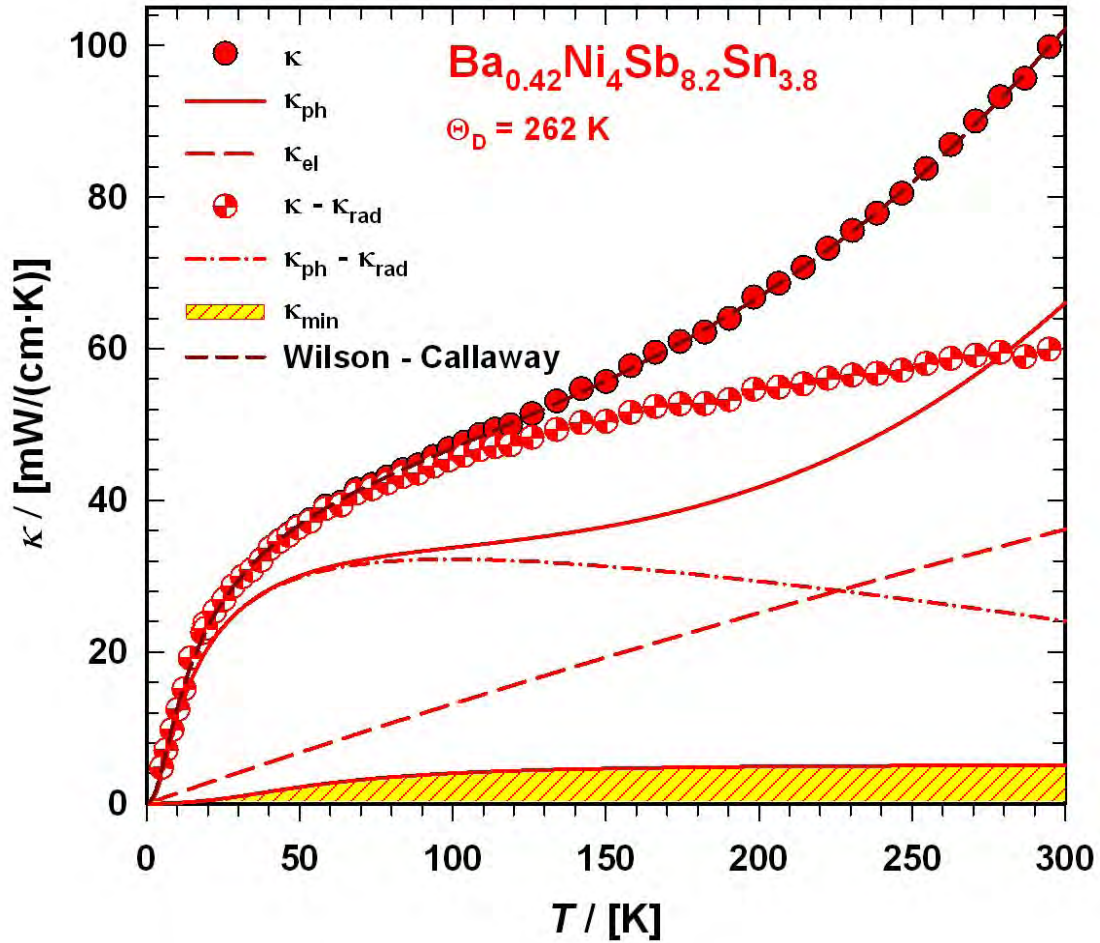

**Figure 6.3c:** Temperature dependent thermal conductivity κ of the skutterudites $Ni_4Sb_{8.2}Sn_{3.8}$, $Ba_{0.42}Ni_4Sb_{8.2}Sn_{3.8}$ and $Ba_{0.92}Ni_4Sb_{6.7}Sn_{5.3}$ at a), c), e) low temperatures and b), d), f) above room temperatures in comparison to the values measured after SPD via HPT. Temperature dependent thermal conductivity κ of the skutterudites $Ni_4Sb_{8.2}Sn_{3.8}$, $Ba_{0.42}Ni_4Sb_{8.2}Sn_{3.8}$ and $Ba_{0.92}Ni_4Sb_{6.7}Sn_{5.3}$ and the dashed lines are least squares fit according to a combination of the Wilson- and the Callaway-model.

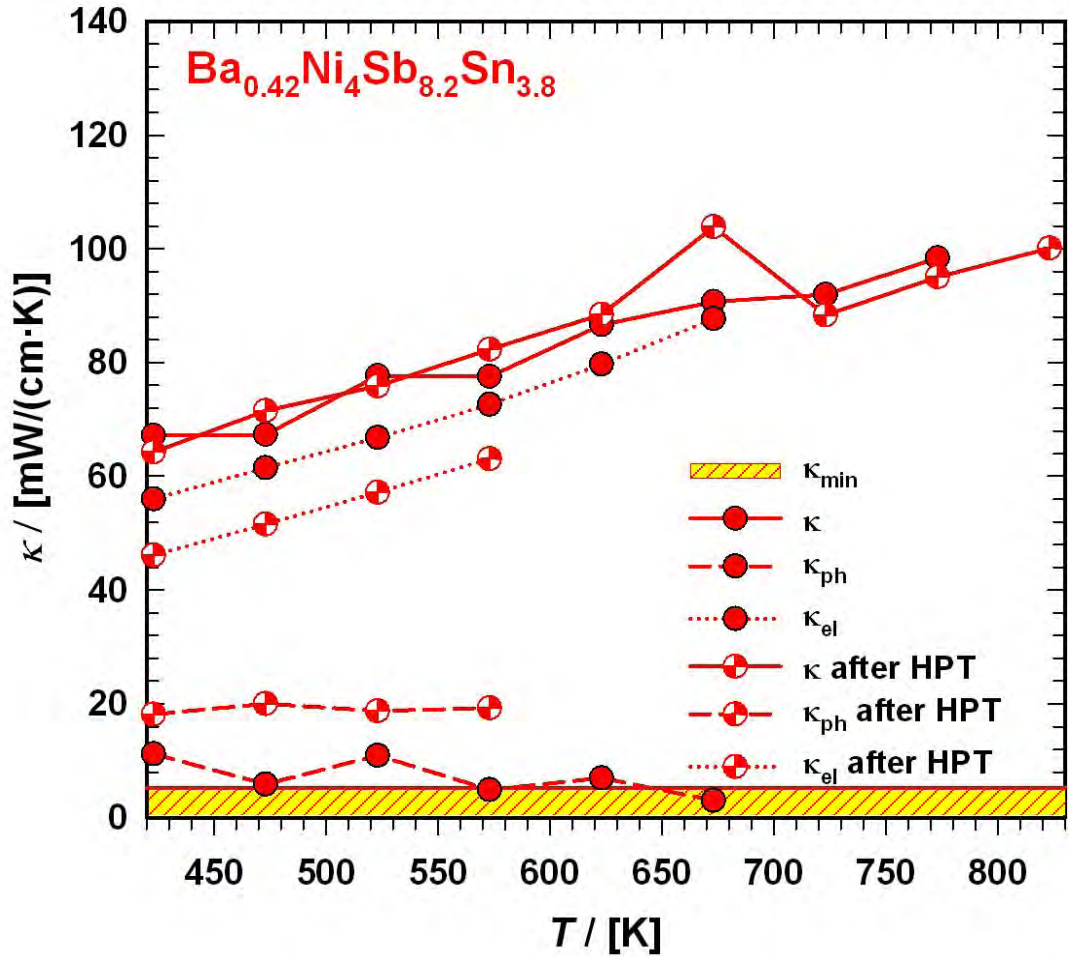

**Figure 6.3d:** Temperature dependent thermal conductivity κ of the skutterudites $Ni_4Sb_{8.2}Sn_{3.8}$, $Ba_{0.42}Ni_4Sb_{8.2}Sn_{3.8}$ and $Ba_{0.92}Ni_4Sb_{6.7}Sn_{5.3}$ at a), c), e) low temperatures and b), d), f) above room temperatures in comparison to the values measured after SPD via HPT. Temperature dependent thermal conductivity κ of the skutterudites $Ni_4Sb_{8.2}Sn_{3.8}$, $Ba_{0.42}Ni_4Sb_{8.2}Sn_{3.8}$ and $Ba_{0.92}Ni_4Sb_{6.7}Sn_{5.3}$ and the dashed lines are least squares fit according to a combination of the Wilson- and the Callaway-model.

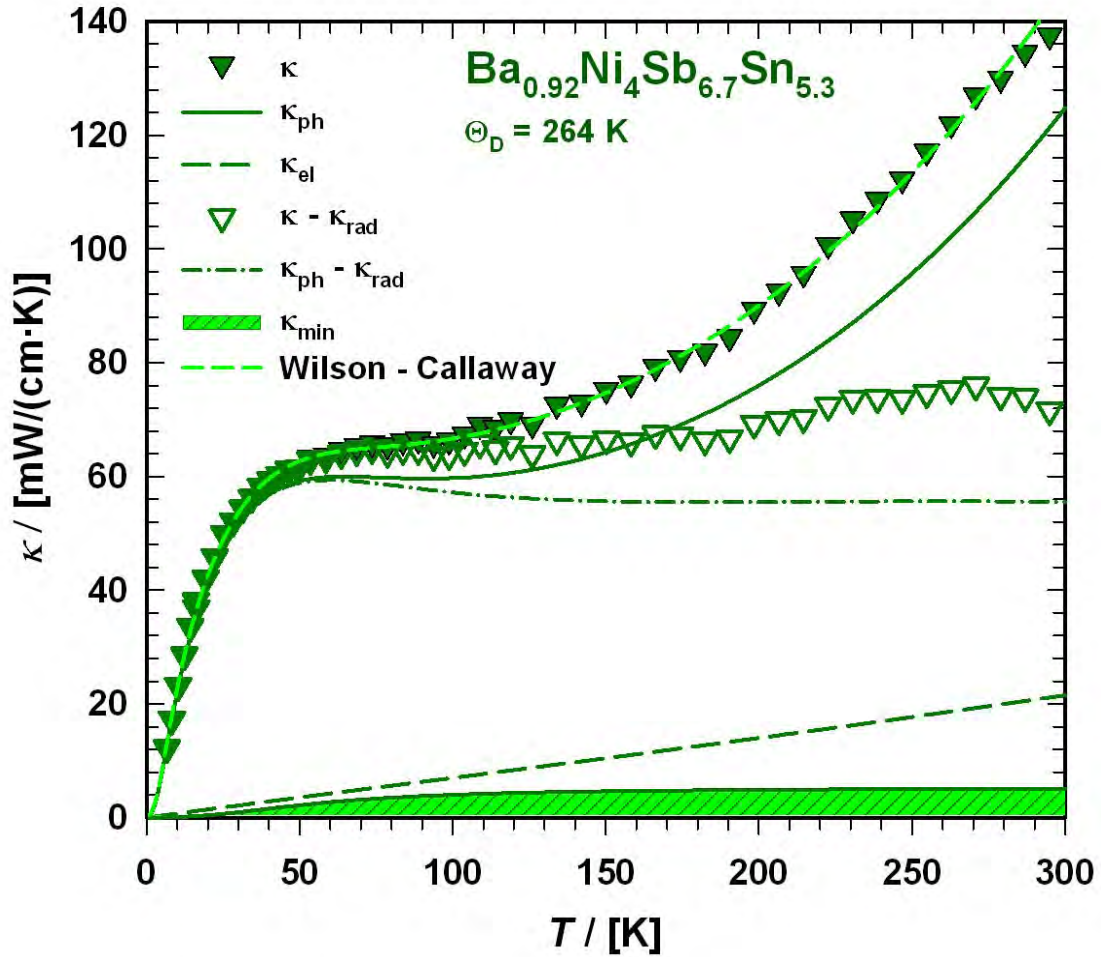

**Figure 6.3e:** Temperature dependent thermal conductivity κ of the skutterudites $Ni_4Sb_{8.2}Sn_{3.8}$, $Ba_{0.42}Ni_4Sb_{8.2}Sn_{3.8}$ and $Ba_{0.92}Ni_4Sb_{6.7}Sn_{5.3}$ at a), c), e) low temperatures and b), d), f) above room temperatures in comparison to the values measured after SPD via HPT. Temperature dependent thermal conductivity κ of the skutterudites $Ni_4Sb_{8.2}Sn_{3.8}$, $Ba_{0.42}Ni_4Sb_{8.2}Sn_{3.8}$ and $Ba_{0.92}Ni_4Sb_{6.7}Sn_{5.3}$ and the dashed lines are least squares fit according to a combination of the Wilson- and the Callaway-model.

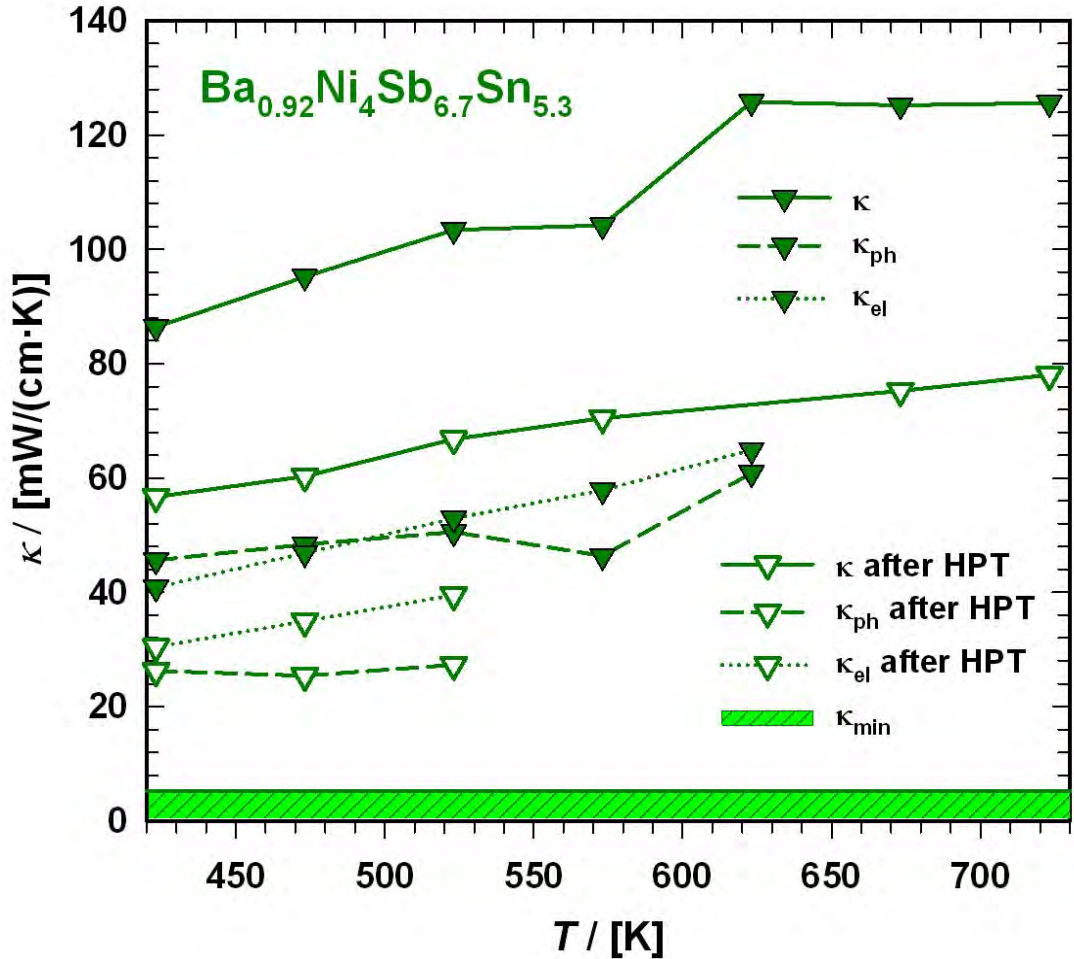

**Figure 6.3f:** Temperature dependent thermal conductivity κ of the skutterudites $Ni_4Sb_{8.2}Sn_{3.8}$, $Ba_{0.42}Ni_4Sb_{8.2}Sn_{3.8}$ and $Ba_{0.92}Ni_4Sb_{6.7}Sn_{5.3}$ at a), c), e) low temperatures and b), d), f) above room temperatures in comparison to the values measured after SPD via HPT. Temperature dependent thermal conductivity κ of the skutterudites $Ni_4Sb_{8.2}Sn_{3.8}$, $Ba_{0.42}Ni_4Sb_{8.2}Sn_{3.8}$ and $Ba_{0.92}Ni_4Sb_{6.7}Sn_{5.3}$ and the dashed lines are least squares fit according to a combination of the Wilson- and the Callaway-model.

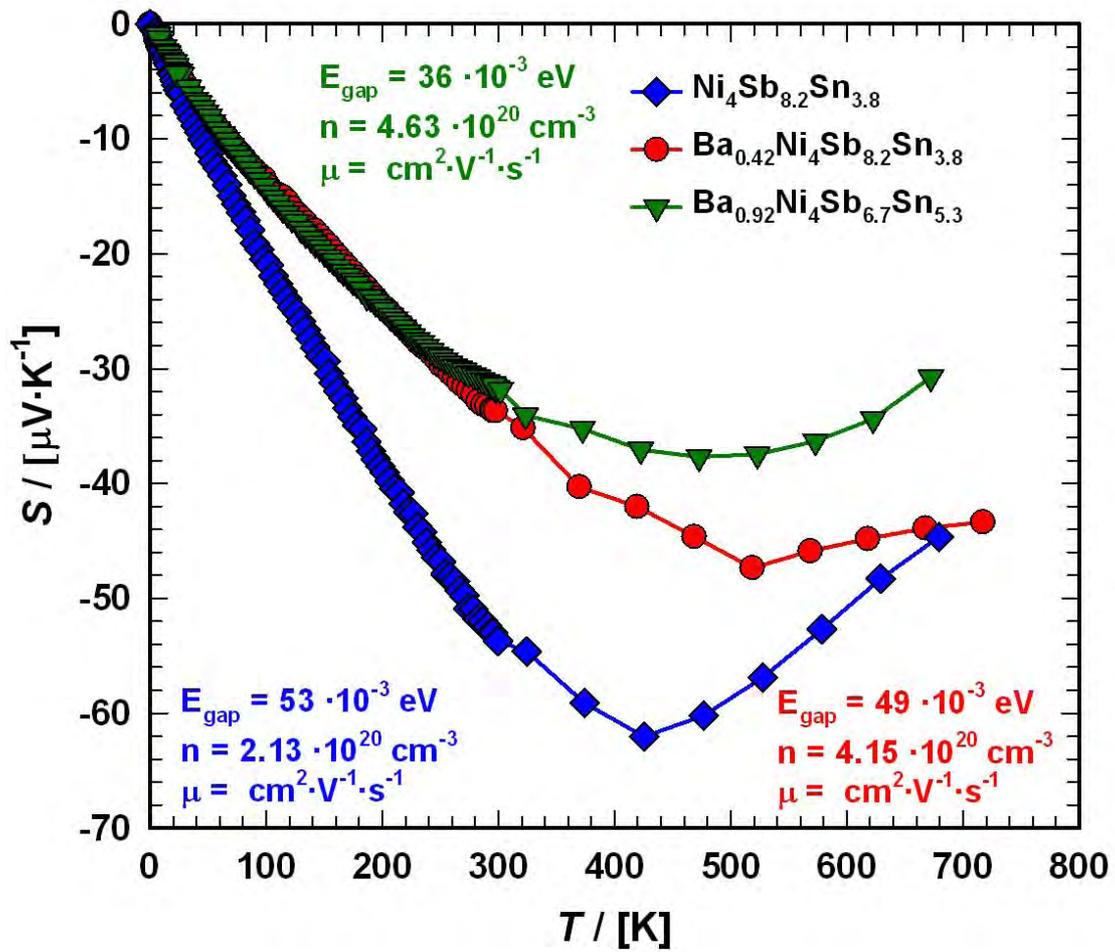

**Figure 6.4a:** Temperature dependent Seebeck S coefficient of the skutterudites $Ni_4Sb_{8.2}Sn_{3.8}$, $Ba_{0.42}Ni_4Sb_{8.2}Sn_{3.8}$ and $Ba_{0.92}Ni_4Sb_{6.7}Sn_{5.3}$ a) in the temperature range from 4 K to 700 K and b) in comparison before and after SPD via HPT above room temperature.

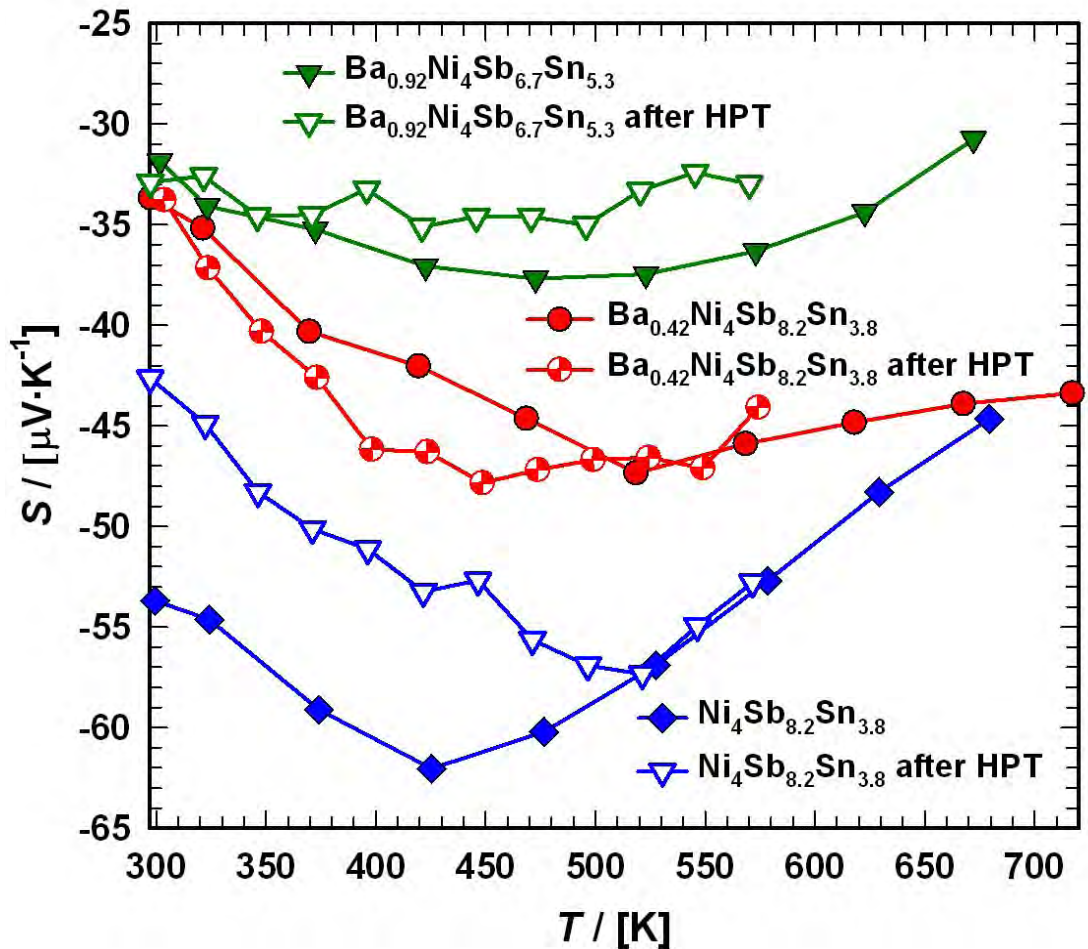

**Figure 6.4b:** Temperature dependent Seebeck S coefficient of the skutterudites $Ni_4Sb_{8.2}Sn_{3.8}$, $Ba_{0.42}Ni_4Sb_{8.2}Sn_{3.8}$ and $Ba_{0.92}Ni_4Sb_{6.7}Sn_{5.3}$ a) in the temperature range from 4 K to 700 K and b) in comparison before and after SPD via HPT above room temperature.

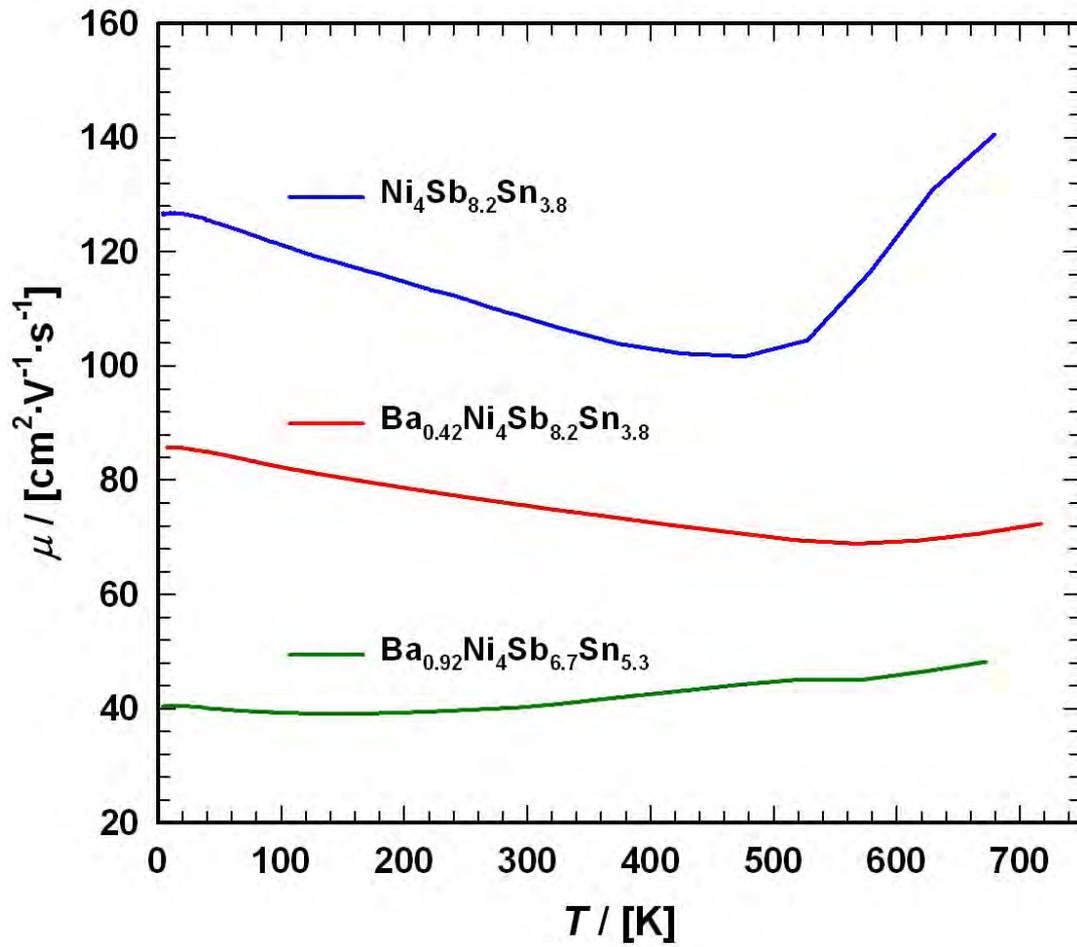

**Figure 6.5:** Temperature dependent charge carrier mobility μ of the skutterudites $Ni_4Sb_{8.2}Sn_{3.8}$, $Ba_{0.42}Ni_4Sb_{8.2}Sn_{3.8}$ and $Ba_{0.92}Ni_4Sb_{6.7}Sn_{5.3}$.

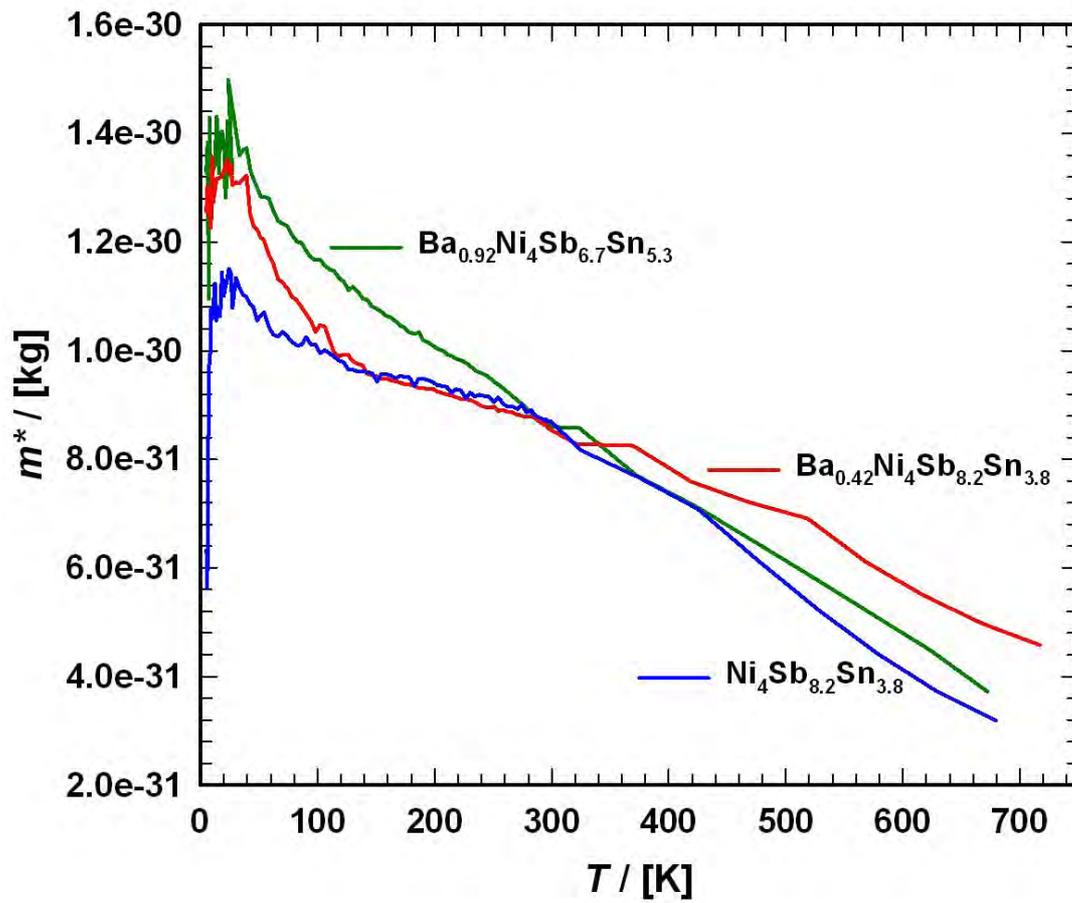

**Figure 6.6:** Temperature dependent charge carrier effective mass $m^*$ of the skutterudites $Ni_4Sb_{8.2}Sn_{3.8}$, $Ba_{0.42}Ni_4Sb_{8.2}Sn_{3.8}$ and $Ba_{0.92}Ni_4Sb_{6.7}Sn_{5.3}$.

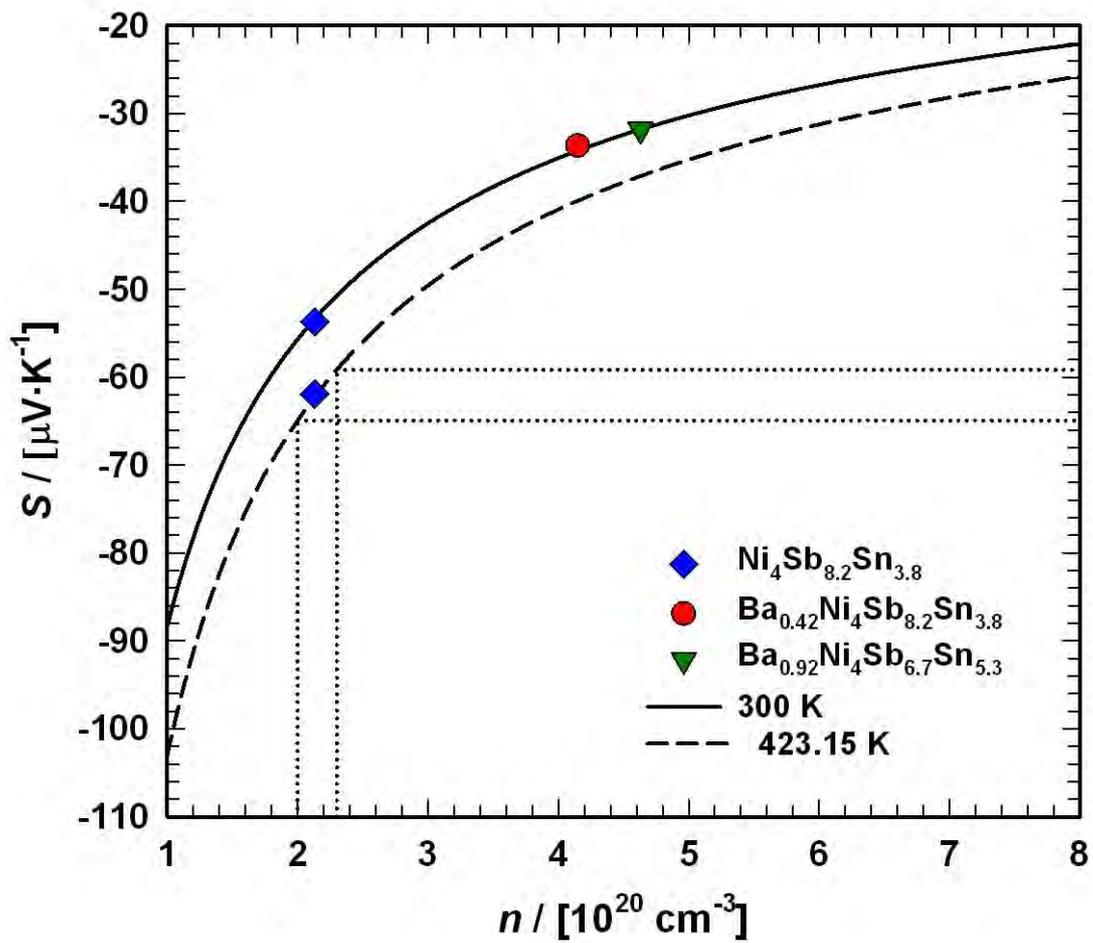

**Figure 6.7:** Pisarenko-plot of the Seebeck coefficient S versus number of charge carriers n.

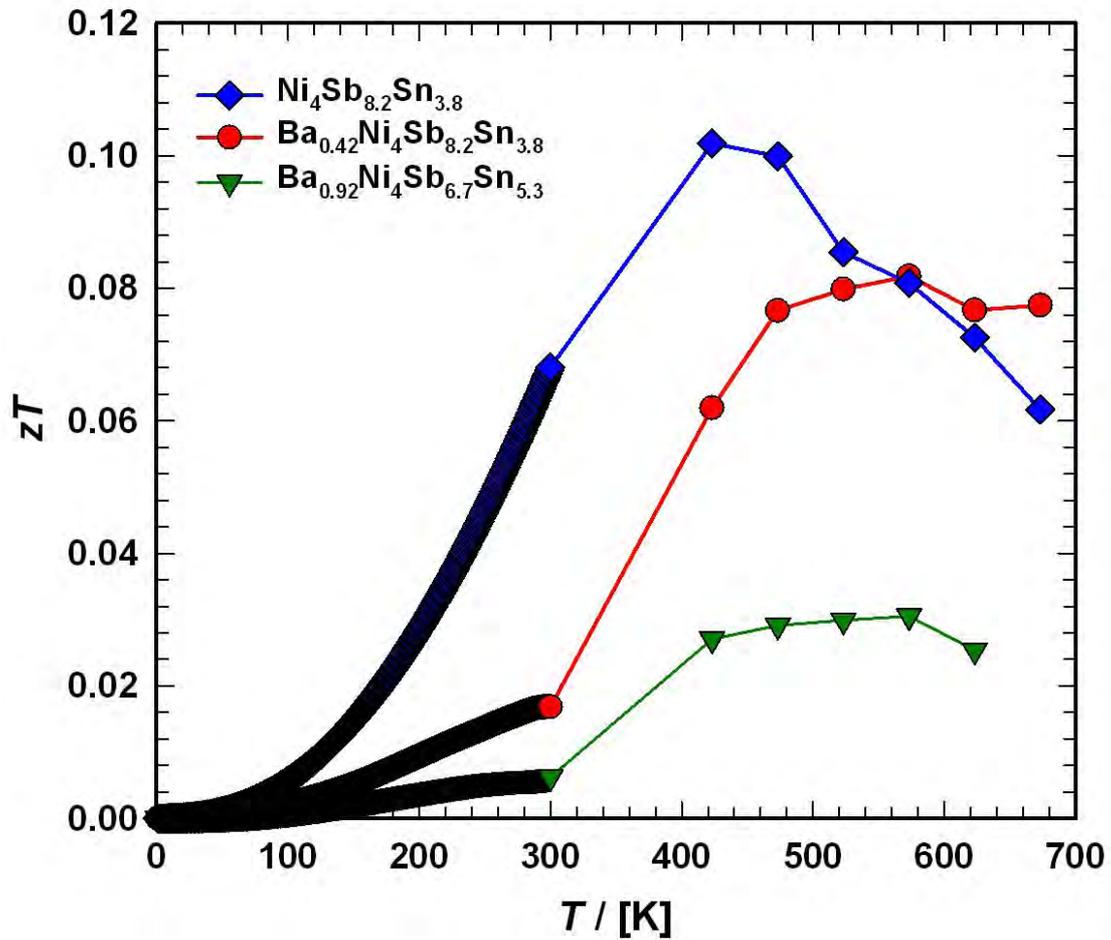

**Figure 6.8a:** Temperature dependent figure of merit zT of the skutterudites Ni$_4$Sb$_{8.2}$Sn$_{3.8}$, Ba$_{0.42}$Ni$_4$Sb$_{8.2}$Sn$_{3.8}$ and Ba$_{0.92}$Ni$_4$Sb$_{6.7}$Sn$_{5.3}$ a) in the temperature range from 4 K to 700 K and b) in comparison before and after SPD via HPT above room temperature.

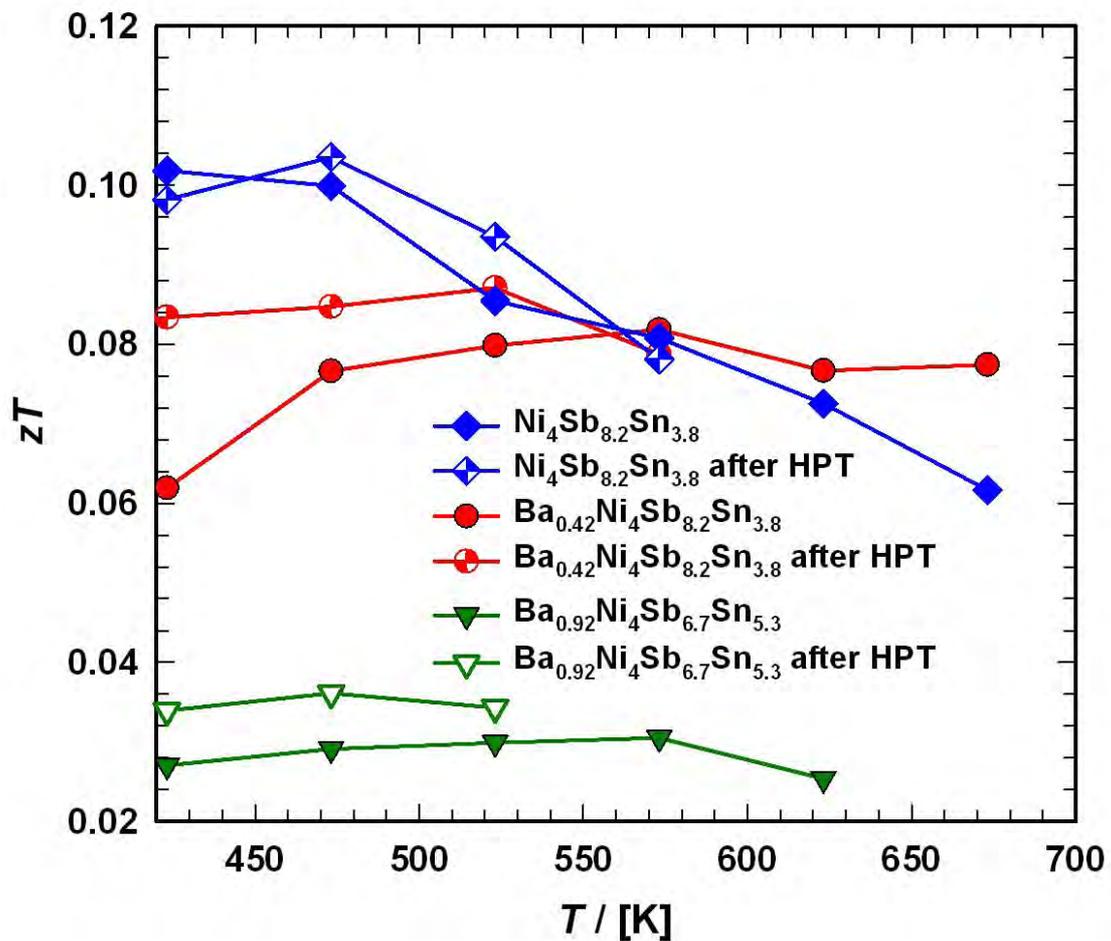

**Figure 6.8b:** Temperature dependent figure of merit zT of the skutterudites $Ni_4Sb_{8.2}Sn_{3.8}$, $Ba_{0.42}Ni_4Sb_{8.2}Sn_{3.8}$ and $Ba_{0.92}Ni_4Sb_{6.7}Sn_{5.3}$ a) in the temperature range from 4 K to 700 K and b) in comparison before and after SPD via HPT above room temperature.

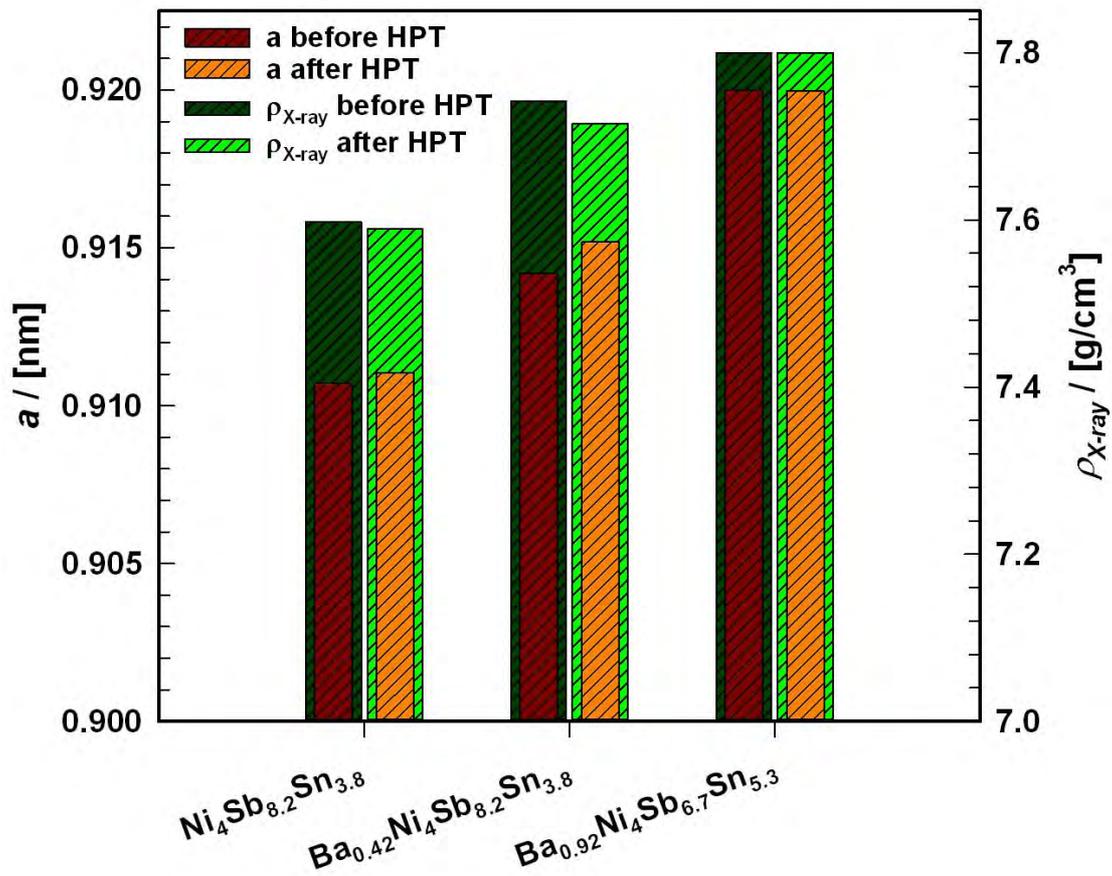

**Figure 7.1:** Comparison of lattice parameters a and corresponding X-ray densities $\rho_{X\text{-ray}}$ for the skutterudites $Ni_4Sb_{82}Sn_{3.8}$, $Ba_{0.42}Ni_4Sb_{8.2}Sn_{3.8}$ and $Ba_{0.92}Ni_4Sb_{6.7}Sn_{5.3}$ before and after SPD via HPT.

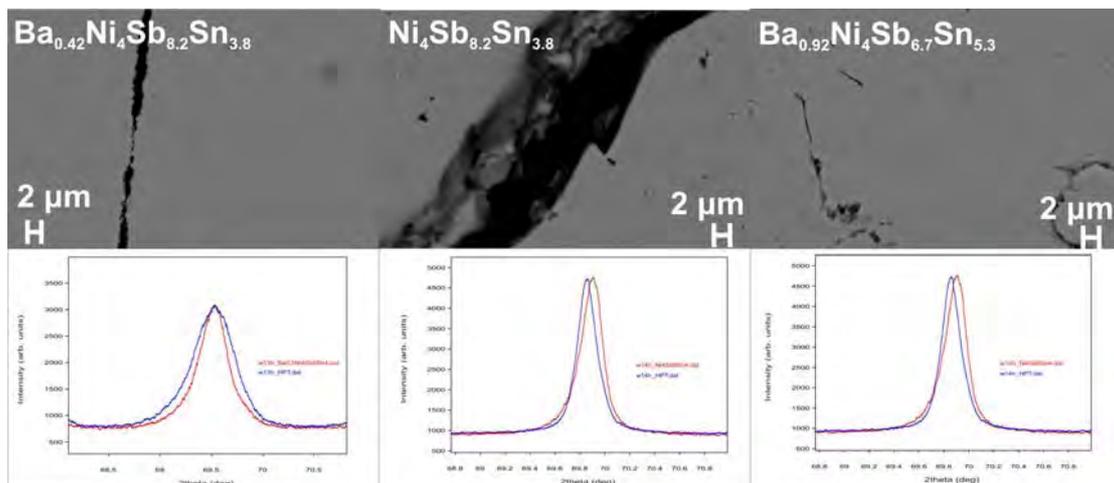

**Figure 7.2:** Microstructure and X-ray profile of the skutterudite samples $Ni_4Sb_{8.2}Sn_{3.8}$, $Ba_{0.42}Ni_4Sb_{8.2}Sn_{3.8}$ and $Ba_{0.92}Ni_4Sb_{6.7}Sn_{5.3}$ after SPD via HPT.

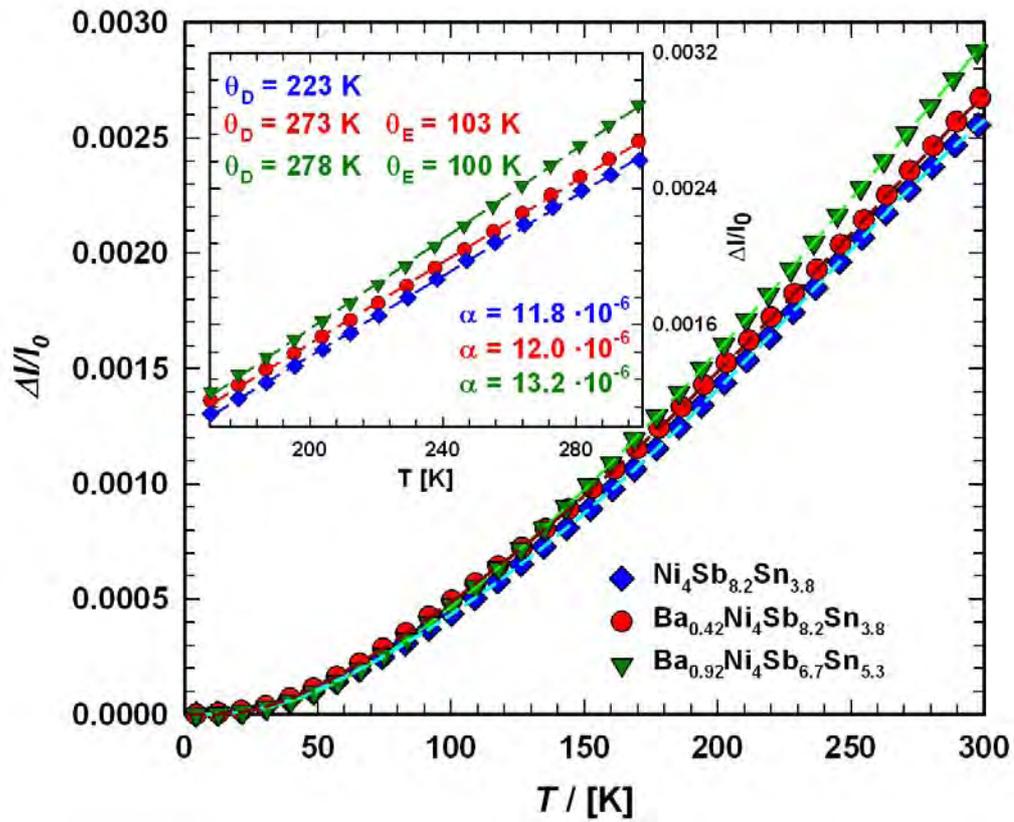

**Figure 8.1:** Temperature dependent thermal expansion $\Delta l/l_0$ of the skutterudites $Ni_4Sb_{8.2}Sn_{3.8}$, $Ba_{0.42}Ni_4Sb_{8.2}Sn_{3.8}$ and $Ba_{0.92}Ni_4Sb_{6.7}Sn_{5.3}$ in the temperature range from 4.2 K to 300 K.

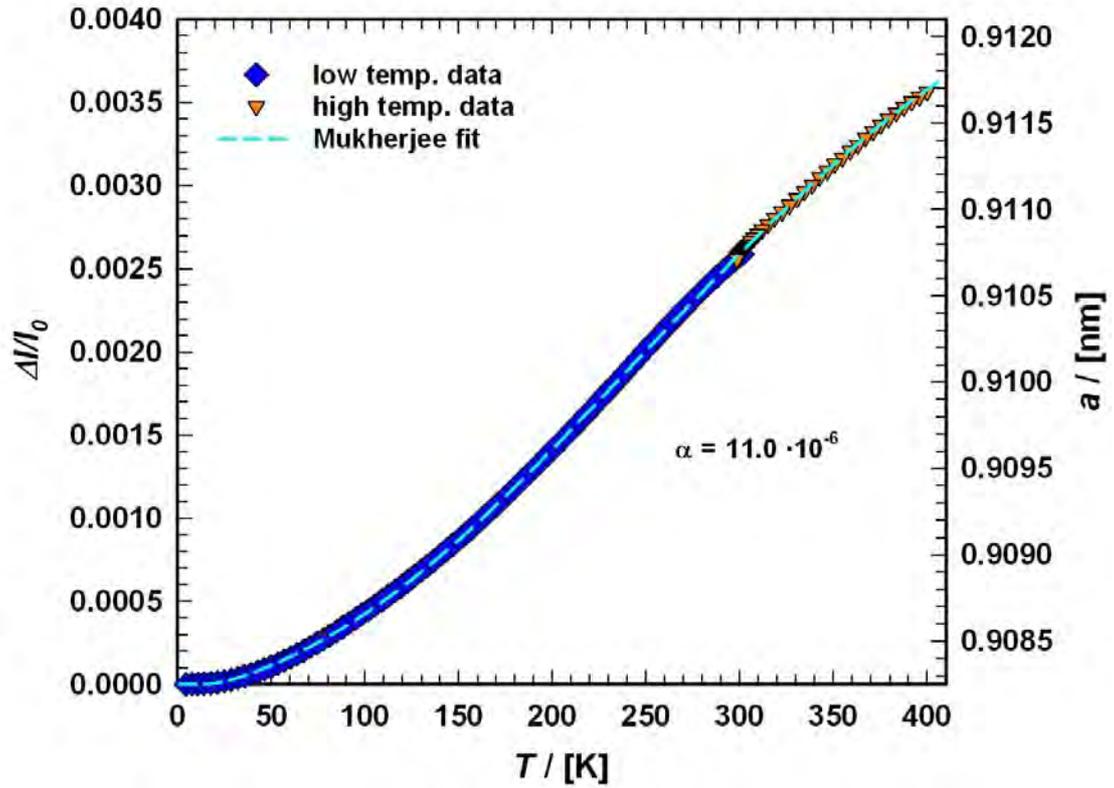

**Figure 8.2a:** Combination of low and high temperature thermal expansion $\Delta l/l_0$ data of the skutterudites a) $Ni_4Sb_{8.2}Sn_{3.8}$, b) $Ba_{0.42}Ni_4Sb_{8.2}Sn_{3.8}$ and c) $Ba_{0.92}Ni_4Sb_{6.7}Sn_{5.3}$. Dashed lines corresponded to least square fits according to Eqn. 8.3.

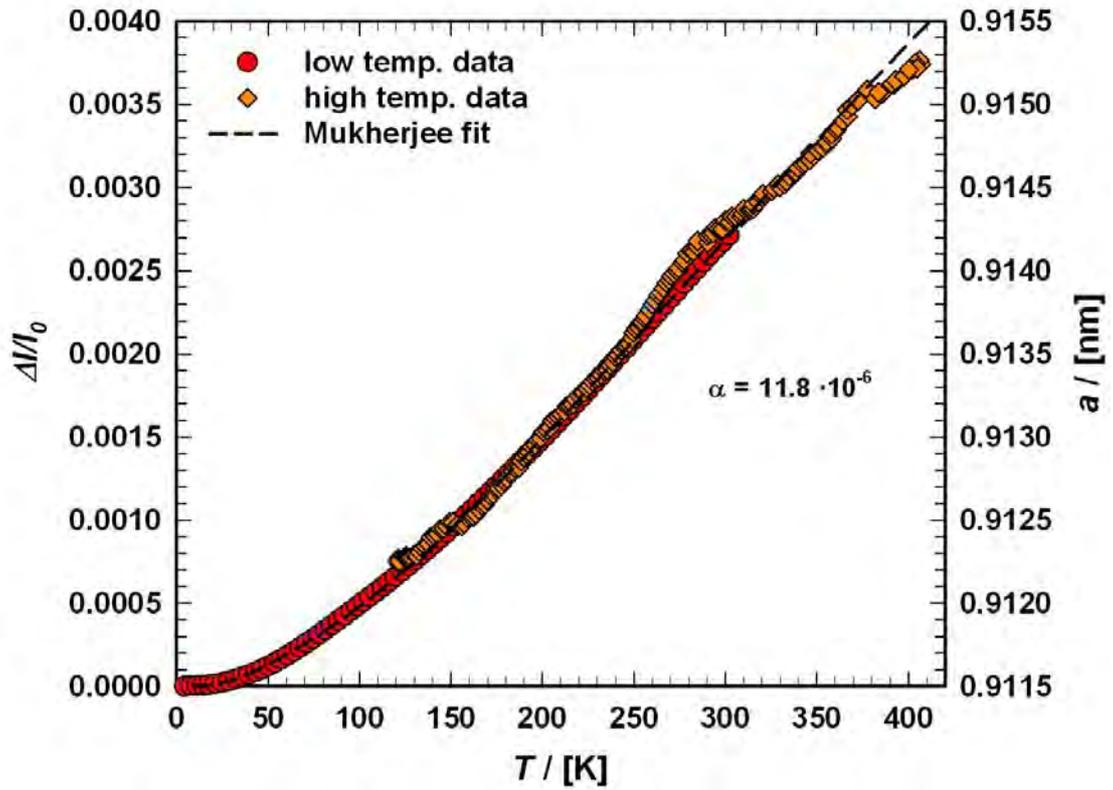

**Figure 8.2b:** Combination of low and high temperature thermal expansion $\Delta l/l_0$ data of the skutterudites a) $Ni_4Sb_{8.2}Sn_{3.8}$, b) $Ba_{0.42}Ni_4Sb_{8.2}Sn_{3.8}$ and c) $Ba_{0.92}Ni_4Sb_{6.7}Sn_{5.3}$. Dashed lines corresponded to least square fits according to Eqn. 8.3.

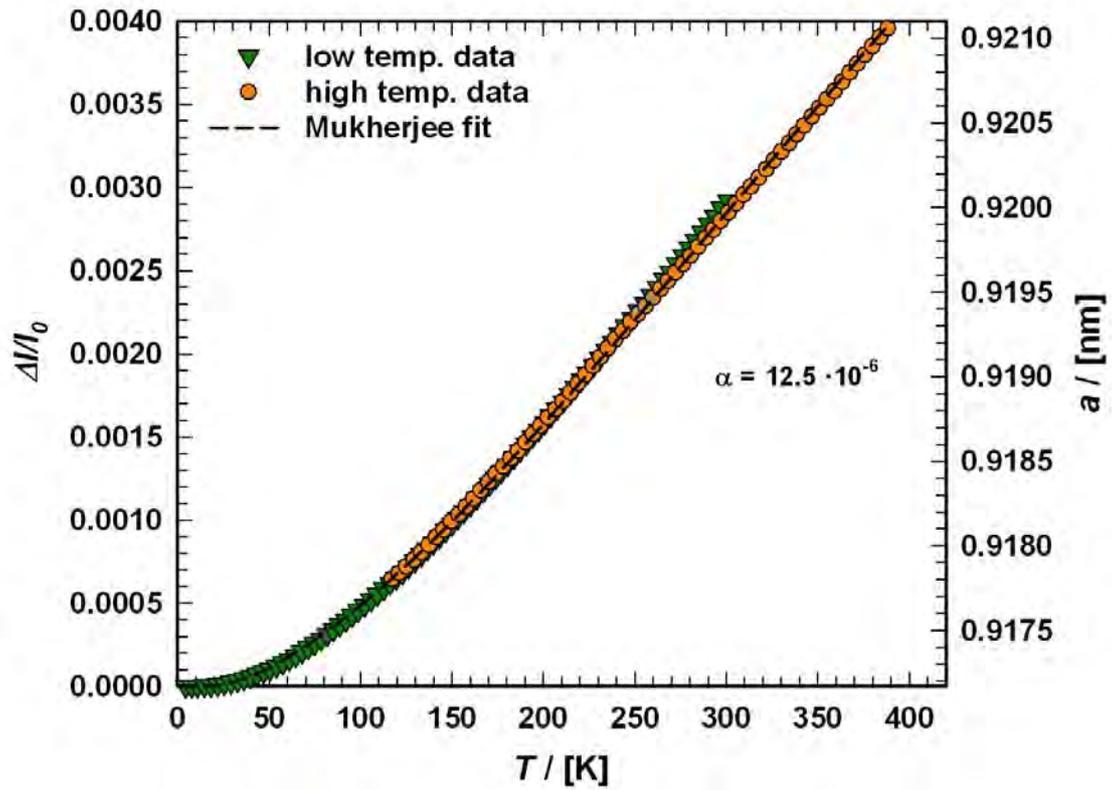

**Figure 8.2c:** Combination of low and high temperature thermal expansion $\Delta l/l_0$ data of the skutterudites a) $Ni_4Sb_{8.2}Sn_{3.8}$, b) $Ba_{0.42}Ni_4Sb_{8.2}Sn_{3.8}$ and c) $Ba_{0.92}Ni_4Sb_{6.7}Sn_{5.3}$. Dashed lines corresponded to least square fits according to Eqn. 8.3.

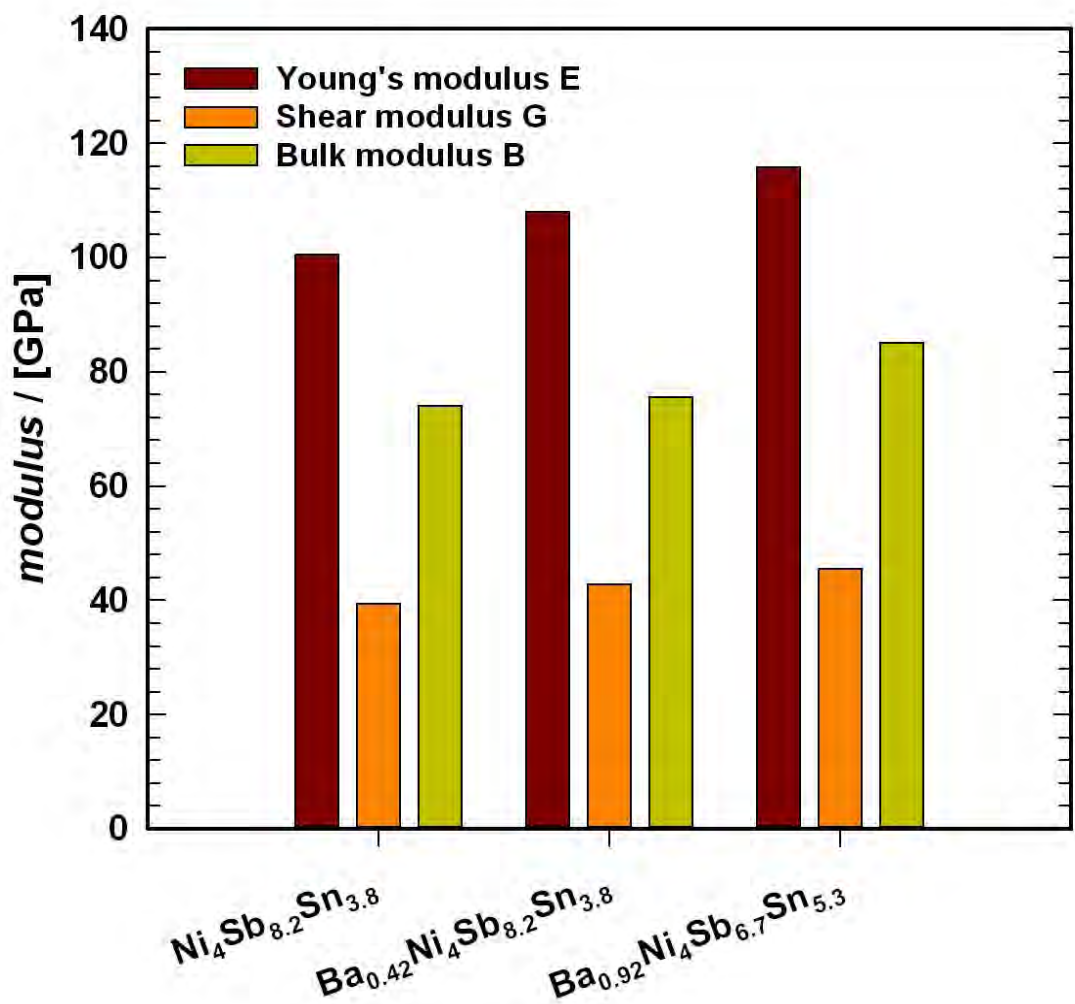

**Figure 9.1:** Comparison of the Young's-, Shear- and Bulk-modulus for the skutterudites $Ni_4Sb_{8.2}Sn_{3.8}$, $Ba_{0.42}Ni_4Sb_{8.2}Sn_{3.8}$ and $Ba_{0.92}Ni_4Sb_{6.7}Sn_{5.3}$.

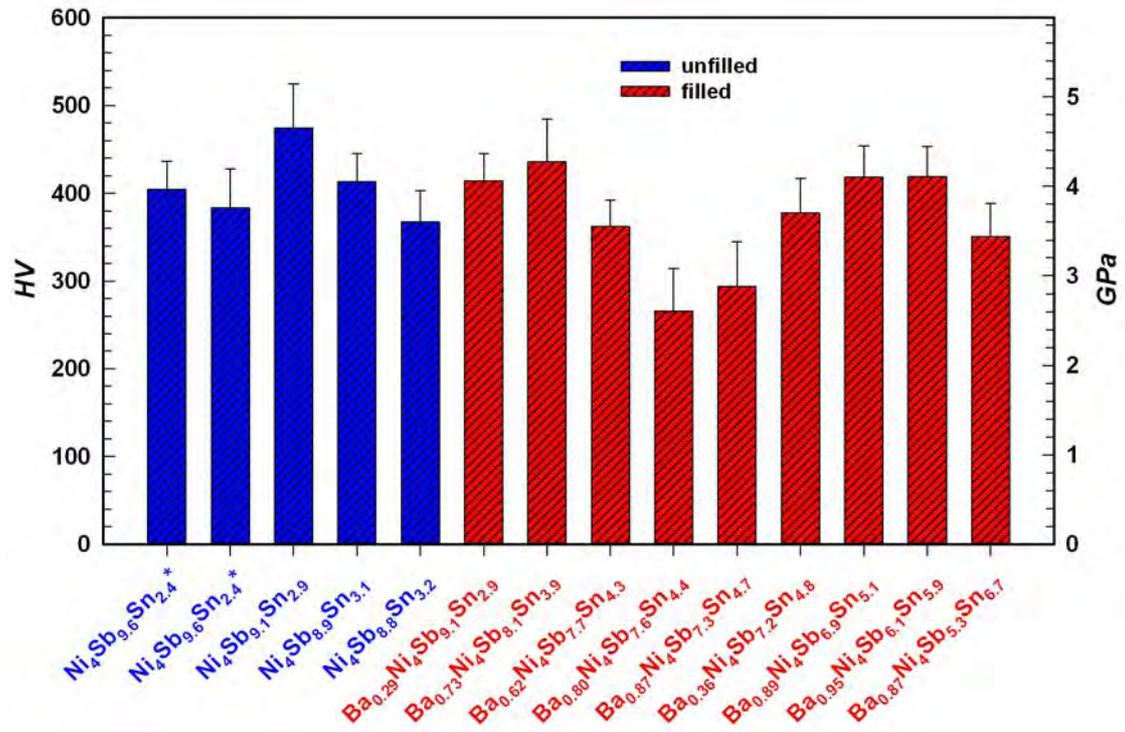

**Figure 10.1:** Comparison of Vicker's hardness HV for various filled and unfilled Ni-Sn-Sb based skutterudites. ([*] skutterudite composition in two different alloys)

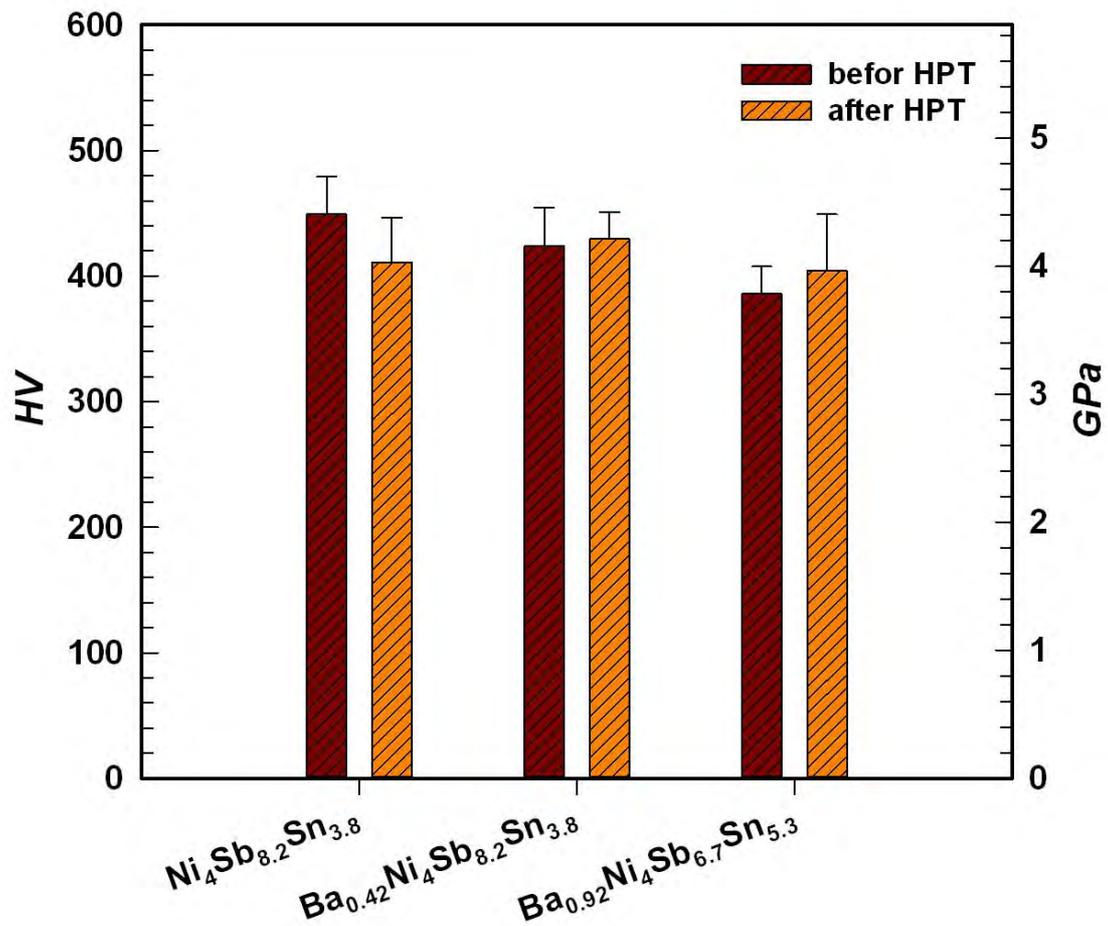

**Figure 10.2:** Comparison of Vicker's hardness HV for the skutterudite single phase samples $Ni_4Sb_{8.2}Sn_{3.8}$, $Ba_{0.42}Ni_4Sb_{8.2}Sn_{3.8}$ and $Ba_{0.92}Ni_4Sb_{6.7}Sn_{5.3}$.

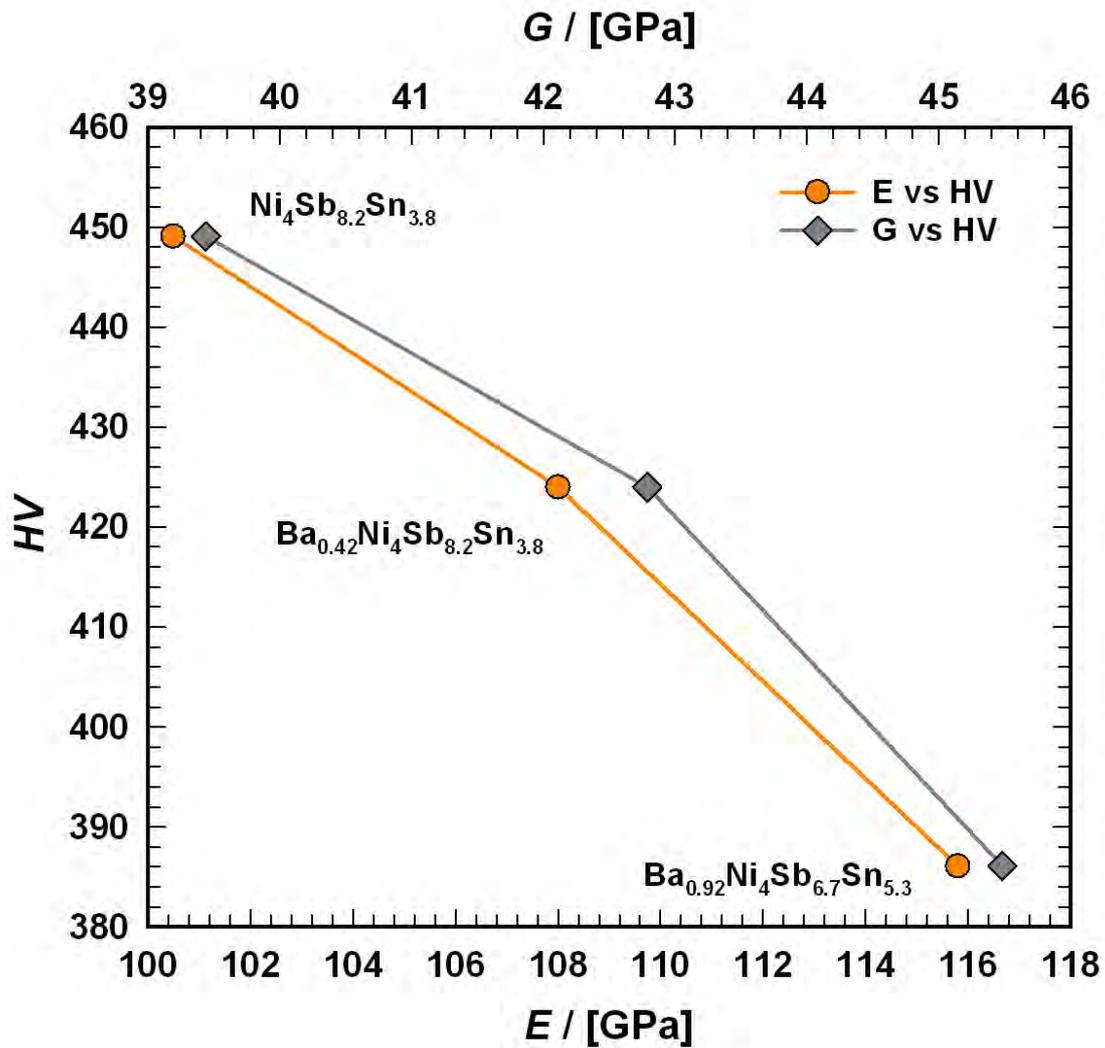

**Figure 10.3:** Comparison of Young's- and Shear-modulus in dependence of Vicker's hardness HV for the skutterudites $Ni_4Sb_{8.2}Sn_{3.8}$, $Ba_{0.42}Ni_4Sb_{8.2}Sn_{3.8}$ and $Ba_{0.92}Ni_4Sb_{6.7}Sn_{5.3}$.

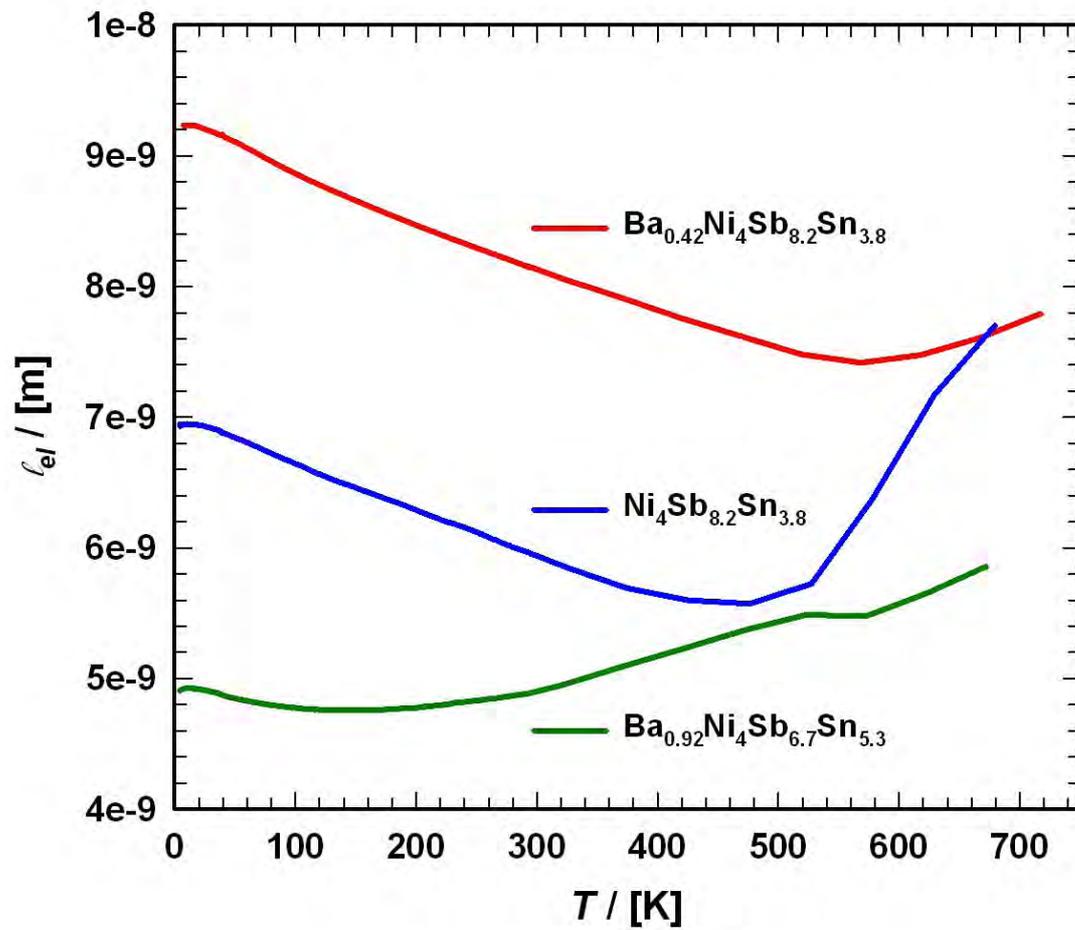

**Figure 11.1:** Temperature dependent electron mean free path lengths $\ell_{el}$ of the skutterudites $Ni_4Sb_{8.2}Sn_{3.8}$, $Ba_{0.42}Ni_4Sb_{8.2}Sn_{3.8}$ and $Ba_{0.92}Ni_4Sb_{6.7}Sn_{5.3}$.

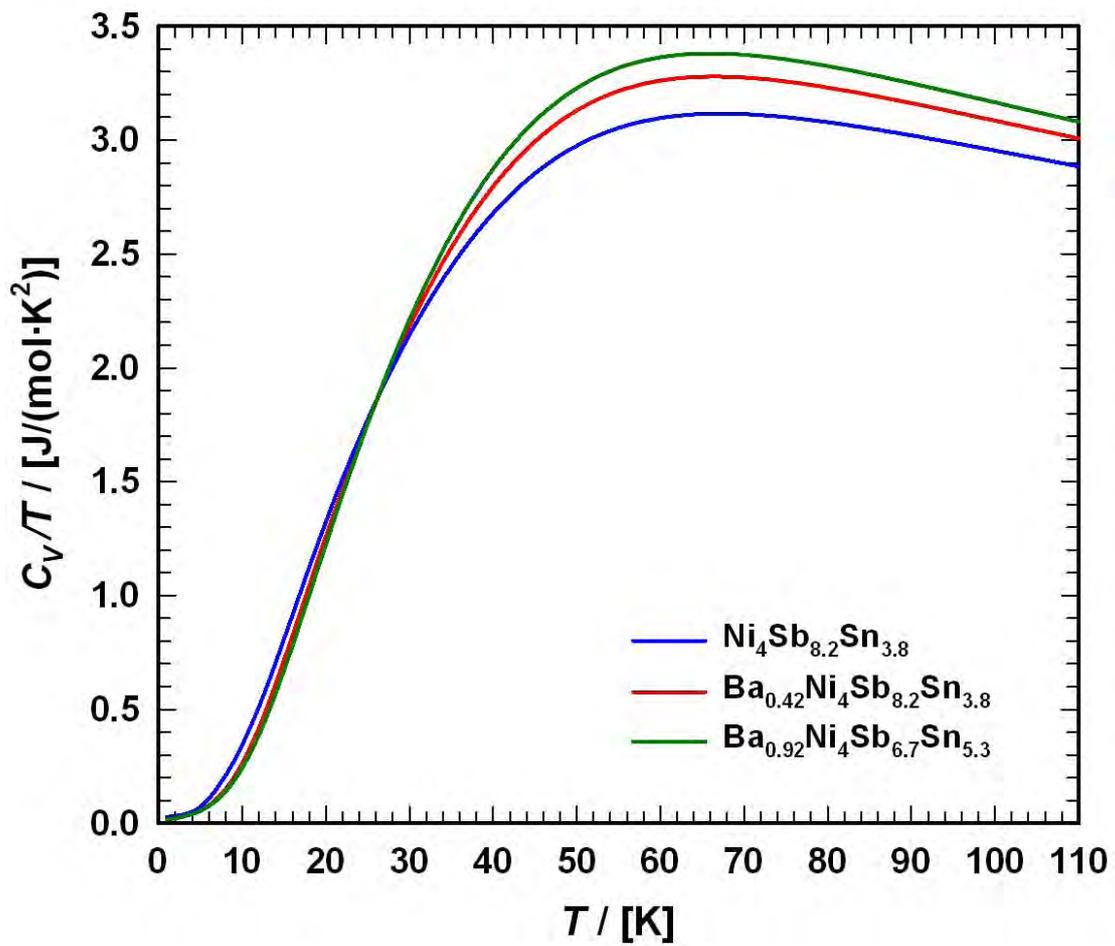

**Figure 11.2:** Temperature-dependent specific heat divided by temperature $C_V/T$ of $Ni_4Sb_{8.2}Sn_{3.8}$, $Ba_{0.42}Ni_4Sb_{8.2}Sn_{3.8}$ and $Ba_{0.92}Ni_4Sb_{6.7}Sn_{5.3}$.

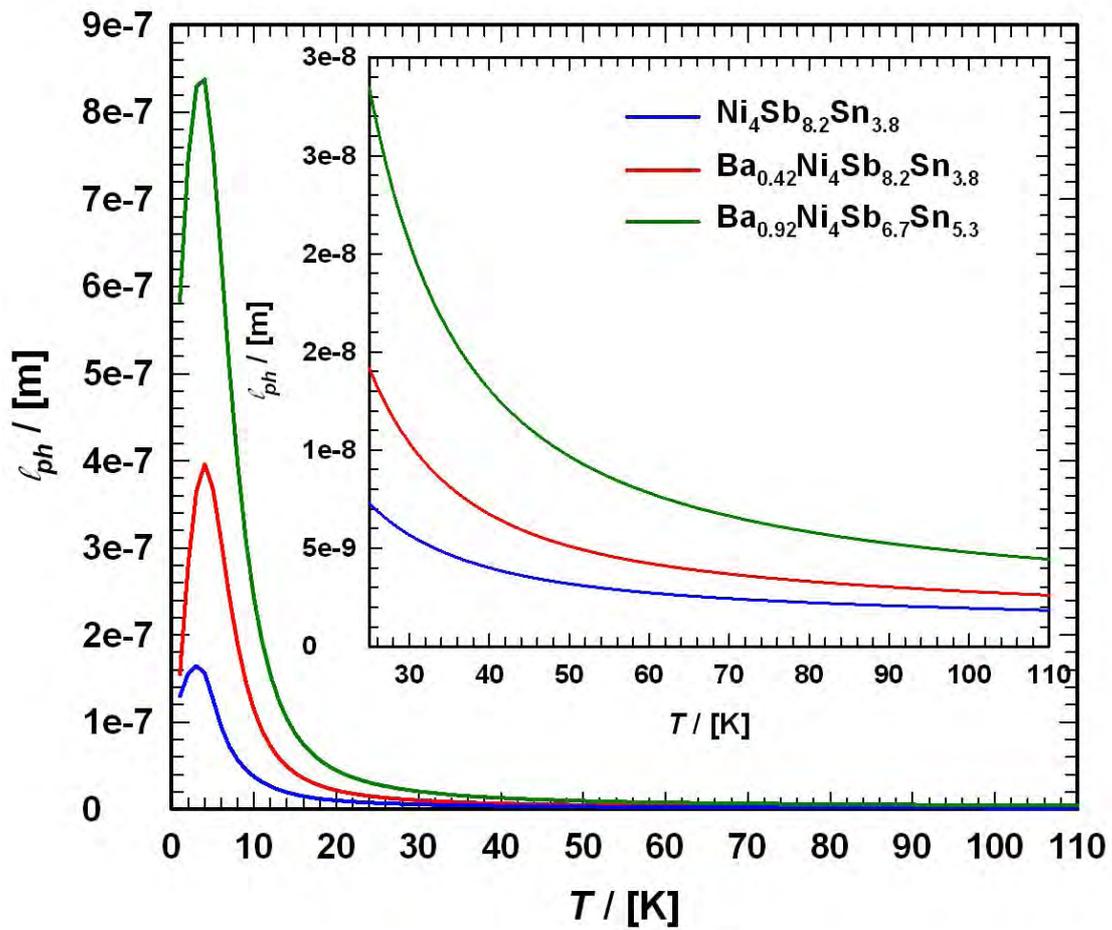

**Figure 11.3:** Temperature dependent electron mean free path lengths $\ell_{ph}$ of the skutterudites $Ni_4Sb_{8.2}Sn_{3.8}$, $Ba_{0.42}Ni_4Sb_{8.2}Sn_{3.8}$ and $Ba_{0.92}Ni_4Sb_{6.7}Sn_{5.3}$.